%% file: main.tex
\documentclass[acmsmall]{acmart}

\AtBeginDocument{%
  \providecommand\BibTeX{{%
\normalfont B\kern-0.5em{\scshape i\kern-0.25em b}\kern-0.8em\TeX}}}

\usepackage{multirow}
\usepackage{makecell}
\usepackage{xspace,mfirstuc,tabulary}
\usepackage{booktabs}
\usepackage{graphicx}
\usepackage{bbding}
\usepackage{xcolor,colortbl}
\usepackage{adjustbox}
\setcitestyle{numbers}
\usepackage{fontawesome5}
\usepackage{hyperref}
\usepackage{pifont}
\usepackage{xcolor}
\definecolor{hidden-black}{rgb}{0,0,0}
\newrobustcmd{\B}{\bfseries}
\newcommand*\colourcheck[1]{%
  \expandafter\newcommand\csname #1check\endcsname{\textcolor{#1}{\ding{52}}}%
}
\newcommand*\colourcross[1]{%
  \expandafter\newcommand\csname #1cross\endcsname{\textcolor{#1}{\ding{55}}}%
}
\colourcheck{green}
\colourcross{red}

\usepackage{tabularx}
\usepackage[most]{tcolorbox}
\newtcolorbox{highlighted}{
  colback=yellow!50!white, colframe=yellow!50!white, boxrule=0pt, sharp corners, left=0pt, right=0pt, top=0pt, bottom=0pt
}

\definecolor{cGreen}{RGB}{0,150,0} % 深绿色
\usepackage{color,soul}
\soulregister{\textcolor}{2}
\soulregister{\cite}7
\soulregister{\citep}7
\soulregister{\citet}7
\soulregister{\ref}7

\usepackage[switch]{lineno}
\sethlcolor{yellow}

\usepackage[edges]{forest}
\definecolor{rootblue}{RGB}{157, 195, 230}
\definecolor{level1green}{RGB}{207, 238, 206}
\definecolor{level2orange}{RGB}{251, 213, 192}
\forestset{
  mindmap style/.style={
    for tree={
      grow'=east,
      parent anchor=east, child anchor=west,
      edge path={
        \noexpand\path[\forestoption{edge}]
        (!u.parent anchor) -- ++(20pt,0) |- (.child anchor)\forestoption{edge label};
      },
      l sep+=2.5em,
      s sep+=1.2em,
      font=\sffamily,
      rounded corners=4pt,
      draw=gray!60,
      line width=0.8pt,
      inner sep=10pt,
      fill=white,
      anchor=west,
      edge={draw=gray!80, line width=1pt},
      text ragged,
      align=left,
    },
    where level=0{
      fill=rootblue,
      text width=14cm,
      align=center,
      font=\sffamily\bfseries\large,
      draw=rootblue!80!black
    }{},
    where level=1{
      fill=level1green,
      text width=4.5cm,
      align=center,
      font=\sffamily\bfseries,
      draw=level1green!80!black
    }{},
    where level=2{
      fill=level2orange,
      text width=4.5cm,
      align=left,
      draw=level2orange!80!black
    }{},
  }
}

\setcopyright{acmlicensed}
\copyrightyear{2026}
\acmYear{2026}
\acmDOI{XXXXXXX.XXXXXXX}
\acmConference[Conference acronym 'XX]{Make sure to enter the correct
conference title from your rights confirmation email}{June 03--05,
2018}{Woodstock, NY}

\begin{document}

\title{The Evolution of Tool Use in LLM Agents: From Single-Tool Call to Multi-Tool Orchestration}

\author{Haoyuan Xu}\authornote{Equal contribution}
\email{hyxu@ir.hit.edu.cn}
\author{Chang Li}\authornotemark[1]
\email{cli@ir.hit.edu.cn}
\affiliation{%
  \institution{Harbin Institute of Technology}
  % \streetaddress{800 Dongchuan Road}
  \city{Harbin}
  \state{Heilongjiang}
  \country{China}
\postcode{150001}}

\author{Xinyan Ma}\authornotemark[1]
\email{xyma@ir.hit.edu.cn}
\author{Xianhao Ou}\authornotemark[1]
\email{25S103427@hit.edu.cn}
\affiliation{%
  \institution{Harbin Institute of Technology}
  % \streetaddress{800 Dongchuan Road}
  \city{Harbin}
  \state{Heilongjiang}
  \country{China}
\postcode{150001}}

\author{Zihan Zhang}
\email{Zihan_zhang@hsph.harvard.edu}
\affiliation{%
  \institution{Harvard University}
  \city{Cambridge}
  \state{MA}
  \country{USA}
\postcode{02138}}

\author{Tao He}
\email{the@ir.hit.edu.cn}
\author{Xiangyu Liu}
\email{2022112441@stu.hit.edu.cn}
\author{Zixiang Wang}
\email{2022112865@stu.hit.edu.cn}
\affiliation{%
  \institution{Harbin Institute of Technology}
  % \streetaddress{800 Dongchuan Road}
  \city{Harbin}
  \state{Heilongjiang}
  \country{China}
\postcode{150001}}

\author{Jiafeng Liang}
\email{jfliang@ir.hit.edu.cn}
\author{Zheng Chu}
\email{zchu@ir.hit.edu.cn}
\author{Runxuan Liu}
\email{rxliu@ir.hit.edu.cn}
\author{Rongchuan Mu}
\email{rcmu@ir.hit.edu.cn}
\affiliation{%
  \institution{Harbin Institute of Technology}
  % \streetaddress{800 Dongchuan Road}
  \city{Harbin}
  \state{Heilongjiang}
  \country{China}
\postcode{150001}}

\author{Dandan Tu}
\email{tudandan@huawei.com}
\affiliation{%
  \institution{Huawei Technologies Co., Ltd.}
  \country{China}
}

\author{Ming Liu}\authornote{Corresponding author}
\email{mliu@ir.hit.edu.cn}
\author{Bing Qin}
\email{qinb@ir.hit.edu.cn}
\affiliation{%
  \institution{Harbin Institute of Technology}
  % \streetaddress{800 Dongchuan Road}
  \city{Harbin}
  \state{Heilongjiang}
  \country{China}
\postcode{150001}}

\renewcommand{\shortauthors}{Xu, et al.}

\begin{abstract}
  Tool use enables large language models (LLMs) to access external information, invoke software systems, and act in digital environments beyond what can be solved from model parameters alone. Early research mainly studied whether a model could select and execute a correct single tool call. As agent systems evolve, however, the central problem has shifted from isolated invocation to multi-tool orchestration over long trajectories with intermediate state, execution feedback, changing environments, and practical constraints such as safety, cost, and verifiability. We comprehensively review recent progress in multi-tool LLM agents and analyze the state of the art in this rapidly developing area. First, we unify task formulations and distinguish single-call tool use from long-horizon orchestration. Then, we organize the literature around six core dimensions: inference-time planning and execution, training and trajectory construction, safety and control, efficiency under resource constraints, capability completeness in open environments, and benchmark design and evaluation. We further summarize representative applications in software engineering, enterprise workflows, graphical user interfaces, and mobile systems. Finally, we discuss major challenges and outline future directions for building reliable, scalable, and verifiable multi-tool agents.
\end{abstract}

\maketitle

\input{1_Introduction}
\input{0_Task}

\input{2_Inference}

\input{3_Train}
\input{4_Safety}
\input{5_Efficiency}
\input{6_Completeness}

\input{7_Benchmark}

\input{8_Application}
\input{10_Conclusion}

\bibliographystyle{ACM-Reference-Format}
\bibliography{custom}

\end{document}

%% file: 1_Introduction.tex
\section{Introduction}

While Large Language Models (LLMs) have demonstrated exceptional reasoning and generation capabilities in natural language processing~\cite{intro_llmsurvey}, their ability to solve complex real-world problems remains constrained by static parametric knowledge, potential hallucination risks, and a lack of interaction with physical or digital environments. Tool learning addresses these limitations by enabling models to invoke external APIs (e.g., search engines, code interpreters), establishing a perception-action loop. Early works such as TALM~\cite{intro_talm}, MRKL~\cite{intro_mrkl}, Toolformer~\cite{intro_toolformer}, and ReAct~\cite{inf_react} laid the foundation for this field by teaching models to recognize single intents and correctly format API requests, effectively internalizing tool usage as an extended language capability.

As task complexity escalates, the linear application of single tools becomes insufficient for real-world challenges. Multi-tool utilization represents an independent research problem that intersects with combinatorial optimization~\cite{intro_surveyllmad}, programmatic semantic constraints, and system scheduling~\cite{intro_ssllm}. The decision space for an autonomous agent undergoes a qualitative shift from simple binary tool selection to solving a series of coupled decisions within a single task. This process includes dynamic tool subset selection, cross-tool dependency modeling, sequential and parallel scheduling, failure recovery, and re-planning. When tool usage extends to long-horizon chains involving state-mutating write operations, maintaining state consistency and managing race conditions under parallel execution emerge as core bottlenecks for system stability.

The primary research objective in this domain has thus transitioned from the correctness of single-point calls to the end-to-end executability and robustness of multi-tool chains in complex environments. We begin by reviewing inference-time reasoning paradigms and architectures, highlighting the transition from serial chain reasoning to structured graph-based execution, alongside dual-system architectures designed to balance long-horizon planning complexity with execution efficiency. We then explore data synthesis and training paradigms, focusing on trajectory synthesis and closed-loop verification methods that address the combinatorial space of multi-tool usage and long-tail dependencies. Furthermore, we analyze safety and robustness, particularly concerning state safety under parallel execution, context drift in extended chains~\cite{intro_agentdrift}, and mitigation strategies for privacy risks. The discussion also encompasses operational efficiency, evaluating end-to-end agent latency, tool invocation costs, and inference budgets~\cite{intro_routellm}. To address system completeness, we review adaptive strategies for incomplete environments where tools or information may be missing. Finally, we trace the evolution of benchmark evaluation standards from single-point functional verification to system-level topological orchestration and interactive closed-loops. Based on these existing challenges, we propose a future research agenda to provide theoretical references and technical pathways for building reliable, efficient, and scalable multi-tool agent systems.

\subsection{Aims and Motivations}

Research on tool-augmented LLMs began with a relatively simple question: can a model select an appropriate tool and produce a valid call? That abstraction is no longer sufficient for agent systems. Many practical tasks require an agent to coordinate multiple tools over long trajectories, preserve intermediate state, recover from failures, and operate under constraints on latency, cost, and safety. In these settings, a key challenge lies not only in tool access, but also in orchestration.

This survey is motivated by two gaps in the current literature. The first is conceptual: terms such as tool use, tool calling, tool retrieval, workflow execution, and orchestration are often used loosely, even though they refer to different capability levels. The second is structural: planning, training, safety, efficiency, benchmarking, and open-environment adaptation are often studied separately, while deployed agent systems depend on their interaction.

\subsection{Related Work}

A growing body of survey work has examined tool use in large language models and related agent topics, yet long-horizon multi-tool orchestration has less often been treated as a primary problem in its own right. Wang et al.~\cite{intro_wang} considered what should count as a tool from the perspective of language models and offered a unified view of external tools. Qu et al.~\cite{intro_qu} surveyed tool learning in LLMs, covering planning, tool selection, tool calling, and response generation. Shen~\cite{intro_shen} reviewed LLM tool use with attention to tool integration, training methods, and the shift from tool use to tool creation. Li~\cite{intro_li} summarized major paradigms of LLM-based agents, including tool use, planning, and feedback learning. Luo et al.~\cite{intro_luo} provided a broader review of LLM agents spanning methodology, applications, and challenges. Chen et al.~\cite{intro_chen} focused on LLM-based multi-agent systems rather than multi-tool orchestration within a single agent. He et al.~\cite{intro_he} surveyed security and privacy risks in LLM agents. Mohammadi et al.~\cite{intro_mo} reviewed the evaluation and benchmarking of LLM agents.

Our survey differs from this literature in several ways. It takes multi-tool orchestration, rather than tool use in general or agent systems more broadly, as the main unit of analysis. It also organizes the field around six connected dimensions: inference, training and trajectory construction, safety and control, efficiency, capability completeness, and evaluation. In addition, it draws clearer boundaries among concepts that are often conflated in prior work, including tool invocation, tool retrieval, orchestration, and toolset expansion. Finally, it relates methodological progress to the recent shift from call-level correctness to system-level reliability in both benchmarks and applications.

\begin{figure}[ht]
  \centering
  \makebox[\textwidth][c]{%
    \includegraphics[width=1\textwidth]{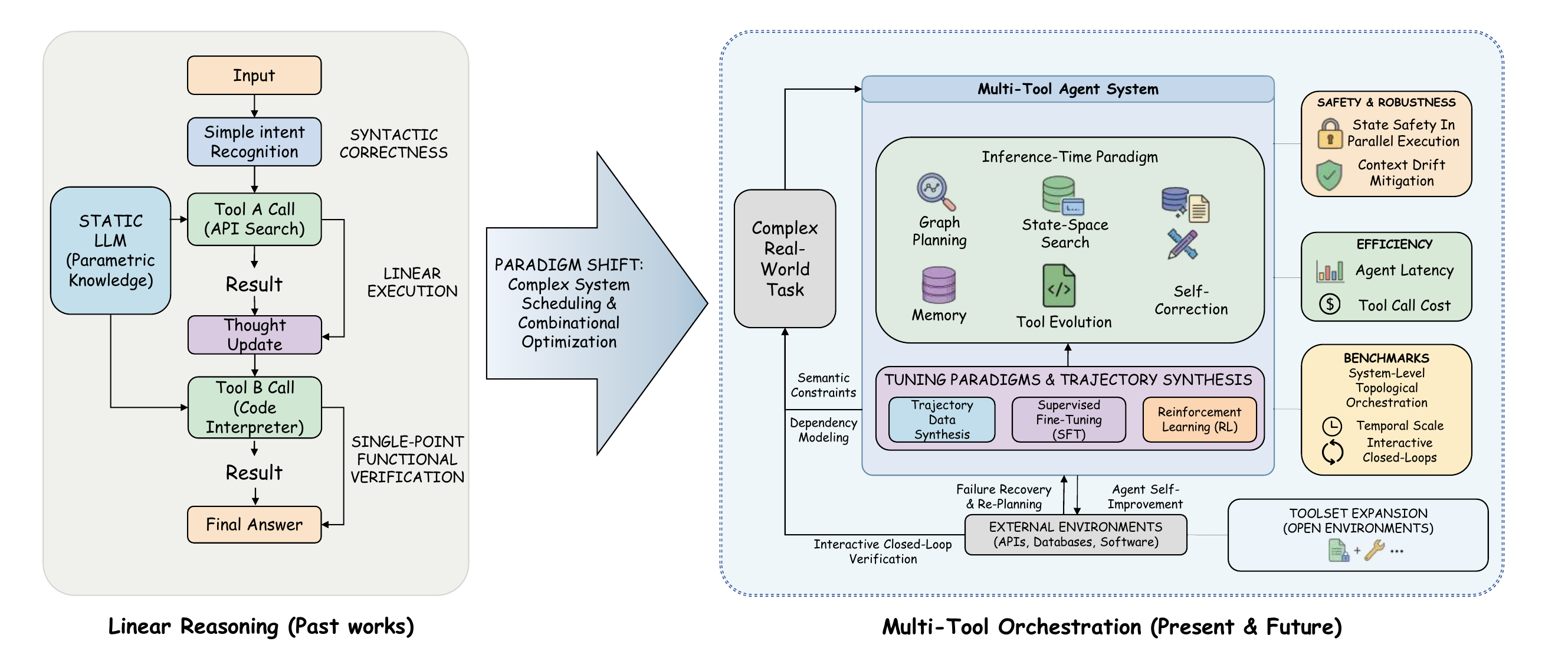}%
  }
  \caption{Paradigm Shift in LLM Tool Use: From Single-Tool Call to Multi-Tool Orchestration}
  \label{fig:mainstruct}
\end{figure}

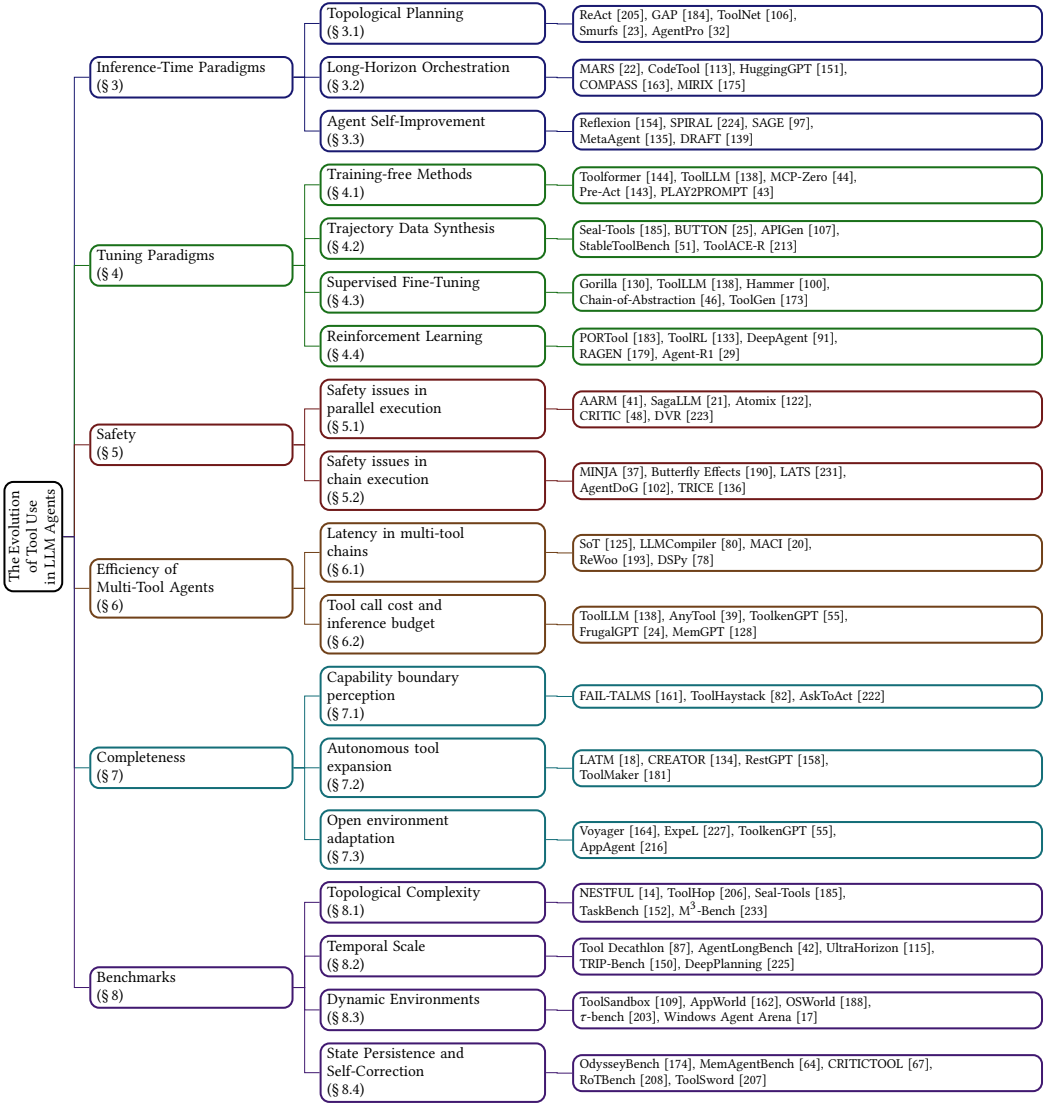
\begin{figure*}[ht]
  \centering

  % 颜色定义
  \definecolor{connect-line}{RGB}{0,0,0}
  \definecolor{middle-color}{RGB}{255,255,255}
  \definecolor{leaf-color}{RGB}{255,255,255}
  \definecolor{black}{RGB}{0,0,0}

  % 为6个主要分支定义6种主色调（边框和连线色）
  \definecolor{c-inf}{RGB}{25,25,112}   % 藏青 - Inference
  \definecolor{c-tune}{RGB}{25,112,25}  % 墨绿 - Tuning
  \definecolor{c-safe}{RGB}{112,25,25}  % 暗红 - Safety
  \definecolor{c-eff}{RGB}{112,65,25}   % 棕橙 - Efficiency
  \definecolor{c-ben}{RGB}{65,25,112}   % 暗紫 - Benchmarks
  \definecolor{c-com}{RGB}{20,110,120}  % 青蓝 - Completeness

  % 节点样式定义 (Middle 和 Leaf)
  \tikzstyle{inf-middle}=[draw=c-inf, rounded corners, minimum height=1em, fill=middle-color!40, text opacity=1, align=left, fill opacity=.5, text=black, font=\scriptsize, inner xsep=3pt, inner ysep=1pt]
  \tikzstyle{inf-leaf}=[draw=c-inf, rounded corners, minimum height=1em, fill=leaf-color!40, text opacity=1, align=left, fill opacity=.5, text=black, font=\tiny, inner xsep=3pt, inner ysep=1pt]

  \tikzstyle{tune-middle}=[draw=c-tune, rounded corners, minimum height=1em, fill=middle-color!40, text opacity=1, align=left, fill opacity=.5, text=black, font=\scriptsize, inner xsep=3pt, inner ysep=1pt]
  \tikzstyle{tune-leaf}=[draw=c-tune, rounded corners, minimum height=1em, fill=leaf-color!40, text opacity=1, align=left, fill opacity=.5, text=black, font=\tiny, inner xsep=3pt, inner ysep=1pt]

  \tikzstyle{safe-middle}=[draw=c-safe, rounded corners, minimum height=1em, fill=middle-color!40, text opacity=1, align=left, fill opacity=.5, text=black, font=\scriptsize, inner xsep=3pt, inner ysep=1pt]
  \tikzstyle{safe-leaf}=[draw=c-safe, rounded corners, minimum height=1em, fill=leaf-color!40, text opacity=1, align=left, fill opacity=.5, text=black, font=\tiny, inner xsep=3pt, inner ysep=1pt]

  \tikzstyle{eff-middle}=[draw=c-eff, rounded corners, minimum height=1em, fill=middle-color!40, text opacity=1, align=left, fill opacity=.5, text=black, font=\scriptsize, inner xsep=3pt, inner ysep=1pt]
  \tikzstyle{eff-leaf}=[draw=c-eff, rounded corners, minimum height=1em, fill=leaf-color!40, text opacity=1, align=left, fill opacity=.5, text=black, font=\tiny, inner xsep=3pt, inner ysep=1pt]

  \tikzstyle{ben-middle}=[draw=c-ben, rounded corners, minimum height=1em, fill=middle-color!40, text opacity=1, align=left, fill opacity=.5, text=black, font=\scriptsize, inner xsep=3pt, inner ysep=1pt]
  \tikzstyle{ben-leaf}=[draw=c-ben, rounded corners, minimum height=1em, fill=leaf-color!40, text opacity=1, align=left, fill opacity=.5, text=black, font=\tiny, inner xsep=3pt, inner ysep=1pt]

  \tikzstyle{com-middle}=[draw=c-com, rounded corners, minimum height=1em, fill=middle-color!40, text opacity=1, align=left, fill opacity=.5, text=black, font=\scriptsize, inner xsep=3pt, inner ysep=1pt]
  \tikzstyle{com-leaf}=[draw=c-com, rounded corners, minimum height=1em, fill=leaf-color!40, text opacity=1, align=left, fill opacity=.5, text=black, font=\tiny, inner xsep=3pt, inner ysep=1pt]

  % 如果树图过宽，取消下一行注释以自适应页面宽度
  \resizebox{\textwidth}{!}{%
    \begin{forest}
      for tree={
        forked edges,
        grow=east,
        reversed=true,
        anchor=base west,
        parent anchor=east,
        child anchor=west,
        base=middle,
        font=\scriptsize,
        rectangle,
        line width=0.9pt,
        draw=connect-line,
        rounded corners,
        align=left,
        minimum width=2em,
        s sep=5pt,
        inner xsep=3pt,
        inner ysep=1pt,
      },
      where level=1{text width=8.5em}{},  % 第一级子节点宽度
      where level=2{text width=9.5em}{},  % 第二级子节点宽度
      where level=3{text width=20.5em}{},   % 叶子节点（引文列表）宽度
      [The Evolution\\of Tool Use\\in LLM Agents, black, rotate=90, anchor=north, align=center, edge=connect-line
        % 分支 1: Inference-Time Reasoning (藏青)
        [{\hyperref[sec:inference]{Inference-Time Paradigms}\\\hyperref[sec:inference]{(\S\,\ref*{sec:inference})}}, inf-middle, edge=c-inf
          [{\hyperref[subsec:topological]{Topological Planning}\\\hyperref[subsec:topological]{(\S\,\ref*{subsec:topological})}}, inf-middle, edge=c-inf
            [{ReAct~\cite{inf_react}, GAP~\cite{inf_gap}, ToolNet~\cite{inf_toolnet},\\ Smurfs~\cite{inf_smurfs}, AgentPro~\cite{inf_agentpro}}, inf-leaf, edge=c-inf]
          ]
          [{\hyperref[subsec:long_horizon]{Long-Horizon Orchestration}\\\hyperref[subsec:long_horizon]{(\S\,\ref*{subsec:long_horizon})}}, inf-middle, edge=c-inf
            [{MARS~\cite{inf_mars}, CodeTool~\cite{inf_codetool}, HuggingGPT~\cite{inf_hugg},\\ COMPASS~\cite{inf_comp}, MIRIX~\cite{inf_mirix}}, inf-leaf, edge=c-inf]
          ]
          [{\hyperref[subsec:improvement]{Agent Self-Improvement}\\\hyperref[subsec:improvement]{(\S\,\ref*{subsec:improvement})}}, inf-middle, edge=c-inf
            [{Reflexion~\cite{inf_reflex}, SPIRAL~\cite{inf_spiral}, SAGE~\cite{inf_sage},\\ MetaAgent~\cite{inf_meta}, DRAFT~\cite{inf_draft}}, inf-leaf, edge=c-inf]
          ]
        ]
        % 分支 2: Tuning Paradigms (墨绿)
        [{\hyperref[sec:tuning]{Tuning Paradigms}\\\hyperref[sec:tuning]{(\S\,\ref*{sec:tuning})}}, tune-middle, edge=c-tune
          [{\hyperref[subsec:tuningfree]{Training-free Methods}\\\hyperref[subsec:tuningfree]{(\S\,\ref*{subsec:tuningfree})}}, tune-middle, edge=c-tune
            [{Toolformer~\cite{intro_toolformer}, ToolLLM~\cite{ben_ToolBench}, MCP-Zero~\cite{tune_mcp-zero},\\ Pre-Act~\cite{tune_pre-act}, PLAY2PROMPT~\cite{tune_PLAY2PROMPT}}, tune-leaf, edge=c-tune]
          ]
          [{\hyperref[subsec:trajectory]{Trajectory Data Synthesis}\\\hyperref[subsec:trajectory]{(\S\,\ref*{subsec:trajectory})}}, tune-middle, edge=c-tune
            [{Seal-Tools~\cite{ben_SealTools}, BUTTON~\cite{tune_Button}, APIGen~\cite{tune_apigen},\\ StableToolBench~\cite{ben_stabletool}, ToolACE-R~\cite{tune_toolace-r}}, tune-leaf, edge=c-tune]
          ]
          [{\hyperref[subsec:sft]{Supervised Fine-Tuning}\\\hyperref[subsec:sft]{(\S\,\ref*{subsec:sft})}}, tune-middle, edge=c-tune
            [{Gorilla~\cite{tune_gorilla}, ToolLLM~\cite{ben_ToolBench}, Hammer~\cite{tune_hammer},\\ Chain-of-Abstraction~\cite{tune_chain_of_abstraction}, ToolGen~\cite{tune_toolgen}}, tune-leaf, edge=c-tune]
          ]
          [{\hyperref[subsec:rl]{Reinforcement Learning}\\\hyperref[subsec:rl]{(\S\,\ref*{subsec:rl})}}, tune-middle, edge=c-tune
            [{PORTool~\cite{tune_portool}, ToolRL~\cite{tune_toolrl}, DeepAgent~\cite{inf_deepagent},\\ RAGEN~\cite{tune_ragen}, Agent-R1~\cite{tune_agent-r1}}, tune-leaf, edge=c-tune]
          ]
        ]
        % 分支 3: Safety (暗红)
        [{\hyperref[sec:safety]{Safety}\\\hyperref[sec:safety]{(\S\,\ref*{sec:safety})}}, safe-middle, edge=c-safe
          [{\hyperref[subsec:safety_parallel]{Safety issues in}\\\hyperref[subsec:safety_parallel]{parallel execution}\\\hyperref[subsec:safety_parallel]{(\S\,\ref*{subsec:safety_parallel})}}, safe-middle, edge=c-safe
            [{AARM~\cite{safe_ARRM}, SagaLLM~\cite{safe_SagaLLM}, Atomix~\cite{safe_Atomix},\\ CRITIC~\cite{safe_CRITIC}, DVR~\cite{safe_DVR}}, safe-leaf, edge=c-safe]
          ]
          [{\hyperref[subsec:safety_chain]{Safety issues in}\\\hyperref[subsec:safety_chain]{chain execution}\\\hyperref[subsec:safety_chain]{(\S\,\ref*{subsec:safety_chain})}}, safe-middle, edge=c-safe
            [{MINJA~\cite{safe_minja}, Butterfly Effects~\cite{safe_Butterfly}, LATS~\cite{safe_LATS},\\ AgentDoG~\cite{safe_AgentDoG}, TRICE~\cite{safe_TRICE}}, safe-leaf, edge=c-safe]
          ]
        ]
        % 分支 4: Efficiency (棕橙)
        [{\hyperref[sec:efficiency]{Efficiency of}\\\hyperref[sec:efficiency]{Multi-Tool Agents}\\\hyperref[sec:efficiency]{(\S\,\ref*{sec:efficiency})}}, eff-middle, edge=c-eff
          [{\hyperref[ssec: Latency in multi-tool chains]{Latency in multi-tool}\\\hyperref[ssec: Latency in multi-tool chains]{chains}\\\hyperref[ssec: Latency in multi-tool chains]{(\S\,\ref*{ssec: Latency in multi-tool chains})}}, eff-middle, edge=c-eff
            [{SoT~\cite{eff_sot}, LLMCompiler~\cite{eff_LLMCompiler}, MACI~\cite{eff_maci},\\ ReWoo~\cite{eff_rewoo}, DSPy~\cite{eff_dspy}}, eff-leaf, edge=c-eff]
          ]
          [{\hyperref[ssec: Tool call cost and inference budget]{Tool call cost and}\\\hyperref[ssec: Tool call cost and inference budget]{inference budget}\\\hyperref[ssec: Tool call cost and inference budget]{(\S\,\ref*{ssec: Tool call cost and inference budget})}}, eff-middle, edge=c-eff
            [{ToolLLM~\cite{ben_ToolBench}, AnyTool~\cite{tune_anytool}, ToolkenGPT~\cite{eff_toolkengpt},\\ FrugalGPT~\cite{eff_frugalgpt}, MemGPT~\cite{eff_memgpt}}, eff-leaf, edge=c-eff]
          ]
        ]
        % 分支 5: Completeness (青蓝)
        [{\hyperref[sec:completeness]{Completeness}\\\hyperref[sec:completeness]{(\S\,\ref*{sec:completeness})}}, com-middle, edge=c-com
          [{\hyperref[ssec: Capability boundary perception]{Capability boundary}\\\hyperref[ssec: Capability boundary perception]{perception}\\\hyperref[ssec: Capability boundary perception]{(\S\,\ref*{ssec: Capability boundary perception})}}, com-middle, edge=c-com
            [{FAIL-TALMS~\cite{com_failtalms}, ToolHaystack~\cite{com_toolhaystack}, AskToAct~\cite{com_asktoact}}, com-leaf, edge=c-com]
          ]
          [{\hyperref[ssec: Autonomous tool expansion]{Autonomous tool}\\\hyperref[ssec: Autonomous tool expansion]{expansion}\\\hyperref[ssec: Autonomous tool expansion]{(\S\,\ref*{ssec: Autonomous tool expansion})}}, com-middle, edge=c-com
            [{LATM~\cite{com_latm}, CREATOR~\cite{com_creator}, RestGPT~\cite{com_restgpt},\\ ToolMaker~\cite{com_toolmaker}}, com-leaf, edge=c-com]
          ]
          [{\hyperref[ssec: Open environment adaptation]{Open environment}\\\hyperref[ssec: Open environment adaptation]{adaptation}\\\hyperref[ssec: Open environment adaptation]{(\S\,\ref*{ssec: Open environment adaptation})}}, com-middle, edge=c-com
            [{Voyager~\cite{com_voyager}, ExpeL~\cite{com_expel}, ToolkenGPT~\cite{eff_toolkengpt},\\ AppAgent~\cite{com_appagent}}, com-leaf, edge=c-com]
          ]
        ]
        % 分支 6: Benchmarks (暗紫)
        [{\hyperref[sec:benchmarks]{Benchmarks}\\\hyperref[sec:benchmarks]{(\S\,\ref*{sec:benchmarks})}}, ben-middle, edge=c-ben
          [{\hyperref[subsec:topo_complexity]{Topological Complexity}\\\hyperref[subsec:topo_complexity]{(\S\,\ref*{subsec:topo_complexity})}}, ben-middle, edge=c-ben
            [{NESTFUL~\cite{ben_NESTFUL}, ToolHop~\cite{ben_ToolHop}, Seal-Tools~\cite{ben_SealTools},\\ TaskBench~\cite{ben_TaskBench}, M$^3$-Bench~\cite{ben_m3bench}}, ben-leaf, edge=c-ben]
          ]
          [{\hyperref[subsec:temporal_scale]{Temporal Scale}\\\hyperref[subsec:temporal_scale]{(\S\,\ref*{subsec:temporal_scale})}}, ben-middle, edge=c-ben
            [{Tool Decathlon~\cite{ben_tooldec}, AgentLongBench~\cite{ben_along}, UltraHorizon~\cite{ben_ultra},\\ TRIP-Bench~\cite{ben_trip}, DeepPlanning~\cite{ben_deepplan}}, ben-leaf, edge=c-ben]
          ]
          [{\hyperref[subsec:dynamic_env]{Dynamic Environments}\\\hyperref[subsec:dynamic_env]{(\S\,\ref*{subsec:dynamic_env})}}, ben-middle, edge=c-ben
            [{ToolSandbox~\cite{ben_toolsandbox}, AppWorld~\cite{ben_appworld}, OSWorld~\cite{ben_osworld},\\ $\tau$-bench~\cite{ben_tau}, Windows Agent Arena~\cite{ben_windows}}, ben-leaf, edge=c-ben]
          ]
          [{\hyperref[subsec:state_persistence]{State Persistence and}\\\hyperref[subsec:state_persistence]{Self-Correction}\\\hyperref[subsec:state_persistence]{(\S\,\ref*{subsec:state_persistence})}}, ben-middle, edge=c-ben
            [{OdysseyBench~\cite{ben_odyssey}, MemAgentBench~\cite{ben_memagentbench}, CRITICTOOL~\cite{ben_criticaltool},\\ RoTBench~\cite{ben_rot}, ToolSword~\cite{ben_toolsword}}, ben-leaf, edge=c-ben]
          ]
        ]
      ]
    \end{forest}
  }

  \vspace{0.15in}
  \caption{The main structure of this paper.}
  \vspace{0.2in}
  \label{fig:evolution_map}
\end{figure*}

%% file: 0_Task.tex
\section{Problem formulation}

\textbf{Notation.}
A multi-tool learning instance contains:
(i) a query/task $q \in \mathcal{Q}$;
(ii) an interactive environment with latent state $s_t \in \mathcal{S}$ and observation $o_t \in \mathcal{O}$ at step $t$;
(iii) a tool inventory $\mathcal{D}=\{d_i\}_{i=1}^{N}$, where each tool $d$ is described by a schema/signature $\Sigma_d$
(e.g., name, description, constraints, return format).
A tool execution interface is
\[
\textsc{Exec}(d,u; s)\rightarrow (f,s'),
\]
where $u \in \mathcal{U}_d$ is a valid argument/parameter object for tool $d$,
$f \in \mathcal{F}$ is tool feedback (e.g., JSON/text/error code),
and $s'\in\mathcal{S}$ is the post-call environment state .
Let $\pi_\theta$ denote the model/policy with parameters $\theta$.
Let $\tau$ denote a tool-use trajectory (defined below), and $\text{Cost}(\tau)$ any cost functional
(number of calls, latency, API fees, risk, etc.).

We define the interaction history up to time $t$ as
\[
h_t := (q, o_0, a_0, f_0, o_1, a_1, f_1, \ldots, o_t),
\]
where $a_k$ is the action taken at step $k$ and $f_k$ is the feedback returned by $\textsc{Exec}$.
Let $m_t$ denote the model's internal memory/state at step $t$ (e.g., a scratchpad summary or retrieved records),
which can be updated as a function of $(m_{t-1},h_t)$.

\textbf{Multi-tool orchestration.}
At each step $t$, the model selects an action
\[
a_t \in \mathcal{A} := \big(\mathcal{D}\times \mathcal{U}\big)\ \cup\ \{\textsc{Finish}(y)\},
\]
where a tool-call action is $a_t=(d_t,u_t)$ with $d_t\in\mathcal{D}$ and $u_t\in\mathcal{U}_{d_t}$,
and $\textsc{Finish}(y)$ terminates with final answer $y\in\mathcal{Y}$.
The policy acts conditioned on history and memory:
\[
a_t \sim \pi_\theta(\cdot \mid h_t, m_t).
\]
If $a_t=(d_t,u_t)$, the environment returns feedback and transitions:
\[
(f_t,s_{t+1}) \leftarrow \textsc{Exec}(d_t,u_t;s_t),
\qquad
h_{t+1} = h_t \oplus (a_t,f_t,o_{t+1}),
\]
where $\oplus$ appends new interaction records and $o_{t+1}$ is the next observation derived from $s_{t+1}$.
A trajectory is
\[
\tau := \big((a_0,f_0),(a_1,f_1),\ldots,(a_{T-1},f_{T-1}), \textsc{Finish}(y)\big),
\]
with variable horizon $T$.
A generic cost-aware objective is
\[
\max_{\theta}\; \mathbb{E}\Big[ R_{\text{task}}(q,\tau,y) - \lambda \cdot \text{Cost}(\tau) \Big],
\]
where $\lambda\ge 0$ trades off success vs cost.
This captures long-horizon scheduling, failure recovery, and re-planning under tool/environment feedback.

%% file: 2_Inference.tex
\section{Multi-Tool Agent Inference-Time Paradigms} \label{sec:inference}
Multi-tool calling constitutes a fundamental shift in agent planning and reasoning. As the tool space expands to encompass thousands of APIs and non-linear dependencies, traditional autoregressive generation models struggle with context limits and cascading errors. This shift motivates a substantial rethinking of agent inference-time paradigms. Recent research focuses on task topological planning, explicit search algorithms, reflection and self-improvement, and system-level orchestration with persistent memory. This section systematically reviews representative works along these key areas.

\subsection{Topological Planning} \label{subsec:topological}

Traditional paradigms such as ReAct~\cite{inf_react} organize tool usage into sequential traces. This structure implicitly assumes that tasks are linearly decomposable. While highly effective for simple single-step retrievals, this linear approach exhibits severe structural mismatches when applied to nested calls, parallel I/O operations, or long-horizon goals. Linear constraints inherently limit the representation of data flow dependencies, preclude the parallel execution of independent operations, and force early decision errors to accumulate throughout the trajectory. Consequently, the field has rapidly moved toward topology-aware planning that utilizes structures with significantly higher expressive power.

As tool spaces expand into the tens of thousands, the primary challenge shifts from binary selection to dependence-aware combinatorial decision-making. One line of work represents tool use as an explicit dependency structure, allowing the agent to reason over asynchronous or partially parallelizable subproblems~\cite{inf_graphen, inf_gap}. Related systems such as ToolNet and StructuredAgent encode reusable transition structure through directed graphs or AND/OR decompositions, reducing the effective search space while preserving non-linear task dependencies~\cite{inf_toolnet, inf_struct}. AutoTool further shows that historical regularities in tool transitions can be used to bypass unnecessary inference for predictable parts of the workflow~\cite{inf_autotool}. Taken together, these approaches treat tool use as graph-structured execution rather than a flat token sequence.

To address goal drift in long-horizon tasks, monolithic planning algorithms are increasingly being replaced by hierarchical delegation. HIPLAN proposes a bi-level architecture where an upper-level planner generates a macroscopic milestone space, leaving a lower-level executor to resolve specific state gaps~\cite{inf_hiplan}. ADaPT introduces a recursive representation that defaults to flat execution, invoking high-level graph decomposition only upon encountering localized failures~\cite{inf_adapt}. Pushing abstraction further, AFlow represents the reasoning process as executable node configurations and control flow edges~\cite{inf_aflow}. Dynamic planners like D-PoT~\cite{inf_dpot}, DyFlow~\cite{inf_dy}, and ReCAP~\cite{inf_recap} further allow real-time subgoal reorganization.

As autoregressive decoding is inherently short-sighted, recent methods cast multi-step tool execution as a state-space search problem. Traditional tree search often suffers from context contamination in language domains. Smurfs introduces context-efficient rollback and branch isolation, reducing contamination from failed trajectories~\cite{inf_smurfs}. AB-MCTS explores adaptive branching under external feedback, while ARTIS separates exploration from commitment through an internal simulator for high-risk actions~\cite{inf_abmcts, inf_artis}. Closely related approaches, including Plan-and-Act, MPO, and AgentPro, also separate abstract planning from execution or use learned evaluators to prune low-quality branches during inference~\cite{inf_paa, inf_mpo, inf_agentpro}. These methods reflect a broader shift from "generate once and execute" to "search, verify, and then commit".

\subsection{Long-Horizon Orchestration} \label{subsec:long_horizon}
When task complexity exceeds the cognitive limitations of a single model, reasoning shifts to system-level architecture. This axis emphasizes cognitive division of labor, standardized semantic routing, and long-term dynamic memory to ensure robust performance across open-ended environments.

Inspired by dual-process reasoning, MARS separates fast parsing of large tool outputs from slower deliberative reasoning~\cite{inf_mars}. In domains with strict execution requirements, CodeTool introduces structured intermediate representations that make progress easier to verify~\cite{inf_codetool}. Other systems externalize durable skills or workflows into procedural memory. DeepAgent integrates tool discovery into a continuous reasoning stream~\cite{inf_deepagent}, while MACLA and HiAgent explicitly maintain reusable procedural abstractions across long tasks and repeated sessions~\cite{inf_macla, inf_hiagent}. The common goal is to avoid placing the full burden of orchestration inside a single transient context window.

A related challenge is dynamic scheduling over large tool ecosystems. Early frameworks such as HuggingGPT illustrated the value of centralized planning and semantic routing across heterogeneous expert models~\cite{inf_hugg}. Subsequent work showed that decomposing the workflow into planner, caller, and summarizer roles can allow smaller models to compete effectively with larger monolithic agents~\cite{inf_small}. As tool access becomes more standardized, schedulers such as Gradientsys coordinate multiple specialized executors under capacity and observability constraints~\cite{inf_grad}. Tool-to-Agent Retrieval extends this idea by jointly indexing tools and their parent agents, preserving both invocation granularity and execution context during retrieval~\cite{inf_t2ar}. These systems suggest that long-horizon competence depends not only on reasoning quality, but also on how execution responsibilities are routed and monitored.

Memory is the third core component of long-horizon orchestration. Raw trajectories are too large and noisy to remain fully useful over extended interaction, so practical systems compress, structure, or externalize execution history. COMPASS and Acon summarize completed subtasks into compact progress representations to maintain alignment with the global objective~\cite{inf_comp, inf_acon}. Other approaches use graph-structured or typed memory to support retrieval and update over longer timescales. A-Mem builds semantic links among historical states~\cite{inf_amem}, MIRIX decomposes memory into multiple specialized stores for persistent multimodal interaction~\cite{inf_mirix}, and SEDM treats memory as a shared resource for multi-agent collaboration~\cite{inf_sedm}. Across these systems, the main design question is not whether to store more history, but how to retain the parts that remain relevant for decisions while suppressing noise and outdated branches.

\subsection{Agent Self-Improvement} \label{subsec:improvement}

Inference-time capability is increasingly shaped not only by planning and memory, but also by mechanisms for self-improvement. In multi-tool settings, errors often arise from misbound parameters, invalid assumptions, or ineffective recovery strategies. As a result, many recent systems explicitly learn to diagnose and revise their own trajectories during or after execution.

A first family of methods focuses on self-correction. Reflexion stores verbalized lessons from prior failures and uses them to guide later attempts~\cite{inf_reflex}. The Failure Makes the Agent Stronger framework turns diagnosis and repair into explicit actions trained with reward-based optimization over perturbed trajectories~\cite{inf_fmas}. SPIRAL and PreFlect further constrain or anticipate reflective behavior to reduce unproductive loops and catch likely failures before execution~\cite{inf_spiral, inf_preflect}. SAGE and Plan2Evolve extend this idea toward continual improvement by combining reflection with memory and planning repair~\cite{inf_sage, inf_p2e}.

A second family of methods treats improvement as capability evolution rather than isolated correction. MetaAgent distills accumulated experience into reusable tools and internal knowledge, enabling the agent to improve its future reasoning and tool-use strategies without additional model training~\cite{inf_meta}. Test-Time Tool Evolution pushes this direction further by synthesizing and validating new executable tools during inference~\cite{inf_tte}. DRAFT and SWE-Adept similarly revise tool descriptions or tool-memory interfaces in response to execution feedback~\cite{inf_draft, inf_swead}. This line of work suggests that inference is increasingly becoming an adaptive loop in which the agent not only uses a tool ecosystem, but also reorganizes and improves that ecosystem while solving tasks.

%% file: 3_Train.tex
\section{Multi-Tool Agent Tuning Paradigms} \label{sec:tuning}
\begin{figure}[t]
  \centering
  \includegraphics[width=\linewidth]{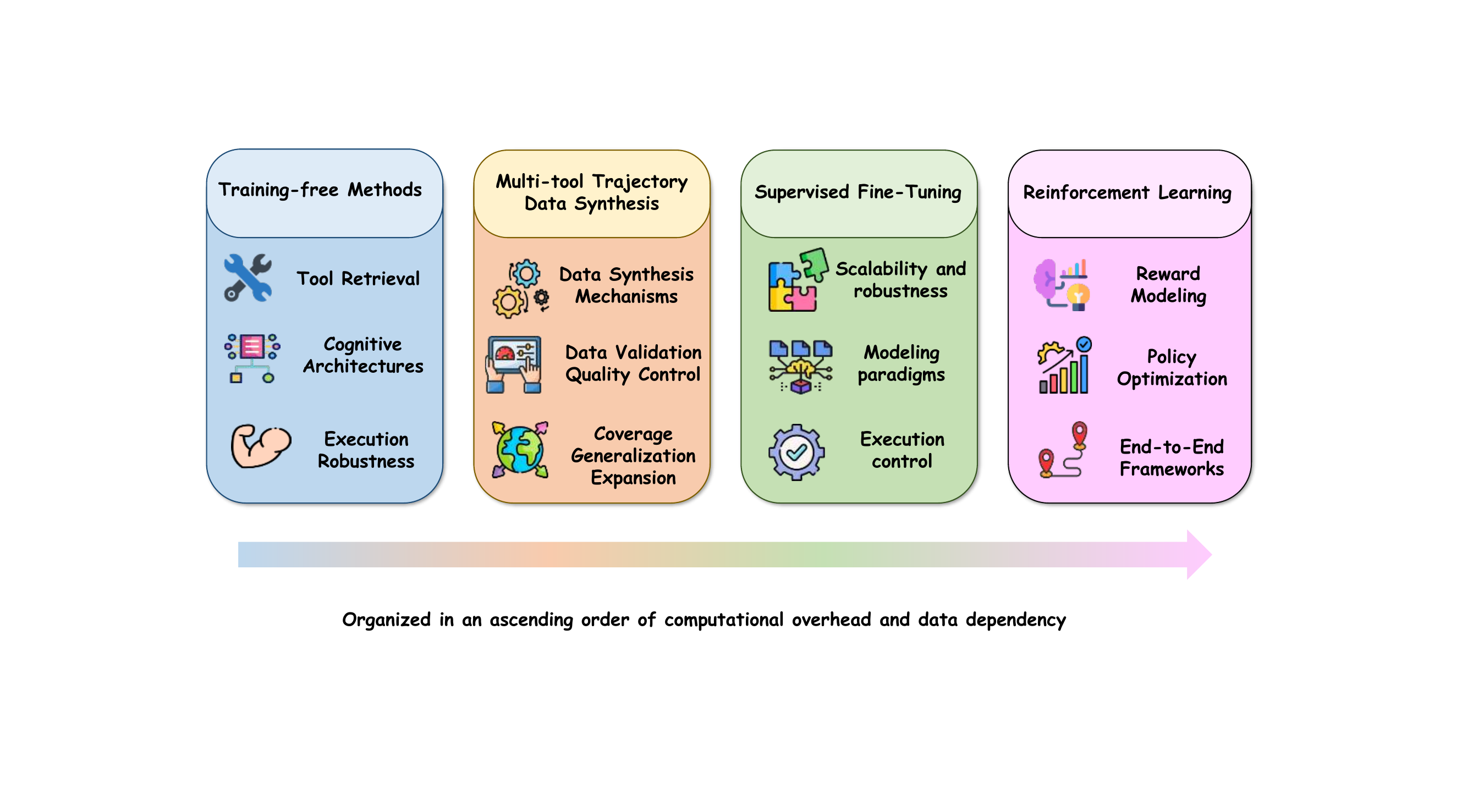}
  \caption{Taxonomy of Methodologies for Multi-tool Agent Tuning Paradigms}
  \label{fig:Tuning-Paradiagms}
\end{figure}

Despite the remarkable generalization capabilities exhibited by LLMs across broad natural language tasks, deploying them as reliable agents in multi-tool orchestration scenarios remains a formidable challenge. Vanilla pre-trained models or general-purpose models frequently exhibit deficiencies in schema compliance, parameter precision, and long-horizon planning, which can easily lead to execution failures or hallucinations in complex multi-tool tasks. To bridge the gap between general-purpose models and specialized multi-tool agents, various adaptation strategies have been proposed to bolster the model’s proficiency in tool perception, selection, and precise execution. As shown in Figure~\ref{fig:Tuning-Paradiagms}, this chapter provides a systematic taxonomy of methodologies for Multi-tool Agent Tuning, organized in an ascending order of computational overhead and data dependency: we first investigate Training-free Methods that activate tool-calling capabilities without parameter updates; then introduce multi-tool trajectory data synthesis as the essential data foundation; followed by supervised fine-tuning (SFT) to align models with specific tool-calling syntaxes and planning patterns; and finally discuss reinforcement learning (RL) frameworks that exploit environmental feedback to improve agent self-evolution and decision-making robustness.

\subsection{Training-free Methods} \label{subsec:tuningfree}
Training-free methods enhance multi-tool agents without updating LLM parameters, offering low deployment overhead and strong adaptability to newly introduced tools. Rather than altering the agent's internal reasoning mechanism, these methods improve tool use through external augmentation, primarily along three directions: tool retrieval and discovery, prompt engineering, and non-parametric memory augmentation.

A major line of work studies tool retrieval and discovery. As the number of available APIs grows, exhaustively including tool documentation in the prompt becomes impractical. To address this issue, ToolLLM~\cite{ben_ToolBench} and AnyTool~\cite{tune_anytool} employ retrieval-based mechanisms to dynamically select relevant tool descriptions from large tool repositories. ToolDreamer~\cite{tune_tooldreamer} further improves retrieval quality by generating idealized tool descriptions as search queries, thereby better aligning user intents with real-world APIs. MCP-Zero~\cite{tune_mcp-zero} extends passive retrieval to active discovery, enabling agents to request tool schemas on demand and thus reducing context redundancy. Re-Invoke~\cite{tune_re-invoke} similarly improves zero-shot tool selection through multi-perspective query rewriting.
Another line of work focuses on prompt and exemplar construction for tool use. PLAY2PROMPT~\cite{tune_PLAY2PROMPT} allows agents to autonomously explore tool behavior and synthesize usage instructions together with few-shot exemplars, thereby improving tool invocation without SFT.
Finally, some methods enhance robustness through non-parametric memory augmentation. Memento~\cite{tune_Memento} equips agents with episodic memory by retrieving successful historical trajectories to support current decision-making. GenTool~\cite{tune_gentool} improves robustness through simulation under diverse conditions, such as missing or updated tools, thereby helping agents generalize to dynamic tool environments.

\subsection{Multi-tool trajectory Data Synthesis} \label{subsec:trajectory}
High-quality training data serves as the cornerstone for constructing robust multi-tool agents. However, the acquisition of real-world tool execution logs is constrained by privacy regulations, access permissions, and the scarcity of diverse scenarios, making large-scale manual annotation impractical. Consequently, synthetic data generation has emerged as the dominant paradigm. Recent literature reveals a systematic evolution toward a "Synthesis-Validation-Expansion" framework, marking a strategic shift in research focus from basic data usability to the comprehensive coverage of the multi-tool composition space. This paradigm aims to synthesize training trajectories that encompass long-horizon interactions, cross-tool dependencies, and error recovery mechanisms in complex scenarios, thereby bolstering the agent's capability to navigate intricate tasks.

Regarding data synthesis mechanisms, Early self-bootstrapping methods such as Seal-Tools~\cite{ben_SealTools} and Luo et al.~\cite{tune_luo} show that LLMs can generate large amounts of tool-use data without expert demonstrations. More recent approaches improve structure and compositionality rather than scale alone. BUTTON~\cite{tune_Button} combines bottom-up instruction construction with top-down trajectory generation to synthesize compositional interaction traces, while ASTRA~\cite{tune_ASTRA} uses tool-call graph topology to generate structurally grounded trajectories in rule-verifiable environments.

In terms of data validation and quality control, recent advancements focus on execution-based iterative verification. As synthetic data grows in size, filtering by format alone becomes insufficient. APIGen~\cite{tune_apigen} therefore verifies data through staged checking that includes syntax, actual execution, and semantic validation. StableToolBench~\cite{ben_stabletool} reduces instability from volatile APIs through simulators and caching. ToolACE-DEV~\cite{tune_toolace-dev} and ToolACE-R~\cite{tune_toolace-r} move further toward iterative repair, using self-improvement and model-aware refinement to adapt synthetic data to the learner's evolving capability.

Concerning coverage and generalization expansion, numerous approaches aim to push the boundaries of the tool compositional space to tackle real-world complexity. ToolMind~\cite{tune_toolmind} advances data engineering for compositional space coverage by constructing large-scale, reasoning-intensive datasets. To enhance robustness against execution errors, LoopTool~\cite{tune_looptool} institutionalizes failure-driven synthesis, creating a data closed-loop that mirrors real-world failure distributions. For complex orchestration, OrchDAG~\cite{tune_orchdag} models tool execution as Directed Acyclic Graphs (DAGs) with controllable complexity, providing structured benchmarks for multi-turn interactions. Collectively, these advancements reflect a new consensus: the critical metric for synthetic data is no longer mere quantity, but the systematic coverage of long-tail effects, complex interaction effects, and multi-step error patterns.

\subsection{Supervised Fine-Tuning} \label{subsec:sft}
Supervised Fine-Tuning (SFT) represents the foundational alignment phase for endowing Large Language Models with tool-use capabilities. Amidst the paradigm shift from atomic tool invocation to multi-tool orchestration, the objectives of SFT have undergone a fundamental transformation. Early methodologies primarily emphasized syntax alignment—ensuring the structural integrity of outputs (e.g., JSON schema compliance). Pioneering works such as Gorilla~\cite{tune_gorilla} and GPT4Tools~\cite{tune_GPT4Tools} demonstrated that fine-tuning on high-quality (Instruction, API Call) pairs enables models to effectively acquire API semantic matching and parameter filling skills. However, mere data accumulation is no longer sufficient to address complex orchestration tasks. Consequently, recent advancements have pivoted toward more specialized objectives, such as tool selection from massive repositories and inter-tool dependency planning.

To address large-scale and noisy tool selection, ToolLLM~\cite{ben_ToolBench} extends SFT to massive toolsets by training on Depth-First Search decision trajectories, enabling the model to internalize exploration and trial-and-error strategies for unresolvable scenarios. Hammer~\cite{tune_hammer} introduces Function Masking during fine-tuning to penalize hallucinations and heighten sensitivity to irrelevant tools, establishing a new benchmark for robust on-device function calling. To optimize the interplay between reasoning and action, Chain-of-Abstraction~\cite{tune_chain_of_abstraction} proposes a dual-stage decoding strategy; it trains the model to first generate abstract reasoning chains with placeholders and subsequently populate them with concrete tool outputs, effectively decoupling general reasoning from domain-specific knowledge retrieval. ToolGen~\cite{tune_toolgen} unifies tool retrieval and parameter generation into a sequence of generative virtual tokens, restructuring the SFT objective to allow the simultaneous learning of tool semantics and parameter structures in a single autoregressive pass, which significantly mitigates cascading errors. Finally, addressing task granularity, the Granite-Function Calling Model~\cite{tune_granite-function} adopts a multi-task learning framework that decomposes tool usage into fine-grained subtasks—such as tool detection, parameter generation, and response formatting—thereby achieving precise control over the entire execution lifecycle. Collectively, these developments signify that SFT has evolved beyond simple pattern imitation toward the construction of autonomous systems capable of high-level decision-making and rigorous execution.

\subsection{Reinforcement Learning} \label{subsec:rl}
Although SFT effectively aligns models with fundamental tool-use syntax, it remains inherently constrained by the Behavioral Cloning paradigm. Models trained via SFT are often limited to myopic imitation of training trajectories, struggling to self-correct during long-horizon tasks or to explore novel paths within open-ended tool spaces. Consequently, the research focus is shifting toward Reinforcement Learning and environmental feedback to optimize high-level decision-making. As the field transitions from single-tool triggering to multi-tool orchestration, RL encounters formidable challenges: sparse reward signals, difficult credit assignment, and training instability. Thus, recent advances have transcended standard algorithms like PPO or GRPO, developing specialized paradigms centered on process reward design, turn-level policy optimization, and end-to-end agent frameworks. These innovations aim to evolve agents from passive executors into adaptive planners capable of robust error recovery and strategic exploration.

Regarding Reward Modeling and Credit Assignment, to bridge the gap between sparse outcome rewards and complex multi-step reasoning, the community has established a comprehensive reward-optimization closed loop. Addressing the sparsity of success signals in long chains, PORTool~\cite{tune_portool} integrates quality signals generated by tree search into policy gradient updates, while Tool-Star~\cite{tune_tool-star} designs hierarchical reward mechanisms to resolve credit assignment issues in multi-step reasoning. Moving toward systematic reward engineering, ToolRL~\cite{tune_toolrl} conducts a comprehensive study on reward granularity, proposing principled reward designs that surpass simple pass/fail metrics to stabilize training. Similarly, ToolRM~\cite{tune_toolrm} trains a dedicated Outcome Reward Model to evaluate tool-based reasoning, outperforming general-purpose verifiers in complex scenarios. To reinforce intermediate steps,
Wei et al.~\cite{tune_wei} introduce verifiable turn-level rewards, extending RL algorithms to handle fine-grained credit assignment in multi-turn search tasks.

In terms of Policy Optimization and Algorithm Adaptation, standard RL algorithms often struggle to adapt to drastic dynamic shifts in agent tasks. DeepAgent~\cite{inf_deepagent} explores ToolPO (Tool Policy Optimization), which addresses the high cost and instability of training with real APIs by utilizing an LLM-simulated API environment for low-cost exploration. It simultaneously introduces Fine-grained Advantage Attribution, precisely allocating reward signals to the specific tokens triggering tool invocations, thereby significantly enhancing stability in open-set tool retrieval. Addressing the limitations of trajectory-level rewards in GRPO, GTPO~\cite{tune_gtpo} innovatively adopts turn-level reward distribution and return-based advantage estimation, effectively preventing training stagnation during iterative reasoning. Furthermore, to mitigate Entropy Spikes following tool interactions, ARPO~\cite{tune_arpo} incorporates an entropy-based adaptive rollback mechanism to dynamically balance global exploration and local exploitation. RAGEN~\cite{tune_ragen} identifies the Echo Trap phenomenon specific to agent training and proposes the StarPO algorithm to stabilize the learning process through trajectory-level optimization.

Concerning End-to-End Frameworks, ToRL~\cite{tune_torl} demonstrates that RL can directly extend tool-use capabilities from a Base Model without a prerequisite SFT phase, unlocking emergent behaviors such as self-regulation. Finally, Agent-R1~\cite{tune_agent-r1} comprehensively defines LLM Agents by extending the Markov Decision Process (MDP) framework, providing a modular and scalable system for end-to-end reinforcement learning.

%% file: 4_Safety.tex
\section{Safety of Multi-Tool Agents} \label{sec:safety}
The security landscape of tool-augmented agents has shifted from mere malicious text generation to the execution of harmful actions within digital or physical environments. While single-tool vulnerabilities—such as indirect prompt injection~\cite{safe_not_what}, unauthorized execution and privilege escalation~\cite{safe_prompt_flow}, privacy leakage, and tool-assisted attacks~\cite{intro_he,safe_design_pattern,safe_survey_on_agentic_security}—are well-documented, the transition to multi-tool orchestration significantly expands the system’s attack surface. In multi-tool systems, security is no longer determined by the legitimacy of isolated API calls but by complex, cross-tool, and cross-timestep execution structures. These interactions breach traditional trust boundaries, where localized malicious inputs or model hallucinations are rarely isolated incidents. Instead, they often induce "state pollution" through the tool-chain context, leading to cascading propagation of attacks. This chapter investigates the novel security challenges and defense mechanisms inherent in multi-tool orchestration. Based on the topological execution logic of tool chains, we categorize multi-tool security risks into two distinct yet interrelated dimensions: (1) security risks in parallel execution, and (2) security risks in long-horizon tool chains.

\subsection{Security Risks in Parallel Execution} \label{subsec:safety_parallel}
In multi-tool orchestration architectures, parallel execution significantly enhances agent efficiency in complex task processing; however, it introduces latent security vulnerabilities. While parallelization is relatively safe for side-effect-free read operations (e.g., web retrieval, information queries), the integration of write operations (e.g., database updates, transaction commits) within tool chains risks race conditions, leading to system state inconsistencies. Consequently, the central challenge of multi-tool parallelization lies in maintaining state integrity within concurrent environments.

Recent work shows that this risk is not merely an implementation detail. Les Dissonances demonstrates that when LLMs invoke interdependent tools concurrently, malicious payloads can propagate across tool boundaries and contaminate later execution~\cite{safe_Les_Dissonances}. Related analysis of stateful tool use in multi-turn dialogue finds that reliability failures often follow a recognizable path from parameter drift to state contamination and then to irrecoverable downstream calls~\cite{safe_Multi-Turn}. SagaLLM makes this issue explicit by modeling tool use as bounded workflows with transactional semantics and rollback logic~\cite{safe_SagaLLM}. Taken together, these studies suggest that safe parallelism requires more than better prompting; it requires explicit control over side effects and execution scope.

To address these challenges, recent studies have begun to introduce transaction management paradigms from distributed systems and software engineering into LLM orchestration frameworks. Current solutions demonstrate an evolutionary trajectory from "pre-execution static constraints" to "in-execution transaction management", and ultimately to "post-execution dynamic verification":

\textit{Pre-execution Static Constraints:}
Intercepting potentially conflicting actions before tools are concurrently triggered serves as the first line of defense. At the system implementation level, the AARM specification ~\cite{safe_ARRM} and AgentSpec ~\cite{safe_agentspec} propose interception-based runtime enforcement mechanisms. These approaches block illegal concurrent requests by deeply injecting intent alignment and security policies prior to execution. For complex environments, Self-Verification Sampling ~\cite{safe_sampling} preemptively filters out high-risk write operations through pre-execution self-verification, thereby mitigating the catastrophic probability of irreversible post-write scenarios.

\textit{In-execution Transaction Management:}
Since concurrent errors are inherently difficult to eradicate entirely during execution, researchers have begun endowing agents with transaction semantics. SagaLLM ~\cite{safe_SagaLLM} and the recent Atomix framework have made breakthroughs by treating tool invocations as bounded workflows. Atomix introduces epoch-based concurrency isolation and resource frontier tracking ~\cite{safe_Atomix}, allowing harmless side effects to undergo a delayed commit while enabling secure rollbacks via compensation logic in the event of an abort. This design mirrors the localized checkpoint mechanism in the Generator-Assistant framework~\cite{safe_Generator-Assistant}, which strictly confines the impact of errors to localized scopes.

\textit{Post-execution Dynamic Verification:}
To address anomalous scenarios beyond the reach of static constraints, post-execution verification logic must be integrated into the system to form a closed-loop feedback mechanism. CRITIC ~\cite{safe_CRITIC} achieves self-correction through interactive critiquing; VerifiAgent~\cite{safe_verifiagent} engineers verification logic into a collaborative Verifier-Executor component architecture; meanwhile, DVR~\cite{safe_DVR} approaches the issue from a graph-structured scheduling perspective, seamlessly embedding verification subgraphs into the primary execution graph. This provides a universal template for the security auditing of parallel execution.

\input{tables/safety_tab}

\subsection{Security Risks in Long-Horizon Tool Chains} \label{subsec:safety_chain}
In long-horizon tool chains, security threats manifest as stealthy propagation and cascading amplification across temporal dimensions. As agents extend multi-step invocation sequences, the continuous introduction of intermediate variables and parameter bindings renders the system susceptible to Agent Drift~\cite{intro_agentdrift}. Here, the pollution of the context window and the accumulation of autoregressive errors cause behavioral patterns to deviate from established safety baselines. In adversarial settings, this "contextual drift" evolves into destructive Cross-Tool State Contamination. Once malicious data are introduced into intermediate results, its indiscriminate reuse in subsequent steps triggers a cascading failure that eventually collapses the trust boundaries of the system.

Centering on this temporal vulnerability, recent literature identifies novel attack modes unique to long-horizon orchestration. Attacks have evolved beyond single-step manipulation to sophisticated memory poisoning and plan injection. For instance, MINJA-style attacks inject malicious content through apparently benign read operations and allow it to persist in memory until it influences later actions~\cite{safe_minja}. Plan Injection attacks instead manipulate the structure of multi-step reasoning itself, steering the agent around guardrails without requiring an obviously malicious final command~\cite{safe_deeper_harm}. Butterfly Effects in Toolchains further shows how small upstream deviations can be amplified through variable binding and parameter reuse, turning downstream tools into confused deputies whose unsafe behavior is rooted in much earlier steps~\cite{safe_Butterfly}.

To mitigate malicious accumulation and state hijacking during long-horizon execution, defense mechanisms are transitioning from single-step result filtering to trajectory-level dynamic intervention and closed-loop auditing. Reflexion provides an early form of revision through verbal feedback memory, but later work places greater emphasis on execution-grounded verification~\cite{inf_reflex}. TRICE, Tool-MVR, and ToolReflection use execution evidence, multi-agent verification, or structured returns to identify and correct errors over longer trajectories~\cite{safe_TRICE, safe_tool-mvr, safe_ToolReflection}. LATS combines search with rollback and circuit-breaking behavior when downstream risks become visible~\cite{safe_LATS}. Other systems focus on diagnosis rather than rejection alone: AgentDoG emphasizes root-cause attribution across the full execution chain~\cite{safe_AgentDoG}, while Thought-Aligner intervenes earlier by correcting risky internal reasoning before unsafe actions are produced~\cite{safe_think}. The overall direction is toward making long-horizon execution inspectable, interruptible, and repairable.

In addition to introducing dynamic auditing during the inference phase, specialized training methodologies tailored for long-horizon multi-tool security are crucial for raising the upper bound of defense from the foundational model level. Because traditional RLHF based on single-turn dialogues struggles to cover the complex state space of long-horizon tool chains, model alignment is evolving toward encompassing the entire multi-step invocation trajectory. The latest agent security alignment frameworks introduce customized Sandbox environments for reinforcement learning ~\cite{safe_agent_safety}, compelling models to execute a "perform-reject-verify" confirmation mechanism when confronted with ambiguous or malicious inputs. Concurrently, to combat memory pollution induced by long-horizon interactions, Episodic Memory Consolidation mechanisms ~\cite{intro_agentdrift} are being employed to periodically compress interaction histories and purge potential malicious payloads. By integrating adversarial tool chains generated via automated red teaming as negative samples for alignment fine-tuning, the underlying LLMs' resilience against interference and the robustness of their parameter extraction—especially when facing unreliable memories and malicious intermediate states—are fundamentally enhanced.

%% file: tables/safety_tab.tex
\begin{table*}[t]
  \centering
  \caption{Representative safety risks and defense mechanisms for multi-tool agents.}
  \label{tab:safety_risk_defense}
  \footnotesize
  \setlength{\tabcolsep}{4.5pt}
  \renewcommand{\arraystretch}{1.05}

  \begin{tabularx}{\textwidth}{@{}>{\raggedright\arraybackslash}p{2.3cm} >{\raggedright\arraybackslash}p{3.7cm} >{\raggedright\arraybackslash}X@{}}
    \toprule
    \textbf{Risk} & \textbf{Failure Mode} & \textbf{Defense} \\
    \midrule

    Concurrent Conflict
    & Race conditions, write interference, and corrupted shared states.
    & Dependency-aware scheduling, policy gating, and pre-dispatch verification
    \cite{safe_Les_Dissonances,safe_ARRM,safe_agentspec}. \\
    \addlinespace[2pt]

    Partial Commit
    & Irreversible side effects leave globally inconsistent states.
    & Transaction-style isolation, delayed commit, compensation, and rollback
    \cite{safe_SagaLLM,safe_Atomix,safe_Generator-Assistant}. \\
    \addlinespace[2pt]

    Post-Execution Anomaly
    & Safety violations emerge only after tool execution.
    & Critic/verifier-based auditing and execution-time checking
    \cite{safe_CRITIC,safe_verifiagent,safe_DVR}. \\
    \addlinespace[2pt]

    Memory/Plan Poisoning
    & Malicious context persists across steps and reactivates later.
    & Trajectory-level verification, meta-verification, and evidence-based rejection
    \cite{safe_minja,safe_TRICE,safe_tool-mvr}. \\
    \addlinespace[2pt]

    Long-Horizon Drift
    & Error accumulation, state pollution, and confused-deputy behavior.
    & Lookahead search, rollback, and trace diagnosis
    \cite{safe_Butterfly,safe_LATS,safe_AgentDoG}. \\

    \bottomrule
  \end{tabularx}
\end{table*}

%% file: 5_Efficiency.tex
\section{Efficiency of Multi-Tool Agents} \label{sec:efficiency}
The integration of external tools has substantially enhanced the capability of agents to address complex tasks. Nevertheless, the reliance on multi-step inference and sequential tool invocation introduces significant efficiency bottlenecks. This chapter reviews emerging optimization strategies from two primary perspectives: mitigating latency in multi-tool chains (\S\ref{ssec: Latency in multi-tool chains}) and managing tool call cost and inference budget (\S\ref{ssec: Tool call cost and inference budget}).

\subsection{Latency in multi-tool chains}
\label{ssec: Latency in multi-tool chains}
The strict dependency of LLM inference on the completion and return of previous tool results leads to a linear or even exponential accumulation of end-to-end system latency with the number of task steps~\cite{eff_LLMCompiler}. To address this efficiency bottleneck, recent research has focused on refactoring the execution mechanism, evolving into three core optimization paradigms: parallel execution (\S\ref{sssec:Parallel execution}), asynchronous decoupling (\S\ref{sssec:Asynchronous decoupling}), and speculative reasoning (\S\ref{sssec:Speculative reasoning}).

\subsubsection{Parallel execution}
\label{sssec:Parallel execution}
To address the cumulative delays and inefficiencies caused by sequential execution in long-chain multi-tool calls, parallel execution~\cite{eff_para} has become a key optimization direction. Its core idea is to decompose tasks, identify subtasks with no interdependencies, and execute them simultaneously. SoT~\cite{eff_sot} accelerates inference by expanding multiple skeleton branches in parallel instead of following a fully sequential decoding pattern. At the tool level, LLMCompiler~\cite{eff_LLMCompiler} explicitly plans task dependencies so that independent tool calls can run concurrently. For longer workflows, M1-Parallel~\cite{eff_M1Parallel} further decomposes sequential tasks into independent subtasks coordinated by multiple agents. Parallelized planning-acting~\cite{eff_ppl} extend the same idea by overlapping planning and execution, allowing later steps to be prepared while current actions are still running. Taken together, these approaches show that parallel execution is most effective when dependency structure is explicit and side effects are sufficiently controlled to avoid invalid concurrent actions.

\subsubsection{Asynchronous decoupling}
\label{sssec:Asynchronous decoupling}
Traditional agent architectures generally follow a synchronous observation-thinking-action loop. LLMs must block and wait for I/O results after calling external tools. This coupling of computation and execution leads to severe latency accumulation~\cite{inf_react}. To break this serial bottleneck, recent studies have proposed asynchronously decoupling the planning and acting of agents.
To overcome the I/O blocking limitations and strict serial dependencies faced by traditional agents during long toolchain calls, asynchronous decoupling significantly improves concurrent throughput by separating computationally intensive inference planning from I/O-intensive tool execution at the system architecture level. In recent years, as models have been increasingly applied in dynamic environments, this decoupling has gradually evolved into the core foundation for environment-aware task planning and multi-node decentralized inference. For example, ~\cite{eff_ppl} is proposed that completely decouples macro-level environment-aware planning from low-level action execution, and uses a centralized agent memory system for asynchronous state synchronization, thoroughly mitigating inference latency caused by dynamic environmental changes. When dealing with complex constraint logic, MACI~\cite{eff_maci} has demonstrated that separating task planning from external tool verification effectively prevents attention shift and constraint drift in large models during long-term calls. This decoupling paradigm, also applied in cybersecurity attack and defense and GUI, Incalmo~\cite{eff_incal} reveals that decoupling macro-level strategic intent from low-level shell script execution is key to preventing agents from getting stuck in local optima. Temac~\cite{eff_temac} specifically designed a Planner-Actor heterogeneous mechanism, using independent executors to asynchronously call interface probing tools.

\subsubsection{Speculative reasoning}
\label{sssec:Speculative reasoning}
Unlike asynchronous decoupling, which optimizes from the system architecture perspective, Speculative Execution borrows the branch prediction concept from computer architecture, aiming to mask waiting latency in long toolchains through algorithmic forward prediction. This mechanism is underpinned by speculative decoding~\cite{eff_spedec}, which allows the model to speculatively generate subsequent action identifiers without waiting for a complete autoregressive loop. In the tool invocation domain, ReWoo~\cite{eff_rewoo} is a typical example of this approach. By decoupling inference from observation, it speculatively generates a complete execution graph containing expected tool outputs, committing execution only when the speculation is correct, thus significantly reducing redundant latency when interacting with external tool environments. To improve the accuracy of speculation, LATS~\cite{safe_LATS} introduces a speculative simulation mechanism based on lookahead search. Before actually invoking an irreversible API, the agent infers and evaluates multiple possible tool interaction trajectories. Similarly, ToT~\cite{eff_tot} utilizes speculative branch exploration to predict the final benefits of different tool combinations, significantly reducing the backtracking time cost caused by tool invocation failures in complex tasks. For retrieval tools that suffer from particularly high latency, REST~\cite{eff_rest} proposes a retrieval-based speculative decoding framework that can predict the subsequent generated content of a large model in advance based on cached results from an external knowledge base. At the engineering implementation and pipeline optimization level, the DSPy~\cite{eff_dspy} framework uses a compile-level declarative pipeline to speculatively route and merge multiple tool call requests, minimizing runtime latency across multiple toolchains without sacrificing task success rate.

\subsection{Tool call cost and inference budget}
\label{ssec: Tool call cost and inference budget}
Besides time latency, multi-tool agent invocation also faces high inference costs and limited context budgets. Frequent invocation of large-parameter models to process simple tools or filling massive tool descriptions into the context all at once leads to wasted resources and performance degradation~\cite{eff_eco}. To address these pain points, existing research mainly focuses on optimization along three core dimensions: dynamic tool retrieval (\S\ref{sssec:Dynamic tool search}), adaptive model routing (\S\ref{sssec:Adaptive Model Routing}), and intelligent caching and memory (\S\ref{sssec:Intelligent caching and memory}).

\subsubsection{Dynamic tool search}
\label{sssec:Dynamic tool search}
Faced with hundreds or thousands of external tool libraries, hardcoding all API descriptions into prompts at once not only consumes a huge input token budget but also easily leads to the Lost in the Middle phenomenon caused by long texts~\cite{eff_lost}. To efficiently organize tools within a limited context window, the current mainstream paradigm has shifted to query-based dynamic retrieval and representation compression. For example, the ToolLLM~\cite{ben_ToolBench} framework introduces a neural API retrieval engine. Before executing a task, the agent retrieves the Top-K most relevant tool descriptions based on the user's instructions, thus avoiding irrelevant tools crowding out the context space. For ultra-large API libraries with tens of thousands of entries, AnyTool~\cite{tune_anytool} further designs a hierarchical API retrieval mechanism, using a categorized directory tree for progressive filtering, significantly reducing the retrieval cost and context overhead of a single call. Furthermore, at the representation compression level, ToolkenGPT~\cite{eff_toolkengpt} proposes an innovative vocabulary expansion strategy that compresses and maps the lengthy tool invocation logic and its parameter descriptions into a single special term. This method allows the agent to trigger the tool simply by predicting the corresponding Toolken at the output, fundamentally bypassing the rigid limitation of using lengthy descriptions as input prompts.

\subsubsection{Adaptive Model Routing}
\label{sssec:Adaptive Model Routing}
To control computational costs, multi-agent architectures have gradually abandoned the crude approach of a single giant model scheduling everything, shifting towards a computational allocation strategy that dynamically matches model size to task difficulty~\cite{eff_orche}. Model cascading mechanisms, exemplified by FrugalGPT~\cite{eff_frugalgpt}, prioritize low-cost models, only routing to expensive large models when confidence is insufficient, significantly reducing budget without sacrificing accuracy. With further refinement, RouteLLM~\cite{intro_routellm} trained a dedicated network that adaptively distributes requests among models of different sizes based on task complexity. In multi-step tasks, SwiftSage~\cite{eff_swiftsage} borrows from the human dual-system approach, layering its architecture into lightweight edge modules handling routine operations and a central large model responsible for complex inference, achieving globally optimal allocation of computational resources.

\subsubsection{Intelligent caching and memory}
\label{sssec:Intelligent caching and memory}
In complex environments with high-frequency interactions, repeated erroneous attempts and highly similar API requests are the core factors consuming inference budgets, while building a structured Agent Memory and caching layer provides key support for breaking this bottleneck. Firstly, regarding reducing trial-and-error costs, Reflexion~\cite{inf_reflex} framework proposes an efficient alternative that does not require updating model weights. Through Verbal Reinforcement Learning, it allows the agent to reflect on its failures and persistently store the resulting improvement strategies in Episodic Memory. In subsequent attempts, the agent can accurately avoid historical errors by calling this structured memory, significantly reducing the number of interaction rounds and total computational overhead required to reach a successful task state. To address the memory capacity bottleneck in long-cycle tasks, MemGPT~\cite{eff_memgpt} introduces the concept of virtual memory management from operating systems into large models. By autonomously paging and swapping memory between the fast context and external database, the agent can maintain an infinite memory length with an extremely low token budget. Furthermore, at the level of API call result reuse, semantic caching systems such as GPTCache~\cite{eff_gptcache} perform vectorized comparison of input queries and tool parameters. When semantically similar duplicate requests are detected, historical results are directly returned from the local machine, completely blocking redundant LLM inference and external network request billing.

%% file: 6_Completeness.tex
\section{Completeness}
\label{sec:completeness}
In complex open-domain scenarios, multi-tool agents are gradually transitioning from closed, static API executions to dynamic interactions within real-world environments~\cite{eff_5ai}. However, the inherent unpredictability of these environments frequently leads to functional incompleteness, wherein predefined toolkits fail to fully satisfy emerging task requirements~\cite{com_latm}. This chapter systematically reviews the challenges and corresponding solutions regarding the tool completeness of multi-tool agents. These solutions are conceptualized as a continuous cognitive loop comprising problem diagnosis, dynamic compensation, and long-term evolution. Specifically, this loop progresses sequentially through three core phases. The first phase is capability boundary perception (\S\ref{ssec: Capability boundary perception}), designed to identify tool deficiencies and operational failures. The second phase is autonomous tool expansion (\S\ref{ssec: Autonomous tool expansion}) is achieved through code generation and tool synthesis. Ultimately, the process culminates in open environment adaptation (\S\ref{ssec: Open environment adaptation}), which leverages experience and memory mechanisms for sustained operation.

\begin{figure}[t]
    \centering
    \includegraphics[width=\linewidth]{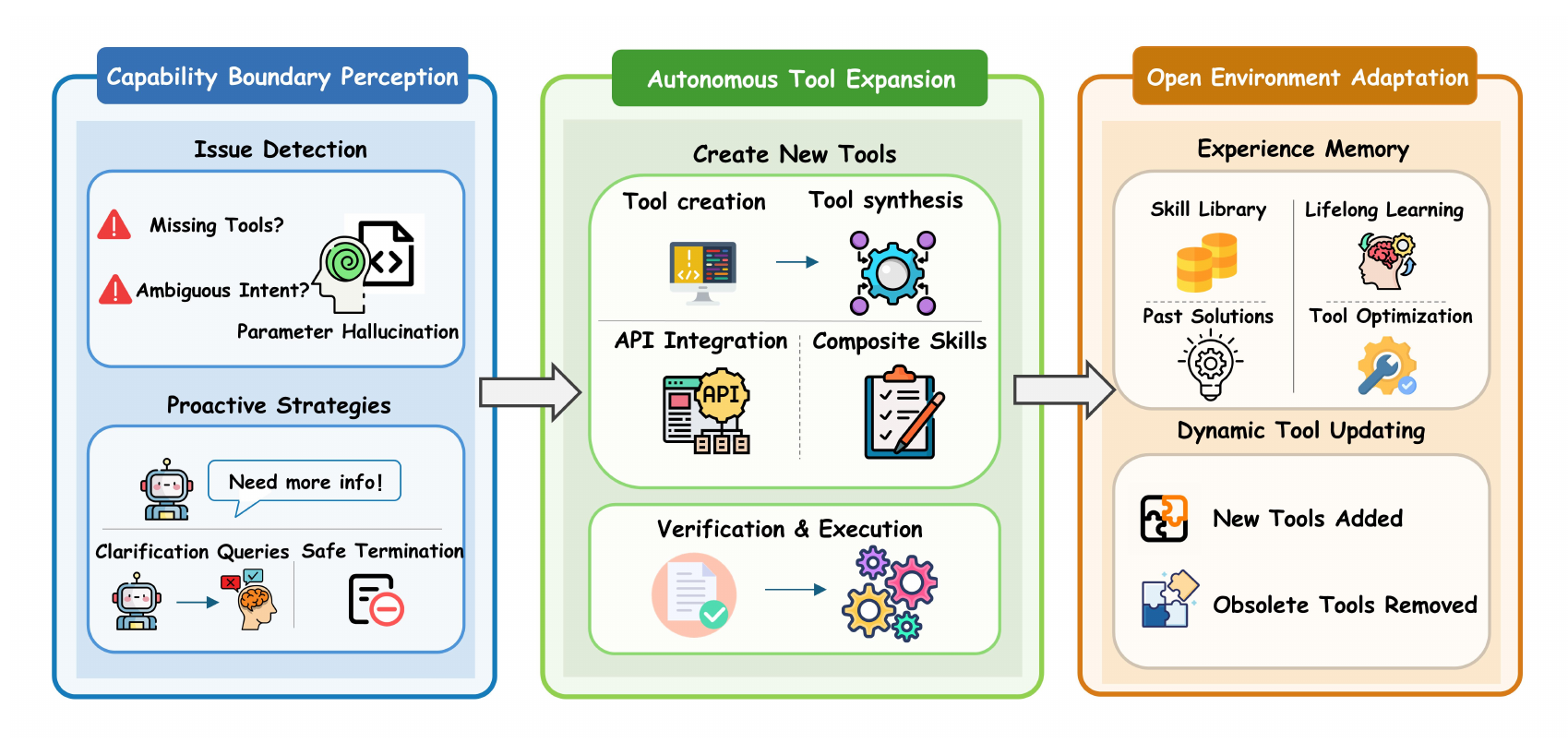}
    \caption{Overview of the completeness evolution in multi-tool agents. The framework characterizes functional completeness as a continual process with three sequential yet interrelated stages: capability boundary perception, which identifies missing tools, ambiguous intents, and parameter anomalies through clarification and safe termination; autonomous tool expansion, which compensates for capability gaps via tool creation, tool synthesis, API integration, and composite skill construction; and open environment adaptation, which consolidates successful tools and interaction traces into reusable experience memory to support dynamic updating and long-term evolution in changing environments.}
    \label{fig:Completeness}
\end{figure}

\subsection{Capability boundary perception}
\label{ssec: Capability boundary perception}
The logical starting point for the completeness of a multi-tool agent lies in its ability to accurately identify intent denial and detect anomalies, that is, to perceive its own capability boundaries. Traditional language models, when faced with tasks beyond the capabilities of the toolkit, are prone to hallucinate tools, creating fictitious API nodes or parameter formats. Therefore, accurately identifying incomplete conditions is the core of ensuring system robustness. Several studies have already quantified the limitations of models in terms of the boundaries of perception capabilities. The FAIL-TALMS~\cite{com_failtalms} demonstrates that large language models frequently lack boundary awareness when dealing with ambiguous intentions or missing tools, rendering them prone to blind trial and error. Furthermore, ToolHaystack~\cite{com_toolhaystack} reveals that during extended complex interactions involving critical information deficits, models are highly susceptible to parameter hallucinations induced by historical noise. To compensate for these deficiencies, the design paradigm of multi-tool systems is shifting from passive execution toward active intervention through two primary approaches. The first approach involves proactive clarification. Frameworks such as AskToAct~\cite{com_asktoact} exemplify this by dynamically triggering multi-turn clarification mechanisms upon detecting missing parameters. This strategy prevents error accumulation via self-correction and substantially improves both the accuracy and efficiency of tool invocations. The second approach relies on proactive termination to counteract the inherent action compulsion observed in large models. By equipping agents with safety mechanisms to identify uncertainty and autonomously abort execution in high-risk or highly ambiguous environments, systems can achieve substantial security improvements with minimal task completion overhead.

\subsection{Autonomous tool expansion}
\label{ssec: Autonomous tool expansion}

When an agent successfully perceives that the existing toolset cannot meet its current needs, the system needs to undergo a paradigm shift from tool user to tool creator. Autonomous tool extension leverages the low-level code generation capabilities or multi-step planning capabilities of large language models to dynamically construct temporary tools or composite skills at runtime, thereby filling capability gaps. A breakthrough in this field lies in decoupling abstract reasoning from concrete execution. The LATM~\cite{com_latm} framework innovatively divides agents into tool creators and tool users. Faced with new tasks, creators generate customized tools by writing Python functions and automatically generate test cases for verification. Once verified, the tools are executed by the user, achieving zero-sample tool creation. The CREATOR~\cite{com_creator} follows the same logic, transforming abstract problems such as mathematical reasoning and table processing into concrete Python scripts, enabling agents to bypass the limitations of static APIs and autonomously build dynamic computation tools. Besides creating tools from scratch, assembling fine-grained basic tools into high-level composite skills is also an effective path for capability extension. RestGPT~\cite{com_restgpt} explores how to use large models to plan and connect multiple real RESTful APIs to complete complex long-range tasks, combining isolated network interfaces into super tools with macroscopic semantics. ToolMaker~\cite{com_toolmaker} further enhances this process. Agents can not only dynamically synthesize tool code to solve specific subtasks, but also reuse these newly created tools within the current context, greatly improving invocation efficiency in multi-round interactions. By using the code interpreter as a Meta-tool, Agents break free from the constraints of predefined interfaces, enabling on-demand dynamic scaling of capabilities.

\subsection{Open environment adaptation}
\label{ssec: Open environment adaptation}

If boundary awareness and tool expansion address the lack of completeness in a single task, then open environment adaptation aims to solve the evolution problem of multi-tool agents over time. With the iterative updates of external environment API versions, the continuous integration of new tools, and the obsolescence of older tools, agents must possess memory mechanisms and lifelong learning capabilities to accumulate temporarily generated tools or successful call traces into long-term experience~\cite{com_review}.

In this field, deeply integrating the agent's memory architecture with tool calling has become a mainstream trend. Voyager~\cite{com_voyager} first demonstrated this in an open environment. Through continuous exploration and feedback, the agent stores successfully executed code into an ever-expanding skill library, which functions as an external memory based on a vector database. When faced with new challenges, the agent can retrieve and combine historical skills, achieving a leapfrog evolution in tool usage capabilities. The ExpeL~\cite{com_expel} framework systematizes the experience learning process. The agent autonomously collects successful or failed tool call traces during training tasks and extracts cross-task insights using natural language, solidifying them into experience memory to effectively avoid repeating mistakes in subsequent API pitfalls.

In the dynamic addition and deletion adaptation of massive tools, ToolkenGPT~\cite{eff_toolkengpt} represents each newly added tool as a special tool token. When facing a new environment, the model only needs to optimize the embedding of these tool tokens, without needing to fine-tune the huge base model, achieving plug-and-play and lightweight adaptation of the tool library. AppAgent~\cite{com_appagent} demonstrates how, in the highly dynamic graphical interaction environment of smartphones, the agent updates its interaction memory for different apps through continuous exploration and clicks to adapt to UI or application version updates. In summary, by introducing long-term and short-term experience pools and dynamic skill memory, the agent can transform the instantaneous tool creation of an individual into the system's experience inheritance, ultimately maintaining true functional completeness in an ever-evolving external environment.

%% file: 7_Benchmark.tex
\section{Benchmarks}
\label{sec:benchmarks}

The evaluation paradigm for multi-tool agents has undergone a decisive shift: from measuring isolated API invocation correctness to assessing system-level intelligence in sustaining, adapting, and repairing extended tool-use trajectories under structural, temporal, and environmental constraints. This evolution reflects a fundamental recognition—long-horizon tool orchestration is not merely the linear accumulation of atomic calls, but a holistic capability demanding topological reasoning, persistent state management, and dynamic adaptation. Early benchmarks like API-Bank~\cite{ben_APIBank}, MetaTool~\cite{ben_MetaTool} and ToolBench~\cite{ben_ToolBench} established foundational capabilities in intent detection and tool retrieval; however, contemporary research prioritizes four interdependent dimensions that define long-horizon intelligence of agents: (1) \textbf{topological complexity} of tool dependencies, (2) \textbf{temporal scale} of execution trajectories, (3) \textbf{environmental dynamism} during interaction, and (4) \textbf{state persistence} with self-correction mechanisms.

\input{tables/benchmarks_longhorizon}

\subsection{Topological Complexity} \label{subsec:topo_complexity}
One line of work examines whether agents can reason over non-linear tool dependencies rather than simply follow sequential scripts. The main challenge is that the correct execution order is determined by latent causal structure: intermediate outputs must be produced, preserved, and passed to later calls under branching, nesting, or multi-hop dependencies. In this setting, failure often arises not from choosing the wrong tool in isolation, but from constructing the wrong dependency graph.

Recent benchmarks expose this issue in several ways. Some emphasize nested and compositional structure, where the output of one call must be embedded into another, sometimes under strict parameter constraints~\cite{ben_NESTFUL, ben_SealTools, ben_ComplexFuncBench}. Others evaluate graph-shaped execution, including branching decisions and parallel paths, and require agents to determine which calls can proceed concurrently and which must remain causally ordered~\cite{ben_TRAJECTBench, ben_ELTBench, ben_TaskBench}. A further group focuses on distributed information dependencies, where the parameters required for correct execution are scattered across stages, modalities, or speakers~\cite{ben_MSCoRe, ben_mnm, ben_dice}. Related benchmarks such as M$^3$-Bench, MMAU, and FuncBenchGen further isolate graph reasoning, multimodal dependency tracking, and robustness to structurally valid distractors~\cite{ben_m3bench, ben_MMAU, ben_FuncBenchGen}.

Taken together, these benchmarks suggest that long-horizon tool use should be evaluated as graph-structured reasoning under execution constraints, rather than as a longer version of single-step tool selection.

\subsection{Temporal Scale} \label{subsec:temporal_scale}
A second dimension concerns how performance changes as trajectories grow longer. Recent work has moved beyond vague references to long-horizon tasks and instead defines horizon length through approximate step counts, token budgets, or accumulated constraints. These settings reveal a distinct class of failures, including memory decay, context saturation, and the way small local errors can accumulate into global task failure.

The Tool Decathlon~\cite{ben_tooldec} establishes a baseline for long-horizon evaluation with tasks averaging 20 interactive rounds across diverse domains. ShortcutsBench~\cite{ben_shortcut} tests real-world sequences by mapping multi-stage OS workflows directly to complex parameter loops, bridging the gap between natural language prompts and executable low-code horizons. HammerBench~\cite{ben_hammer} evaluates multi-turn conversation across 10,000 instances, while AgentLongBench~\cite{ben_along} scales the temporal horizon dramatically, executing dynamic environment rollouts up to 4 million tokens, demonstrating that models fail precipitously at dynamic information synthesis across extreme horizons.

UltraHorizon~\cite{ben_ultra} defines the ultra-long-horizon frontier with trajectories averaging 400+ tool calls and 200K+ tokens, rigorously testing context window saturation and memory decay. TRIP-Bench~\cite{ben_trip} constructs travel planning dialogues exceeding 150 tool invocations with hard constraints (budget, time), exposing planning fragility—where single constraint violations cascade into total failure. This is expanded upon by DeepPlanning~\cite{ben_deepplan}, which forces verifiable global constraint optimization across multiple simulated days of execution, providing controlled evidence that modern agents degrade rapidly when localized reasoning conflicts with long-term budgeting. Critically, these benchmarks show steep degradation as horizon length and global constraints increase, suggesting that temporal scale is a major bottleneck.

% \textbf{Efficiency-Aware Long-Horizon Execution.} CostBench~\cite{bench_costbench} introduces cost-sensitivity as a core dimension, evaluating whether agents minimize API expenses and latency while completing extended tasks. TPS-Bench~\cite{bench_tpsbench} assesses scheduling efficiency under temporal constraints, requiring agents to optimize execution order for compound tasks. This shift from completion to efficient completion reflects maturation in long-horizon evaluation philosophy. RestBench~\cite{bench_restbench} evaluates efficient sequential interactions with real-world RESTful APIs. This shift from completion to efficient completion reflects maturation in long-horizon evaluation philosophy.

\subsection{Dynamic Environments} \label{subsec:dynamic_env}
Real-world tool use occurs in environments that evolve independently of agent actions. Contemporary benchmarks immerse agents in progressively dynamic settings—from stateful simulators to human-in-the-loop systems—testing adaptation to execution feedback, user corrections, and environmental perturbations.

\textbf{Stateful Interactive Environments.} ToolSandbox~\cite{ben_toolsandbox} provides a stateful conversational environment where intermediate tool outputs alter subsequent state space, requiring agents to track evolving context. AppWorld~\cite{ben_appworld} fundamentally shifts the verification paradigm from matching syntactic trajectories to evaluating persistent database state changes across nine heavily interconnected applications, requiring complex interactive coding to mutate a persistent digital world. ToolGym~\cite{ben_toolgym} injects runtime interruptions to simulate real-world uncertainty, evaluating robustness to environmental volatility. StableToolBench~\cite{ben_stabletool} addresses evaluation reproducibility by virtualizing APIs with caching layers, isolating agent performance from external API instability.

\textbf{Human-in-the-Loop Environments.} $\tau$-bench~\cite{ben_tau} and $\tau^2$-bench~\cite{ben_tau2} construct high-fidelity user simulators that issue constraint changes, ambiguity requests, and corrections during multi-turn interactions. Critically, $\tau^2$-bench formalizes a dual-control paradigm where agent and user jointly manipulate a shared world state through tools, requiring agents to negotiate action space and manage user expectations—a capability where even advanced models exhibit >40\% performance degradation versus single-agent settings. MTU-Bench~\cite{ben_mtu} extends this to multi-turn user guidance scenarios with evolving task specifications. T1~\cite{ben_t1} evaluates multi-turn conversational agents with a focus on inter-tool dependencies and replanning, utilizing an integrated caching mechanism for short and long-term memory. COMPASS~\cite{ben_compass} assesses strategic tool use and user preference optimization in multi-turn travel-planning interactions. Crucially, WildToolBench~\cite{ben_wild} grounds evaluation in the nature of human behavior, testing agents against implicit intent, instruction transitions, and messy conversational data.

\textbf{Real-World OS and Web Environments.} OSWorld~\cite{ben_osworld} provides a state-of-the-art interactive computer environment supporting hundreds of open-domain tasks, revealing massive deficiencies in visual GUI grounding. GTA~\cite{ben_gta} deploys human-written queries requiring perception and logic tools across real-world multimodal contexts. Windows Agent Arena~\cite{ben_windows} and Mobile-Bench~\cite{ben_mobile} evaluate cross-application operations within native GUI environments, where visual noise and UI variability challenge perception-action loops. RealWebAssist~\cite{ben_realweb} uses authentic human-recorded web navigation trajectories to assess agents' ability to interpret implicit user intent from sequential actions. VitaBench~\cite{ben_vita} targets diverse real-world applications with heterogeneous tool interfaces, emphasizing adaptation to domain-specific interaction patterns.

\subsection{State Persistence and Self-Correction} \label{subsec:state_persistence}

Sustaining extended trajectories demands not only planning but also mechanisms for state persistence and error recovery. In many practical workflows, success depends on maintaining intermediate artifacts, remembering prior commitments, and detecting when the current trajectory has drifted away from task requirements. Benchmarks along this axis therefore examine not only final success, but also whether the agent can recover from partial failure.

Some benchmarks focus on persistent state management, asking agents to maintain intermediate resources or task variables across extended execution traces~\cite{ben_toolsandbox, ben_appworld, ben_m3bench}. Others explicitly target self-correction and recovery, where agents must identify inconsistencies, revise previous actions, and continue toward task completion rather than restart from scratch~\cite{ben_tau2, ben_t1, ben_wild}. Closely related evaluations also treat process supervision as a first-class target, measuring whether the agent followed a valid path even when the final outcome appears acceptable~\cite{ben_Toolcomp, ben_ComplexFuncBench}.

This line of work highlights an important methodological point: long-horizon evaluation should not rely solely on endpoint success. For multi-tool systems, the more informative question is often whether a benchmark isolates a specific failure mode, exposes the relevant intermediate signals, and distinguishes between accidental success and genuinely robust orchestration.

%% file: tables/benchmarks_longhorizon.tex
\begin{table*}[htbp]
  \centering
  \vspace{-2mm}
  \caption{\textbf{Taxonomy of Long-Horizon Multi-Tool Benchmarks (2023--2026).} Benchmarks are categorized by four core dimensions defining long-horizon intelligence. Key metrics include tool space size (Number of Tools), evaluation scale (Number of Instances). \textit{Env Type}: \textbf{S}=Static Dataset, \textbf{I}=Interactive Simulator, \textbf{H}=Human/Dual-Control, \textbf{R}=Real-World OS/Web.}
  \vspace{-2mm}
  \label{tab:benchmark_taxonomy}
  \scriptsize
  \setlength{\tabcolsep}{3pt}
  \renewcommand{\arraystretch}{0.78}

  \newcolumntype{C}{>{\centering\arraybackslash}X}

  \begin{tabularx}{\textwidth}{@{} l | C | C | C | C | C @{}}
    \toprule
    \textbf{Benchmark} & \textbf{\# Tools} & \textbf{\# Instances} & \textbf{Year} & \textbf{Link} & \textbf{Env Type} \\
    \midrule

    % \multicolumn{6}{@{}l}{\textbf{Topological Complexity}} \\
    % \midrule
    NESTFUL~\cite{ben_NESTFUL} & 900+ & 1,800+ & 2024 & \href{https://huggingface.co/datasets/ibm-research/nestful}{\footnotesize\faHome} & S \\
    ToolHop~\cite{ben_ToolHop} & 3,912 & 995 & 2025 & \href{https://huggingface.co/datasets/bytedance-research/ToolHop}{\footnotesize\faHome} & S \\
    TRAJECT-Bench~\cite{ben_TRAJECTBench} & 1,228 & 5,670 & 2025 & \href{https://huggingface.co/datasets/bigboss24/TRAJECT-Bench}{\footnotesize\faHome} & I \\
    ELT-Bench~\cite{ben_ELTBench} & 200 & 100 & 2025 & \href{https://github.com/uiuc-kang-lab/ELT-Bench}{\footnotesize\faGithub} & I \\
    Toolcomp~\cite{ben_Toolcomp} & 11 & 493 & 2025 & \href{https://github.com/vaskar-nath/toolcomp}{\footnotesize\faGithub} & S \\
    ComplexFuncBench~\cite{ben_ComplexFuncBench} & 43 & 1,000 & 2025 & \href{https://huggingface.co/papers/2501.10132}{\footnotesize\faHome} & S \\
    ToolBench~\cite{ben_ToolBench} & 3,451 & 126,486 & 2023 & \href{https://github.com/OpenBMB/ToolBench}{\footnotesize\faGithub} & R/I \\
    TaskBench~\cite{ben_TaskBench} & 103 & 28,271 & 2023 & \href{https://github.com/microsoft/JARVIS}{\footnotesize\faGithub} & S \\
    Seal-Tools~\cite{ben_SealTools} & 4,076 & 14,076 & 2024 & \href{https://github.com/fairyshine/Seal-Tools}{\footnotesize\faGithub} & S \\
    m\&m's~\cite{ben_mnm} & 33 & 4,427 & 2024 & \href{https://mnms-project.github.io/}{\footnotesize\faHome} & I \\
    FuncBenchGen~\cite{ben_FuncBenchGen} & Varies & Varies & 2025 & \href{https://github.com/megagonlabs/FuncBenchGen}{\footnotesize\faGithub} & S \\
    M$^3$-Bench~\cite{ben_m3bench} & 231 & 208 & 2025 & \href{https://github.com/EtaYang10th/Open-M3-Bench}{\footnotesize\faGithub} & I \\
    MSCoRe~\cite{ben_MSCoRe} & Varies & 126k+ & 2025 & \href{https://huggingface.co/datasets/032564yn/MSCoRe}{\footnotesize\faHome} & S \\
    MMAU~\cite{ben_MMAU} & Varies & 3k+ & 2024 & \href{https://github.com/apple/axlearn/tree/main/docs/research/mmau}{\footnotesize\faGithub} & S \\
    DICE-BENCH~\cite{ben_dice} & 124 & 1607 & 2025 & \href{https://huggingface.co/datasets/OfficerChul/DICE-BENCH}{\footnotesize\faHome} & S \\
    % \midrule

    % \multicolumn{6}{@{}l}{\textbf{Temporal Scale}} \\
    % \midrule
    Tool Decathlon~\cite{ben_tooldec} & 604 & 108 & 2025 & \href{https://github.com/hkust-nlp/Toolathlon}{\footnotesize\faGithub} & R \\
    TRIP-Bench~\cite{ben_trip} & 18 & 40+ & 2026 & \href{https://arxiv.org/pdf/2602.01675}{\footnotesize\faHome} & R \\
    UltraHorizon~\cite{ben_ultra} & 400+ & 168 & 2025 & \href{https://github.com/StarDewXXX/UltraHorizon}{\footnotesize\faGithub} & I \\
    AgentLongBench~\cite{ben_along} & Varies & Varies & 2026 & \href{https://github.com/euReKa025/AgentLongBench}{\footnotesize\faGithub} & I \\
    DeepPlanning~\cite{ben_deepplan} & 24 & 240 & 2026 & \href{https://qwenlm.github.io/Qwen-Agent/en/benchmarks/deepplanning/}{\footnotesize\faHome} & I \\
    ShortcutsBench~\cite{ben_shortcut} & 1,414 & 7,627 & 2024 & \href{https://eachsheep.space/ShortcutsBench/}{\footnotesize\faHome} & R \\
    HammerBench~\cite{ben_hammer} & 1,063 & 6,531 & 2024 & \href{https://github.com/MadeAgents/HammerBench}{\footnotesize\faGithub} & I \\
    % \midrule

    % \multicolumn{6}{@{}l}{\textbf{Environmental Dynamism}} \\
    % \midrule
    $\tau$-bench~\cite{ben_tau} & 28 & 165 & 2024 & \href{https://github.com/sierra-research/tau-bench}{\footnotesize\faGithub} & H \\
    $\tau^2$-bench~\cite{ben_tau2} & 68 & 279 & 2025 & \href{https://github.com/sierra-research/tau2-bench}{\footnotesize\faGithub} & H \\
    ToolSandbox~\cite{ben_toolsandbox} & 34 & 1,032 & 2024 & \href{https://github.com/apple/ToolSandbox}{\footnotesize\faGithub} & I \\
    ToolGym~\cite{ben_toolgym} & 5,571 & 1,170 & 2026 & \href{https://huggingface.co/ToolGym}{\footnotesize\faHome} & I \\
    VitaBench~\cite{ben_vita} & 66 & 400 & 2025 & \href{https://github.com/meituan-longcat/vitabench}{\footnotesize\faGithub} & I \\
    OSWorld~\cite{ben_osworld} & Varies & 369 & 2024 & \href{https://os-world.github.io/}{\footnotesize\faHome} & R \\
    AppWorld~\cite{ben_appworld} & 457 & 750 & 2024 & \href{https://appworld.dev/}{\footnotesize\faHome} & I \\
    GTA~\cite{ben_gta} & 14 & 229 & 2024 & \href{https://github.com/open-compass/GTA}{\footnotesize\faGithub} & R \\
    COMPASS~\cite{ben_compass} & 18 & 281 & 2025 & \href{https://github.com/sunnytqin/compass.git}{\footnotesize\faGithub} & I \\
    T1~\cite{ben_t1} & 14 & 13,500 & 2025 & \href{https://github.com/CapitalOne-Research/T1}{\footnotesize\faGithub} & S \\
    MTU-Bench~\cite{ben_mtu} & 136 & 159k & 2024 & \href{https://github.com/MTU-Bench-Team/MTU-Bench}{\footnotesize\faGithub} & S \\
    WildToolBench~\cite{ben_wild} & 1,600 & 1,024 & 2026 & \href{https://github.com/yupeijei1997/WildToolBench}{\footnotesize\faGithub} & S/H \\
    Windows Agent Arena~\cite{ben_windows} & Varies & 150+ & 2024 & \href{https://microsoft.github.io/WindowsAgentArena/}{\footnotesize\faHome} & R \\
    Mobile-Bench~\cite{ben_mobile} & 103 & 832 & 2024 & \href{https://github.com/XiaoMi/MobileBench}{\footnotesize\faGithub} & R \\
    RealWebAssist~\cite{ben_realweb} & Varies & 1,885 & 2025 & \href{https://scai.cs.jhu.edu/projects/RealWebAssist/}{\footnotesize\faHome} & R \\
    StableToolBench~\cite{ben_stabletool} & 16k+ & 126k+ & 2024 & \href{https://zhichengg.github.io/stb.github.io}{\footnotesize\faHome} & I \\
    % \midrule

    % \multicolumn{6}{@{}l}{\textbf{State \& Domain Specialization}} \\
    % \midrule
    CRITICTOOL~\cite{ben_criticaltool} & 87 & 2,740 & 2025 & \href{https://shellorley0513.github.io/CriticTool}{\footnotesize\faHome} & S \\
    OdysseyBench~\cite{ben_odyssey} & Office & 602 & 2025 & \href{https://github.com/microsoft/OdysseyBench}{\footnotesize\faGithub} & I \\
    MemAgentBench~\cite{ben_memagentbench} & 289 & 6,340 & 2025 & \href{https://huggingface.co/datasets/ai-hyz/MemoryAgentBench}{\footnotesize\faHome} & I \\
    ColorBench~\cite{ben_color} & 21 & 175 & 2025 & \href{https://github.com/MadeAgents/ColorBench}{\footnotesize\faGithub} & I \\
    ASTRA~\cite{ben_astra} & Varies & 2,413 & 2026 & \href{https://arxiv.org/abs/2603.01357}{\footnotesize\faHome} & I \\
    RoTBench~\cite{ben_rot} & 568 & 105 & 2024 & \href{https://github.com/Junjie-Ye/RoTBench}{\footnotesize\faGithub} & I \\
    ToolSword~\cite{ben_toolsword} & 100 & 440 & 2024 & \href{https://github.com/Junjie-Ye/ToolSword}{\footnotesize\faGithub} & I \\
    \bottomrule
  \end{tabularx}
  \vspace{-2mm}
\end{table*}

%% file: 8_Application.tex
\section{Application}

Following the analysis of reasoning, training, and core system properties, this chapter examines multi-tool LLM agent deployment in industrial settings. The primary challenge extends beyond basic tool invocation to achieving end-to-end objectives under practical engineering constraints. Since real-world workloads feature heterogeneous components and evolving states, execution strategies must seamlessly manage tool errors and distribution shifts. Consequently, effective deployment demands closed-loop orchestration based on continuous iterative execution, coupled with strict verification of all intermediate and final outcomes.

This approach builds upon neuro-symbolic architectures that enhance language models with external knowledge and discrete reasoning. Recent research integrating reasoning with actions highlights the necessity of interactive execution and systematic exception handling, while scalable tool-use learning has driven protocols that better reflect complex industrial realities. Consequently, rather than providing a domain-centric inventory, this chapter analyzes diverse applications as concrete instances of multi-tool orchestration. By characterizing specific tool ecosystems and abstracting their orchestration patterns, we examine methodologies for selecting, composing, and validating tool invocations. This approach highlights trade-offs among reliability, safety, and cost, establishing a unified framework to compare industrial practices.

\subsection{A Unified Abstraction for Application-Side Multi-Tool Orchestration}

% A unified abstraction for multi-tool orchestration frames tasks as executable processes driving an external system toward a target state.

\begin{figure}[t]
  \centering
  \includegraphics[width=\linewidth]{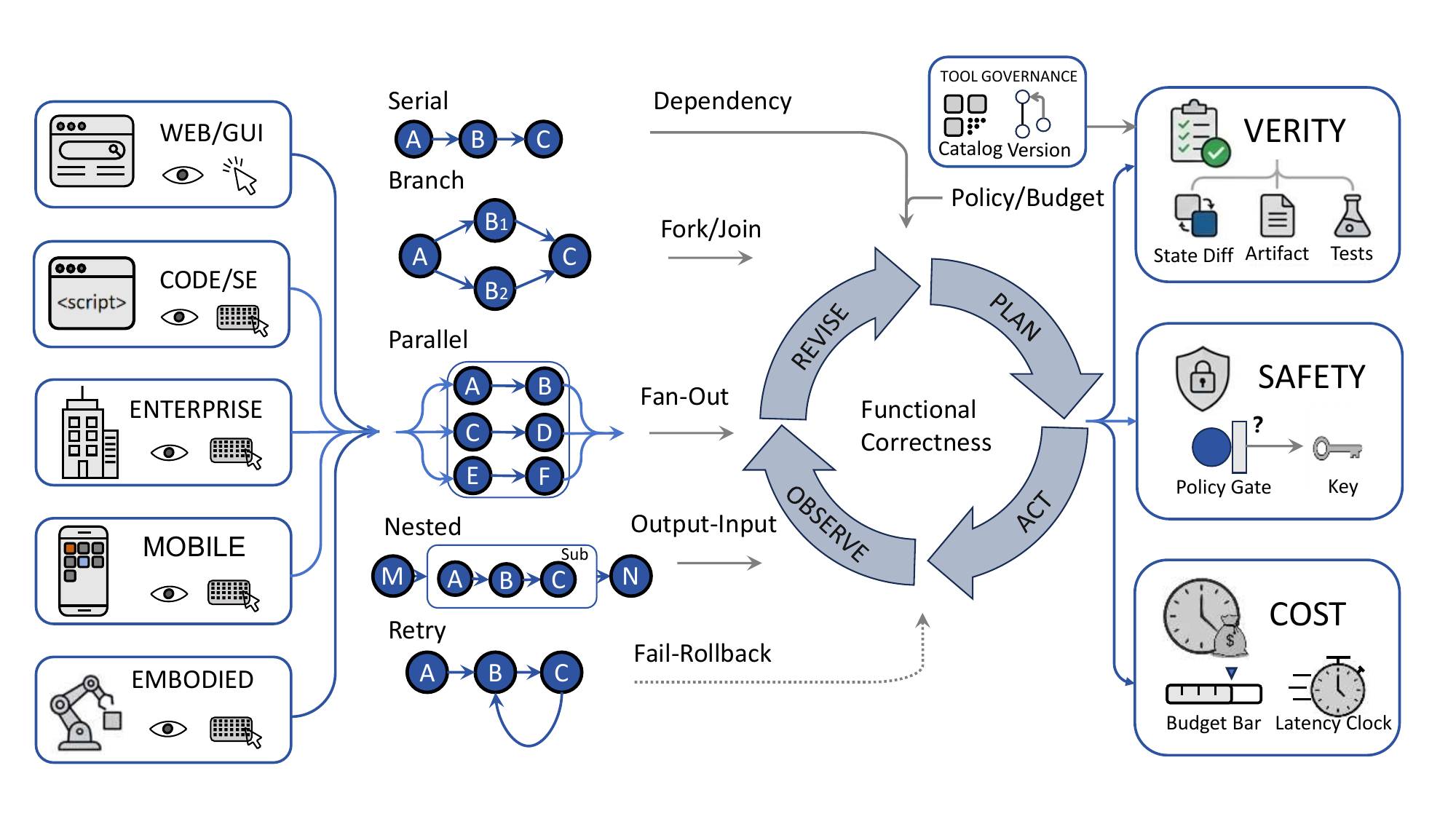}
  \caption{A unified abstraction of application-side multi-tool orchestration. Real-world agent systems operate over heterogeneous environments, instantiate different orchestration topologies, and rely on iterative plan-act-observe-revise loops under verification, safety, and cost constraints.}
  \label{fig:application-orchestration}
\end{figure}

% Figure~\ref{fig:application-orchestration} summarizes this perspective by organizing application-side multi-tool orchestration into execution environments, orchestration topologies, closed-loop control, and operational constraints.

A unified abstraction for multi-tool orchestration frames tasks as executable processes driving an external system toward a target state. As summarized in Figure~\ref{fig:application-orchestration}, this perspective organizes application-side orchestration into execution environments, topologies, closed-loop control, and operational constraints. Within real operating systems and web environments, tool use and low-level interface operations are subsumed into a unified sequential decision-making problem. Orchestrating these applications necessitates explicitly modeling the observation-action-state transition loop to capture the tight coupling of visual cues and actions in multimodal contexts~\cite{ben_osworld, app_visualwebarena}. Concurrently, growing evidence suggests that action grounding, rather than high-level planning, is the primary execution bottleneck. Robust performance in vision-language models relies on accurately mapping natural language intents to actionable interface elements and pixel-level regions to maintain stable tool pipelines~\cite{app_som_gpt4v}. Advancing this direction, general-purpose visual grounding solutions such as UGround~\cite{app_uground} improve graphical user interface agent accuracy across varying resolutions and heterogeneous elements.

Beyond individual action grounding, orchestrating these operations into extended tool chains introduces strict systems-level engineering constraints. As workflows scale, critical vulnerabilities emerge in state management, parameter binding, and variable passing across pipeline topologies~\cite{ben_ToolHop, ben_NESTFUL}. Maintaining stable execution under these implicit constraints requires integrating rigorous process supervision, treating invocation path correctness as a primary objective rather than a mere byproduct of terminal success~\cite{ben_ComplexFuncBench, ben_Toolcomp}. Furthermore, real-world deployments must manage highly heterogeneous ecosystems via standardized protocols to process continuous online feedback~\cite{app_mcpbench}. Crucially, as agents utilize powerful read and write primitives, security transitions into a foundational requirement; orchestration frameworks must proactively prevent unauthorized access and mitigate high-risk operations without compromising task completion~\cite{app_stwebagentbench}.

Motivated by these dynamics, the remainder of this chapter analyzes representative domains by focusing on the sources of dominant operational constraints, such as state verification, rollback mechanisms, and access control. Rather than re-evaluating previously established structural configurations, the subsequent analysis investigates how these strict operational boundaries directly dictate application-side orchestration strategies and necessitate robust engineering implementations across diverse industrial ecosystems. Ultimately, this chapter conceptualizes application orchestration through two complementary abstractions: executable tasks within interactive environments and execution workflows governed by explicit control flow. Building upon this framework, Section~9.2 categorizes application scenarios by their deployment environments to compare operational demands across tool sets, verification mechanisms, and stringent safety constraints, translating tool integration from a theoretical algorithmic capability into actionable workflow design.

\subsection{Orchestration Paradigms in Representative Interactive Environments}\label{sec:8.2}

This section analyzes multi-tool orchestration across diverse interactive environments, focusing on how stringent system constraints dictate execution policies. In graphical environments, managing non-stationary interfaces and visual grounding poses a significantly greater bottleneck than high-level planning~\cite{app_webvoyager, app_screenagent}. Orchestrating these multimodal loops requires prioritizing engineered representations and stable state transitions to handle user interface volatility and operational latency effectively~\cite{app_seeact, app_agente, app_agentoccam, app_weblinx, app_osworldhuman}.

Conversely, software engineering demands rigorous execution loops where success is strictly grounded in deterministic tests~\cite{app_swebenchverified, app_utboost}. The primary deployment bottleneck shifts from isolated code synthesis to the complex coupling of context retrieval, patch generation, and computationally expensive test verification~\cite{app_agentless, app_repobench, app_usebench, app_swepolybench}. To mitigate these high verification costs, systems-level optimizations including tool parallelization, intelligent caching, and overlapping language model decoding with partial tool execution become critical for maximizing throughput and minimizing end-to-end latency~\cite{app_conveyor}.

Enterprise orchestration manifests as continuous transactional workflows encompassing retrieval, analysis, execution, and write-back operations across distributed systems~\cite{app_workarena, app_workarenapp}. The primary engineering bottlenecks are state maintenance and transactional consistency under long-context constraints~\cite{app_spider2}. Consequently, enterprise deployment necessitates rigorous constraint systematization, elevating auditing, permission boundaries, and cost budgets to primary design variables. Agents must justify invocations, log execution trajectories for compliance, and utilize audit logs as verifiable completion signals~\cite{app_wowbench, app_topclass}.

Conversely, mobile environments represent extreme open ecosystems requiring complex cross-application coordination under severe physical and system-level interference~\cite{app_mobileagentbench, app_androidarena, app_androidlab}. Given the high risks of destructive misoperations in these dynamic multimodal settings, orchestration inherently demands explicit user confirmation points, continuous final state validation, and highly conservative recovery strategies rather than autonomous retries~\cite{app_mobilebenchv2, app_uinavbench}.

Across these diverse environments, domain distinctions arise primarily from unique operational bottlenecks rather than structural topologies. Engineering challenges range from high verification costs in software development and interface drift in mobile ecosystems to strict regulatory compliance in enterprise workflows. Grounded in this unified operational perspective, the subsequent section investigates essential industrial deployment factors including tool governance, continuous observability, and strict cost-latency budget management.

\subsection{Engineering Essentials and Governance for Industrial-Grade Multi-Tool Orchestration}\label{sec:8.3}

Industrial deployment necessitates rigorous tool ecosystem governance, as unstable API availability often causes irreproducible pipeline drift, making virtualization and regression validation architectural imperatives~\cite{app_stabletoolbench}. To manage heterogeneous access, standardized protocols mitigate integration costs~\cite{app_mcp, app_mcp_intro}, demanding fully auditable systems with strict semantic contracts and graceful degradation~\cite{app_mcpbench}. Concurrently, mitigating critical vulnerabilities like indirect prompt injections~\cite{app_injecagent} requires externalizing safety to dedicated policy layers that enforce input/output sanitization~\cite{app_stwebagentbench}. Aligning with standards such as the OWASP Top 10~\cite{app_owasp_llm_top10_2025} and NIST guidelines~\cite{app_nist_ai600_1}, architectures must embed least-privilege access, environment partitioning, and comprehensive logging within the execution loop to guarantee non-repudiable auditability and secure invocations.

Beyond security, arresting error cascades during execution demands intermediate checkpoints for decidable verification~\cite{ben_osworld} and comprehensive observability that evaluates process-level constraint satisfaction rather than mere terminal completion~\cite{app_agentops, app_agentic_assess}. Post-deployment, these pipelines must adhere to strict service level agreements and budgets. Because end-to-end latency is heavily dominated by planning cycles~\cite{app_osworldhuman}, robust orchestration requires the joint optimization of quality, time, and cost. Current research addresses this through interface compression to reduce token overhead~\cite{app_focusagent, app_agentoccam}, decoupling reasoning from execution~\cite{eff_rewoo}, and budget-aware routing to dynamically allocate resources to high-uncertainty tasks~\cite{eff_frugalgpt}.

Ultimately, industrial orchestration requires comprehensive systems-level remodeling rather than merely scaling research prototypes. The fundamental challenge lies in directly embedding these operational constraints—spanning governance, safety, verification, and efficiency—into the underlying infrastructure. Because industrial competitiveness heavily relies on capabilities like automated regression and executable acceptance testing, the subsequent section explores open research problems specifically targeting robust interface evolution, verifiability boundaries, and cost-effective long-horizon planning.
\subsection{Open Problems and Future Directions}\label{sec:8.4}

Although multi-tool LLM agents demonstrate encouraging generality, transitioning them from usable prototypes to trusted production systems requires addressing several critical application-side gaps: mitigating behavioral drift caused by continuous interface evolution via adaptive regression mechanisms; establishing process-level verifiability and replayable audits for long-horizon orchestration without prohibitive efficiency loss; developing systematic methodologies to jointly optimize cost, latency, and quality under strict budget constraints; and replacing heuristic failure recovery with composable, adversarial-resistant recovery frameworks. Ultimately, addressing these challenges demands a unified research agenda centered on evolvable tool governance, verifiable closed-loop execution, and adaptive orchestration to ensure these agents can be safely and reliably deployed in demanding professional environments.

%% file: 10_Conclusion.tex
\section{Conclusion}
The development of tool use in LLM agents reflects a broader shift in the field, from isolated function invocation to sustained interaction with complex external systems. Once agents must coordinate multiple tools over long horizons, the problem is no longer only about prompting or API formatting. It also involves planning under dependencies, managing mutable state, controlling execution, and preserving correctness under uncertainty, latency, and safety constraints. This is why progress depends not only on stronger base models, but also on advances in memory, orchestration, verification, and benchmark design. This survey has reviewed work on multi-tool agents across inference-time planning, training, safety, efficiency, capability completeness, benchmarks, and applications. A central observation is that these dimensions are tightly connected. Planning interacts with state management and verifiability. Safety depends on execution control and isolation. Efficiency depends on decomposition, routing, and reuse. Looking ahead, the field needs better abstractions for stateful orchestration, stronger evaluation protocols for dynamic and long-horizon settings, and closer integration between model-level reasoning and system-level guarantees. Progress on these fronts will be necessary if multi-tool agents are to move from promising prototypes to dependable, auditable, and scalable systems for real-world use.

%% file: custom.bib
@article{intro_llmsurvey,
  title={Large language models: A survey},
  author={Minaee, Shervin and Mikolov, Tomas and Nikzad, Narjes and Chenaghlu, Meysam and Socher, Richard and Amatriain, Xavier and Gao, Jianfeng},
  journal={arXiv preprint arXiv:2402.06196},
  year={2024}
}

@article{intro_talm,
  title={Talm: Tool augmented language models},
  author={Parisi, Aaron and Zhao, Yao and Fiedel, Noah},
  journal={arXiv preprint arXiv:2205.12255},
  year={2022}
}

@article{intro_mrkl,
  title={MRKL Systems: A modular, neuro-symbolic architecture that combines large language models, external knowledge sources and discrete reasoning},
  author={Karpas, Ehud and Abend, Omri and Belinkov, Yonatan and Lenz, Barak and Lieber, Opher and Ratner, Nir and Shoham, Yoav and Bata, Hofit and Levine, Yoav and Leyton-Brown, Kevin and others},
  journal={arXiv preprint arXiv:2205.00445},
  year={2022}
}

@article{intro_toolformer,
  title={Toolformer: Language models can teach themselves to use tools},
  author={Schick, Timo and Dwivedi-Yu, Jane and Dess{\`\i}, Roberto and Raileanu, Roberta and Lomeli, Maria and Hambro, Eric and Zettlemoyer, Luke and Cancedda, Nicola and Scialom, Thomas},
  journal={Advances in neural information processing systems},
  volume={36},
  pages={68539--68551},
  year={2023}
}

@article{intro_surveyllmad,
  title={A systematic survey on large language models for algorithm design},
  author={Liu, Fei and Yao, Yiming and Guo, Ping and Yang, Zhiyuan and Lin, Xi and Zhao, Zhe and Tong, Xialiang and Mao, Kun and Lu, Zhichao and Wang, Zhenkun and others},
  journal={ACM Computing Surveys},
  volume={58},
  number={8},
  pages={1--32},
  year={2026},
  publisher={ACM New York, NY}
}

@article{intro_ssllm,
  title={Semantic Scheduling for LLM Inference},
  author={Hua, Wenyue and Ding, Dujian and Gu, Yile and Ren, Yujie and Mei, Kai and Ma, Minghua and Wang, William Yang},
  journal={arXiv preprint arXiv:2506.12204},
  year={2025}
}

@article{intro_agentdrift,
  title={Agent Drift: Quantifying Behavioral Degradation in Multi-Agent LLM Systems Over Extended Interactions},
  author={Rath, Abhishek},
  journal={arXiv preprint arXiv:2601.04170},
  year={2026}
}

@article{intro_routellm,
  title={Routellm: Learning to route llms with preference data},
  author={Ong, Isaac and Almahairi, Amjad and Wu, Vincent and Chiang, Wei-Lin and Wu, Tianhao and Gonzalez, Joseph E and Kadous, M Waleed and Stoica, Ion},
  journal={arXiv preprint arXiv:2406.18665},
  year={2024}
}

@article{intro_wang,
  title={What are tools anyway? a survey from the language model perspective},
  author={Wang, Zhiruo and Cheng, Zhoujun and Zhu, Hao and Fried, Daniel and Neubig, Graham},
  journal={arXiv preprint arXiv:2403.15452},
  year={2024}
}

@article{intro_qu,
  title={Tool learning with large language models: A survey},
  author={Qu, Changle and Dai, Sunhao and Wei, Xiaochi and Cai, Hengyi and Wang, Shuaiqiang and Yin, Dawei and Xu, Jun and Wen, Ji-Rong},
  journal={Frontiers of Computer Science},
  volume={19},
  number={8},
  pages={198343},
  year={2025},
  publisher={Springer}
}

@article{intro_shen,
  title={Llm with tools: A survey},
  author={Shen, Zhuocheng},
  journal={arXiv preprint arXiv:2409.18807},
  year={2024}
}

@inproceedings{intro_li,
  title={A review of prominent paradigms for llm-based agents: Tool use, planning (including rag), and feedback learning},
  author={Li, Xinzhe},
  booktitle={Proceedings of the 31st international conference on computational linguistics},
  pages={9760--9779},
  year={2025}
}

@article{intro_luo,
  title={Large language model agent: A survey on methodology, applications and challenges},
  author={Luo, Junyu and Zhang, Weizhi and Yuan, Ye and Zhao, Yusheng and Yang, Junwei and Gu, Yiyang and Wu, Bohan and Chen, Binqi and Qiao, Ziyue and Long, Qingqing and others},
  journal={arXiv preprint arXiv:2503.21460},
  year={2025}
}

@article{intro_chen,
      title={A survey on llm-based multi-agent system: Recent advances and new frontiers in application},
  author={Chen, Shuaihang and Liu, Yuanxing and Han, Wei and Zhang, Weinan and Liu, Ting},
  journal={arXiv preprint arXiv:2412.17481},
  year={2024}
}

@article{intro_he,
  title={The emerged security and privacy of llm agent: A survey with case studies},
  author={He, Feng and Zhu, Tianqing and Ye, Dayong and Liu, Bo and Zhou, Wanlei and Yu, Philip S},
  journal={ACM Computing Surveys},
  volume={58},
  number={6},
  pages={1--36},
  year={2025},
  publisher={ACM New York, NY}
}

@inproceedings{intro_mo,
  title={Evaluation and benchmarking of llm agents: A survey},
  author={Mohammadi, Mahmoud and Li, Yipeng and Lo, Jane and Yip, Wendy},
  booktitle={Proceedings of the 31st ACM SIGKDD Conference on Knowledge Discovery and Data Mining V. 2},
  pages={6129--6139},
  year={2025}
}

@article{inf_hiplan,
  title={Hiplan: Hierarchical planning for llm-based agents with adaptive global-local guidance},
  author={Li, Ziyue and Chang, Yuan and Yu, Gaihong and Le, Xiaoqiu},
  journal={arXiv preprint arXiv:2508.19076},
  year={2025}
}

@inproceedings{inf_react,
  title={React: Synergizing reasoning and acting in language models},
  author={Yao, Shunyu and Zhao, Jeffrey and Yu, Dian and Du, Nan and Shafran, Izhak and Narasimhan, Karthik R and Cao, Yuan},
  booktitle={The eleventh international conference on learning representations},
  year={2022}
}

@article{inf_graphen,
  title={Graph-enhanced large language models in asynchronous plan reasoning},
  author={Lin, Fangru and La Malfa, Emanuele and Hofmann, Valentin and Yang, Elle Michelle and Cohn, Anthony and Pierrehumbert, Janet B},
  journal={arXiv preprint arXiv:2402.02805},
  year={2024}
}

@article{inf_gap,
  title={GAP: Graph-Based Agent Planning with Parallel Tool Use and Reinforcement Learning},
  author={Wu, Jiaqi and Zhao, Qinlao and Chen, Zefeng and Qin, Kai and Zhao, Yifei and Wang, Xueqian and Yao, Yuhang},
  journal={arXiv preprint arXiv:2510.25320},
  year={2025}
}

@article{inf_toolnet,
  title={Toolnet: Connecting large language models with massive tools via tool graph},
  author={Liu, Xukun and Peng, Zhiyuan and Yi, Xiaoyuan and Xie, Xing and Xiang, Lirong and Liu, Yuchen and Xu, Dongkuan},
  journal={arXiv preprint arXiv:2403.00839},
  year={2024}
}

@article{inf_struct,
  title={STRUCTUREDAGENT: Planning with AND/OR Trees for Long-Horizon Web Tasks},
  author={Lobo, ELita and Chen, Xu and Meng, Jingjing and Xi, Nan and Jiao, Yang and Agarwal, Chirag and Zick, Yair and Gao, Yan},
  journal={arXiv preprint arXiv:2603.05294},
  year={2026}
}

@inproceedings{inf_autotool,
  title={AutoTool: Efficient tool selection for large language model agents},
  author={Jia, Jingyi and Li, Qinbin},
  booktitle={Proceedings of the AAAI Conference on Artificial Intelligence},
  volume={40},
  number={37},
  pages={31265--31273},
  year={2026}
}

@article{inf_aflow,
  title={Aflow: Automating agentic workflow generation},
  author={Zhang, Jiayi and Xiang, Jinyu and Yu, Zhaoyang and Teng, Fengwei and Chen, Xionghui and Chen, Jiaqi and Zhuge, Mingchen and Cheng, Xin and Hong, Sirui and Wang, Jinlin and others},
  journal={arXiv preprint arXiv:2410.10762},
  year={2024}
}

@inproceedings{inf_adapt,
  title={Adapt: As-needed decomposition and planning with language models},
  author={Prasad, Archiki and Koller, Alexander and Hartmann, Mareike and Clark, Peter and Sabharwal, Ashish and Bansal, Mohit and Khot, Tushar},
  booktitle={Findings of the Association for Computational Linguistics: NAACL 2024},
  pages={4226--4252},
  year={2024}
}

@inproceedings{inf_dpot,
  title={Dynamic planning for llm-based graphical user interface automation},
  author={Zhang, Shaoqing and Zhang, Zhuosheng and Chen, Kehai and Ma, Xinbei and Yang, Muyun and Zhao, Tiejun and Zhang, Min},
  booktitle={Findings of the Association for Computational Linguistics: EMNLP 2024},
  pages={1304--1320},
  year={2024}
}

@article{inf_dy,
  title={DyFlow: Dynamic Workflow Framework for Agentic Reasoning},
  author={Wang, Yanbo and Xu, Zixiang and Huang, Yue and Wang, Xiangqi and Song, Zirui and Gao, Lang and Wang, Chenxi and Tang, Xiangru and Zhao, Yue and Cohan, Arman and others},
  journal={arXiv preprint arXiv:2509.26062},
  year={2025}
}

@article{inf_recap,
  title={ReCAP: Recursive Context-Aware Reasoning and Planning for Large Language Model Agents},
  author={Zhang, Zhenyu and Chen, Tianyi and Xu, Weiran and Pentland, Alex and Pei, Jiaxin},
  journal={arXiv preprint arXiv:2510.23822},
  year={2025}
}

@inproceedings{inf_smurfs,
  title={Smurfs: Multi-agent system using context-efficient DFSDT for tool planning},
  author={Chen, Junzhi and Liang, Juhao and Wang, Benyou},
  booktitle={Proceedings of the 2025 Conference of the Nations of the Americas Chapter of the Association for Computational Linguistics: Human Language Technologies (Volume 1: Long Papers)},
  pages={3281--3298},
  year={2025}
}

@article{inf_abmcts,
  title={Wider or deeper? scaling llm inference-time compute with adaptive branching tree search},
  author={Inoue, Yuichi and Misaki, Kou and Imajuku, Yuki and Kuroki, So and Nakamura, Taishi and Akiba, Takuya},
  journal={arXiv preprint arXiv:2503.04412},
  year={2025}
}

@article{inf_artis,
  title={ARTIS: Agentic Risk-Aware Test-Time Scaling via Iterative Simulation},
  author={Zeng, Xingshan and Wang, Lingzhi and Liu, Weiwen and Li, Liangyou and Wang, Yasheng and Shang, Lifeng and Jiang, Xin and Liu, Qun},
  journal={arXiv preprint arXiv:2602.01709},
  year={2026}
}

@article{inf_paa,
  title={Plan-and-act: Improving planning of agents for long-horizon tasks},
  author={Erdogan, Lutfi Eren and Lee, Nicholas and Kim, Sehoon and Moon, Suhong and Furuta, Hiroki and Anumanchipalli, Gopala and Keutzer, Kurt and Gholami, Amir},
  journal={arXiv preprint arXiv:2503.09572},
  year={2025}
}

@article{inf_mpo,
  title={Mpo: Boosting llm agents with meta plan optimization},
  author={Xiong, Weimin and Song, Yifan and Dong, Qingxiu and Zhao, Bingchan and Song, Feifan and Wang, Xun and Li, Sujian},
  journal={arXiv preprint arXiv:2503.02682},
  volume={5},
  number={6},
  pages={7},
  year={2025}
}

@inproceedings{inf_agentpro,
  title={AgentPro: Enhancing LLM Agents with Automated Process Supervision},
  author={Deng, Yuchen and Fan, Shichen and Wang, Naibo and Zhao, Xinkui and Ng, See Kiong},
  booktitle={Proceedings of the 2025 Conference on Empirical Methods in Natural Language Processing},
  pages={9992--10017},
  year={2025}
}

@misc{inf_mars,
      title={MARS: Co-evolving Dual-System Deep Research via Multi-Agent Reinforcement Learning}, 
      author={Guoxin Chen and Zile Qiao and Wenqing Wang and Donglei Yu and Xuanzhong Chen and Hao Sun and Minpeng Liao and Kai Fan and Yong Jiang and Penguin Xie and Wayne Xin Zhao and Ruihua Song and Fei Huang},
      year={2026},
      eprint={2510.04935},
      archivePrefix={arXiv},
      primaryClass={cs.AI},
      url={https://arxiv.org/abs/2510.04935}, 
}

@article{inf_codetool,
  title={Codetool: Enhancing programmatic tool invocation of llms via process supervision},
  author={Lu, Yifei and Ye, Fanghua and Li, Jian and Gao, Qiang and Liu, Cheng and Luo, Haibo and Du, Nan and Li, Xiaolong and Ren, Feiliang},
  journal={arXiv preprint arXiv:2503.20840},
  year={2025}
}

@article{inf_deepagent,
  title={Deepagent: A general reasoning agent with scalable toolsets},
  author={Li, Xiaoxi and Jiao, Wenxiang and Jin, Jiarui and Dong, Guanting and Jin, Jiajie and Wang, Yinuo and Wang, Hao and Zhu, Yutao and Wen, Ji-Rong and Lu, Yuan and others},
  journal={arXiv preprint arXiv:2510.21618},
  year={2025}
}

@article{inf_macla,
  title={Learning Hierarchical Procedural Memory for LLM Agents through Bayesian Selection and Contrastive Refinement},
  author={Forouzandeh, Saman and Peng, Wei and Moradi, Parham and Yu, Xinghuo and Jalili, Mahdi},
  journal={arXiv preprint arXiv:2512.18950},
  year={2025}
}

@inproceedings{inf_hiagent,
  title={Hiagent: Hierarchical working memory management for solving long-horizon agent tasks with large language model},
  author={Hu, Mengkang and Chen, Tianxing and Chen, Qiguang and Mu, Yao and Shao, Wenqi and Luo, Ping},
  booktitle={Proceedings of the 63rd Annual Meeting of the Association for Computational Linguistics (Volume 1: Long Papers)},
  pages={32779--32798},
  year={2025}
}

@article{inf_hugg,
  title={Hugginggpt: Solving ai tasks with chatgpt and its friends in hugging face},
  author={Shen, Yongliang and Song, Kaitao and Tan, Xu and Li, Dongsheng and Lu, Weiming and Zhuang, Yueting},
  journal={Advances in Neural Information Processing Systems},
  volume={36},
  pages={38154--38180},
  year={2023}
}

@inproceedings{inf_small,
  title={Small llms are weak tool learners: A multi-llm agent},
  author={Shen, Weizhou and Li, Chenliang and Chen, Hongzhan and Yan, Ming and Quan, Xiaojun and Chen, Hehong and Zhang, Ji and Huang, Fei},
  booktitle={Proceedings of the 2024 conference on empirical methods in natural language processing},
  pages={16658--16680},
  year={2024}
}

@article{inf_grad,
  title={Gradientsys: A Multi-Agent LLM Scheduler with ReAct Orchestration},
  author={Song, Xinyuan and Wang, Zeyu and Wu, Siyi and Shi, Tianyu and Ai, Lynn},
  journal={arXiv preprint arXiv:2507.06520},
  year={2025}
}

@article{inf_t2ar,
  title={Tool-to-agent retrieval: Bridging tools and agents for scalable llm multi-agent systems},
  author={Lumer, Elias and Nizar, Faheem and Gulati, Anmol and Basavaraju, Pradeep Honaganahalli and Subbiah, Vamse Kumar},
  journal={arXiv preprint arXiv:2511.01854},
  year={2025}
}

@article{inf_comp,
  title={Compass: Enhancing agent long-horizon reasoning with evolving context},
  author={Wan, Guangya and Ling, Mingyang and Ren, Xiaoqi and Han, Rujun and Li, Sheng and Zhang, Zizhao},
  journal={arXiv preprint arXiv:2510.08790},
  year={2025}
}

@article{inf_acon,
  title={Acon: Optimizing context compression for long-horizon llm agents},
  author={Kang, Minki and Chen, Wei-Ning and Han, Dongge and Inan, Huseyin A and Wutschitz, Lukas and Chen, Yanzhi and Sim, Robert and Rajmohan, Saravan},
  journal={arXiv preprint arXiv:2510.00615},
  year={2025}
}

@article{inf_amem,
  title={A-mem: Agentic memory for llm agents},
  author={Xu, Wujiang and Liang, Zujie and Mei, Kai and Gao, Hang and Tan, Juntao and Zhang, Yongfeng},
  journal={arXiv preprint arXiv:2502.12110},
  year={2025}
}

@article{inf_mirix,
  title={Mirix: Multi-agent memory system for llm-based agents},
  author={Wang, Yu and Chen, Xi},
  journal={arXiv preprint arXiv:2507.07957},
  year={2025}
}

@article{inf_sedm,
  title={Sedm: Scalable self-evolving distributed memory for agents},
  author={Xu, Haoran and Hu, Jiacong and Zhang, Ke and Yu, Lei and Tang, Yuxin and Song, Xinyuan and Duan, Yiqun and Ai, Lynn and Shi, Bill},
  journal={arXiv preprint arXiv:2509.09498},
  year={2025}
}

@article{inf_reflex,
  title={Reflexion: Language agents with verbal reinforcement learning},
  author={Shinn, Noah and Cassano, Federico and Gopinath, Ashwin and Narasimhan, Karthik and Yao, Shunyu},
  journal={Advances in neural information processing systems},
  volume={36},
  pages={8634--8652},
  year={2023}
}

@article{inf_fmas,
  title={Failure makes the agent stronger: Enhancing accuracy through structured reflection for reliable tool interactions},
  author={Su, Junhao and Wan, Yuanliang and Yang, Junwei and Shi, Hengyu and Han, Tianyang and Luo, Junfeng and Qiu, Yurui},
  journal={arXiv preprint arXiv:2509.18847},
  year={2025}
}

@inproceedings{inf_spiral,
  title={SPIRAL: Symbolic LLM Planning via Grounded and Reflective Search},
  author={Zhang, Yifan and Ganapavarapu, Giridhar and Jayaraman, Srideepika and Agrawal, Bhavna and Patel, Dhaval and Fokoue, Achille},
  booktitle={Proceedings of the AAAI Conference on Artificial Intelligence},
  volume={40},
  number={43},
  pages={36527--36535},
  year={2026}
}

@article{inf_preflect,
  title={PreFlect: From Retrospective to Prospective Reflection in Large Language Model Agents},
  author={Wang, Hanyu and Cao, Yuanpu and Lin, Lu and Chen, Jinghui},
  journal={arXiv preprint arXiv:2602.07187},
  year={2026}
}

@article{inf_sage,
  title={Sage: Self-evolving agents with reflective and memory-augmented abilities},
  author={Liang, Xuechen and Tao, Meiling and Xia, Yinghui and Wang, Jianhui and Li, Kun and Wang, Yijin and He, Yangfan and Yang, Jingsong and Shi, Tianyu and Wang, Yuantao and others},
  journal={Neurocomputing},
  volume={647},
  pages={130470},
  year={2025},
  publisher={Elsevier}
}

@article{inf_p2e,
  title={Plan2Evolve: LLM Self-Evolution for Improved Planning Capability via Automated Domain Generation},
  author={Huang, Jinbang and Li, Zhiyuan and Zhang, Zhanguang and Quan, Xingyue and Hao, Jianye and Zhang, Yingxue},
  journal={arXiv preprint arXiv:2509.21543},
  year={2025}
}

@article{inf_meta,
  title={MetaAgent: Toward Self-Evolving Agent via Tool Meta-Learning},
  author={Qian, Hongjin and Liu, Zheng},
  journal={arXiv preprint arXiv:2508.00271},
  year={2025}
}

@article{inf_tte,
  title={Beyond Static Tools: Test-Time Tool Evolution for Scientific Reasoning},
  author={Lu, Jiaxuan and Kong, Ziyu and Wang, Yemin and Fu, Rong and Wan, Haiyuan and Yang, Cheng and Lou, Wenjie and Sun, Haoran and Wang, Lilong and Jiang, Yankai and others},
  journal={arXiv preprint arXiv:2601.07641},
  year={2026}
}

@article{inf_draft,
  title={From exploration to mastery: Enabling llms to master tools via self-driven interactions},
  author={Qu, Changle and Dai, Sunhao and Wei, Xiaochi and Cai, Hengyi and Wang, Shuaiqiang and Yin, Dawei and Xu, Jun and Wen, Ji-Rong},
  journal={arXiv preprint arXiv:2410.08197},
  year={2024}
}

@article{inf_swead,
  title={SWE-Adept: An LLM-Based Agentic Framework for Deep Codebase Analysis and Structured Issue Resolution},
  author={He, Kang and Roy, Kaushik},
  journal={arXiv preprint arXiv:2603.01327},
  year={2026}
}

@inproceedings{ben_APIBank,
  title={Api-bank: A comprehensive benchmark for tool-augmented llms},
  author={Li, Minghao and Zhao, Yingxiu and Yu, Bowen and Song, Feifan and Li, Hangyu and Yu, Haiyang and Li, Zhoujun and Huang, Fei and Li, Yongbin},
  booktitle={Proceedings of the 2023 conference on empirical methods in natural language processing},
  pages={3102--3116},
  year={2023}
}

@article{ben_MetaTool,
  title={Metatool benchmark for large language models: Deciding whether to use tools and which to use},
  author={Huang, Yue and Shi, Jiawen and Li, Yuan and Fan, Chenrui and Wu, Siyuan and Zhang, Qihui and Liu, Yixin and Zhou, Pan and Wan, Yao and Gong, Neil Zhenqiang and others},
  journal={arXiv preprint arXiv:2310.03128},
  year={2023}
}

@article{ben_ToolBench,
  title={Toolllm: Facilitating large language models to master 16000+ real-world apis},
  author={Qin, Yujia and Liang, Shihao and Ye, Yining and Zhu, Kunlun and Yan, Lan and Lu, Yaxi and Lin, Yankai and Cong, Xin and Tang, Xiangru and Qian, Bill and others},
  journal={arXiv preprint arXiv:2307.16789},
  year={2023}
}

@inproceedings{ben_NESTFUL,
  title={Nestful: A benchmark for evaluating llms on nested sequences of api calls},
  author={Basu, Kinjal and Abdelaziz, Ibrahim and Kate, Kiran and Agarwal, Mayank and Crouse, Maxwell and Rizk, Yara and Bradford, Kelsey and Munawar, Asim and Kumaravel, Sadhana and Goyal, Saurabh and others},
  booktitle={Proceedings of the 2025 Conference on Empirical Methods in Natural Language Processing},
  pages={33526--33535},
  year={2025}
}

@inproceedings{ben_ToolHop,
  title={ToolHop: A Query-Driven Benchmark for Evaluating Large Language Models in Multi-Hop Tool Use},
  author={Ye, Junjie and Du, Zhengyin and Yao, Xuesong and Lin, Weijian and Xu, Yufei and Chen, Zehui and Wang, Zaiyuan and Zhu, Sining and Xi, Zhiheng and Yuan, Siyu and others},
  booktitle={Proceedings of the 63rd Annual Meeting of the Association for Computational Linguistics (Volume 1: Long Papers)},
  pages={2995--3021},
  year={2025}
}

@article{ben_ComplexFuncBench,
  title={ComplexFuncBench: Exploring multi-step and constrained function calling under long-context scenario},
  author={Zhong, Lucen and Du, Zhengxiao and Zhang, Xiaohan and Hu, Haiyi and Tang, Jie},
  journal={arXiv preprint arXiv:2501.10132},
  year={2025}
}

@inproceedings{ben_SealTools,
  title={Seal-tools: Self-instruct tool learning dataset for agent tuning and detailed benchmark},
  author={Wu, Mengsong and Zhu, Tong and Han, Han and Tan, Chuanyuan and Zhang, Xiang and Chen, Wenliang},
  booktitle={CCF International Conference on Natural Language Processing and Chinese Computing},
  pages={372--384},
  year={2024},
  organization={Springer}
}

@article{ben_Toolcomp,
  title={Toolcomp: A multi-tool reasoning \& process supervision benchmark},
  author={Nath, Vaskar and Raja, Pranav and Yoon, Claire and Hendryx, Sean},
  journal={arXiv preprint arXiv:2501.01290},
  year={2025}
}

@article{ben_TRAJECTBench,
  title={TRAJECT-Bench: A Trajectory-Aware Benchmark for Evaluating Agentic Tool Use},
  author={He, Pengfei and Dai, Zhenwei and He, Bing and Liu, Hui and Tang, Xianfeng and Lu, Hanqing and Li, Juanhui and Ding, Jiayuan and Mukherjee, Subhabrata and Wang, Suhang and others},
  journal={arXiv preprint arXiv:2510.04550},
  year={2025}
}

@article{ben_ELTBench,
  title={Elt-bench: An end-to-end benchmark for evaluating ai agents on elt pipelines},
  author={Jin, Tengjun and Zhu, Yuxuan and Kang, Daniel},
  journal={Proceedings of the VLDB Endowment},
  volume={19},
  number={2},
  pages={84--98},
  year={2025},
  publisher={VLDB Endowment}
}

@article{ben_MSCoRe,
  title={MSCoRe: A Benchmark for Multi-Stage Collaborative Reasoning in LLM Agents},
  author={Lei, Yuzhen and Xie, Hongbin and Zhao, Jiaxing and Liu, Shuangxue and Song, Xuan},
  journal={arXiv preprint arXiv:2509.17628},
  year={2025}
}

@article{ben_TaskBench,
  title={Taskbench: Benchmarking large language models for task automation},
  author={Shen, Yongliang and Song, Kaitao and Tan, Xu and Zhang, Wenqi and Ren, Kan and Yuan, Siyu and Lu, Weiming and Li, Dongsheng and Zhuang, Yueting},
  journal={Advances in Neural Information Processing Systems},
  volume={37},
  pages={4540--4574},
  year={2024}
}

@inproceedings{ben_mnm,
  title={m \& m’s: A benchmark to evaluate tool-use for m ulti-step m ulti-modal tasks},
  author={Ma, Zixian and Huang, Weikai and Zhang, Jieyu and Gupta, Tanmay and Krishna, Ranjay},
  booktitle={European Conference on Computer Vision},
  pages={18--34},
  year={2024},
  organization={Springer}
}

@article{ben_m3bench,
  title={M\^{} 3-Bench: Multi-Modal, Multi-Hop, Multi-Threaded Tool-Using MLLM Agent Benchmark},
  author={Zhou, Yang and Zhao, Mingyu and Wang, Zhenting and Gu, Difei and Guo, Bangwei and Ye, Ruosong and Han, Ligong and Jin, Can and Metaxas, Dimitris N},
  journal={arXiv preprint arXiv:2511.17729},
  year={2025}
}

@inproceedings{ben_MMAU,
  title={Mmau: A holistic benchmark of agent capabilities across diverse domains},
  author={Yin, Guoli and Bai, Haoping and Ma, Shuang and Nan, Feng and Sun, Yanchao and Xu, Zhaoyang and Ma, Shen and Lu, Jiarui and Kong, Xiang and Zhang, Aonan and others},
  booktitle={Findings of the Association for Computational Linguistics: NAACL 2025},
  pages={4737--4765},
  year={2025}
}

@article{ben_FuncBenchGen,
  title={Towards Reliable Benchmarking: A Contamination Free, Controllable Evaluation Framework for Multi-step LLM Function Calling},
  author={Maekawa, Seiji and Hassell, Jackson and Pezeshkpour, Pouya and Mitchell, Tom and Hruschka, Estevam},
  journal={arXiv preprint arXiv:2509.26553},
  year={2025}
}

@inproceedings{ben_dice,
  title={DICE-BENCH: Evaluating the Tool-Use Capabilities of Large Language Models in Multi-Round, Multi-Party Dialogues},
  author={Jang, Kyochul and Lee, Donghyeon and Kim, Kyusik and Heo, Dongseok and Lee, Taewhoo and Kim, Woojeong and Suh, Bongwon},
  booktitle={Findings of the Association for Computational Linguistics: ACL 2025},
  pages={26822--26846},
  year={2025}
}

@article{ben_tooldec,
  title={The tool decathlon: Benchmarking language agents for diverse, realistic, and long-horizon task execution},
  author={Li, Junlong and Zhao, Wenshuo and Zhao, Jian and Zeng, Weihao and Wu, Haoze and Wang, Xiaochen and Ge, Rui and Cao, Yuxuan and Huang, Yuzhen and Liu, Wei and others},
  journal={arXiv preprint arXiv:2510.25726},
  year={2025}
}

@inproceedings{ben_shortcut,
  author       = {Haiyang Shen and
                  Yue Li and
                  Desong Meng and
                  Dongqi Cai and
                  Sheng Qi and
                  Li Zhang and
                  Mengwei Xu and
                  Yun Ma},
  title        = {ShortcutsBench: {A} Large-Scale Real-world Benchmark for API-based
                  Agents},
  booktitle    = {The Thirteenth International Conference on Learning Representations,
                  {ICLR} 2025, Singapore, April 24-28, 2025},
  publisher    = {OpenReview.net},
  year         = {2025},
  url          = {https://openreview.net/forum?id=kKILfPkhSz},
  bibsource    = {dblp computer science bibliography, https://dblp.org}
}

@inproceedings{ben_hammer,
  title={Hammerbench: Fine-grained function-calling evaluation in real mobile assistant scenarios},
  author={Wang, Jun and Zhou, Jiamu and Wang, Xihuai and Mo, Xiaoyun and Zhang, Haoyu and Lin, Qiqiang and Jincheng, Jincheng and Wen, Muning and Zhang, Weinan and Peng, Qiuying},
  booktitle={Findings of the Association for Computational Linguistics: ACL 2025},
  pages={3350--3376},
  year={2025}
}

@article{ben_along,
  title={AgentLongBench: A Controllable Long Benchmark For Long-Contexts Agents via Environment Rollouts},
  author={Fang, Shicheng and Wang, Yuxin and Liu, Xiaoran and Lu, Jiahao and Tan, Chuanyuan and Chen, Xinchi and Zheng, Yining and Huang, Xuanjing and Qiu, Xipeng},
  journal={arXiv preprint arXiv:2601.20730},
  year={2026}
}

@article{ben_ultra,
  title={Ultrahorizon: Benchmarking agent capabilities in ultra long-horizon scenarios},
  author={Luo, Haotian and Zhang, Huaisong and Zhang, Xuelin and Wang, Haoyu and Qin, Zeyu and Lu, Wenjie and Ma, Guozheng and He, Haiying and Xie, Yingsha and Zhou, Qiyang and others},
  journal={arXiv preprint arXiv:2509.21766},
  year={2025}
}

@article{ben_trip,
  title={TRIP-Bench: A Benchmark for Long-Horizon Interactive Agents in Real-World Scenarios},
  author={Shen, Yuanzhe and Huang, Zisu and Wang, Zhengyuan and Tian, Muzhao and Guo, Zhengkang and Zhang, Chenyang and Zhou, Shuaiyu and Hu, Zengjie and Li, Dailin and Xu, Jingwen and others},
  journal={arXiv preprint arXiv:2602.01675},
  year={2026}
}

@article{ben_deepplan,
  title={DeepPlanning: Benchmarking Long-Horizon Agentic Planning with Verifiable Constraints},
  author={Zhang, Yinger and Jiang, Shutong and Li, Renhao and Tu, Jianhong and Su, Yang and Deng, Lianghao and Guo, Xudong and Lv, Chenxu and Lin, Junyang},
  journal={arXiv preprint arXiv:2601.18137},
  year={2026}
}

@inproceedings{ben_toolsandbox,
  title={Toolsandbox: A stateful, conversational, interactive evaluation benchmark for llm tool use capabilities},
  author={Lu, Jiarui and Holleis, Thomas and Zhang, Yizhe and Aumayer, Bernhard and Nan, Feng and Bai, Haoping and Ma, Shuang and Ma, Shen and Li, Mengyu and Yin, Guoli and others},
  booktitle={Findings of the Association for Computational Linguistics: NAACL 2025},
  pages={1160--1183},
  year={2025}
}

@inproceedings{ben_appworld,
  title={Appworld: A controllable world of apps and people for benchmarking interactive coding agents},
  author={Trivedi, Harsh and Khot, Tushar and Hartmann, Mareike and Manku, Ruskin and Dong, Vinty and Li, Edward and Gupta, Shashank and Sabharwal, Ashish and Balasubramanian, Niranjan},
  booktitle={Proceedings of the 62nd Annual Meeting of the Association for Computational Linguistics (Volume 1: Long Papers)},
  pages={16022--16076},
  year={2024}
}

@article{ben_toolgym,
  title={ToolGym: an Open-world Tool-using Environment for Scalable Agent Testing and Data Curation},
  author={Xi, Ziqiao and Liang, Shuang and Liu, Qi and Zhang, Jiaqing and Peng, Letian and Nan, Fang and Nayim, Meshal and Zhang, Tianhui and Mundada, Rishika and Qin, Lianhui and others},
  journal={arXiv preprint arXiv:2601.06328},
  year={2026}
}

@inproceedings{ben_stabletool,
  title={Stabletoolbench-mirrorapi: Modeling tool environments as mirrors of 7,000+ real-world apis},
  author={Guo, Zhicheng and Cheng, Sijie and Niu, Yuchen and Wang, Hao and Zhou, Sicheng and Huang, Wenbing and Liu, Yang},
  booktitle={Findings of the Association for Computational Linguistics: ACL 2025},
  pages={5247--5270},
  year={2025}
}

@article{ben_tau,
  author       = {Shunyu Yao and
                  Noah Shinn and
                  Pedram Razavi and
                  Karthik Narasimhan},
  title        = {{\(\tau\)}-bench: {A} Benchmark for Tool-Agent-User Interaction in
                  Real-World Domains},
  journal      = {CoRR},
  volume       = {abs/2406.12045},
  year         = {2024},
  url          = {https://doi.org/10.48550/arXiv.2406.12045},
  doi          = {10.48550/ARXIV.2406.12045},
  eprinttype   = {arXiv},
  eprint       = {2406.12045},
  bibsource    = {dblp computer science bibliography, https://dblp.org}
}

@article{ben_tau2,
  author       = {Victor Barres and
                  Honghua Dong and
                  Soham Ray and
                  Xujie Si and
                  Karthik Narasimhan},
  title        = {{\(\tau\)}\({}^{\mbox{2}}\)-Bench: Evaluating Conversational Agents
                  in a Dual-Control Environment},
  journal      = {CoRR},
  volume       = {abs/2506.07982},
  year         = {2025},
  url          = {https://doi.org/10.48550/arXiv.2506.07982},
  doi          = {10.48550/ARXIV.2506.07982},
  eprinttype   = {arXiv},
  eprint       = {2506.07982},
  bibsource    = {dblp computer science bibliography, https://dblp.org}
}

@inproceedings{ben_mtu,
  author       = {Pei Wang and
                  Yanan Wu and
                  Noah Wang and
                  Jiaheng Liu and
                  Xiaoshuai Song and
                  Z. Y. Peng and
                  Ken Deng and
                  Chenchen Zhang and
                  Jiakai Wang and
                  Junran Peng and
                  Ge Zhang and
                  Hangyu Guo and
                  Zhaoxiang Zhang and
                  Wenbo Su and
                  Bo Zheng},
  title        = {MTU-Bench: {A} Multi-granularity Tool-Use Benchmark for Large Language
                  Models},
  booktitle    = {The Thirteenth International Conference on Learning Representations,
                  {ICLR} 2025, Singapore, April 24-28, 2025},
  publisher    = {OpenReview.net},
  year         = {2025},
  url          = {https://openreview.net/forum?id=6guG2OlXsr},
  bibsource    = {dblp computer science bibliography, https://dblp.org}
}

@article{ben_t1,
  title={T1: A Tool-Oriented Conversational Dataset for Multi-Turn Agentic Planning},
  author={Chakraborty, Amartya and Dashore, Paresh and Bathaee, Nadia and Jain, Anmol and Das, Anirban and Zhang, Shi-Xiong and Sahu, Sambit and Naphade, Milind and Winata, Genta Indra},
  journal={Advances in Neural Information Processing Systems},
  year={2025}
}

@article{ben_compass,
  title={COMPASS: A Multi-Turn Benchmark for Tool-Mediated Planning \& Preference Optimization},
  author={Qin, Tian and Bai, Felix and Hu, Ting-Yao and Vemulapalli, Raviteja and Koppula, Hema Swetha and Xu, Zhiyang and Jin, Bowen and Cemri, Mert and Lu, Jiarui and Wang, Zirui and others},
  journal={arXiv preprint arXiv:2510.07043},
  year={2025}
}

@inproceedings{ben_wild,
    title={Benchmarking {LLM} Tool-Use in the Wild},
    author={Peijie Yu and Wei Liu and Yifan Yang and Jinjian Li and Zelong Zhang and Xiao Feng and feng zhang},
    booktitle={The Fourteenth International Conference on Learning Representations},
    year={2026},
    url={https://openreview.net/forum?id=yz7fL5vfpn}
}

@article{ben_osworld,
  title={Osworld: Benchmarking multimodal agents for open-ended tasks in real computer environments},
  author={Xie, Tianbao and Zhang, Danyang and Chen, Jixuan and Li, Xiaochuan and Zhao, Siheng and Cao, Ruisheng and Hua, Toh J and Cheng, Zhoujun and Shin, Dongchan and Lei, Fangyu and others},
  journal={Advances in Neural Information Processing Systems},
  volume={37},
  pages={52040--52094},
  year={2024}
}

@article{ben_gta,
  title={GTA: a benchmark for general tool agents},
  author={Wang, Jize and Ma, Zerun and Li, Yining and Zhang, Songyang and Chen, Cailian and Chen, Kai and Le, Xinyi},
  journal={Advances in Neural Information Processing Systems},
  volume={37},
  pages={75749--75790},
  year={2024}
}

@article{ben_windows,
  title={Windows agent arena: Evaluating multi-modal os agents at scale},
  author={Bonatti, Rogerio and Zhao, Dan and Bonacci, Francesco and Dupont, Dillon and Abdali, Sara and Li, Yinheng and Lu, Yadong and Wagle, Justin and Koishida, Kazuhito and Bucker, Arthur and others},
  journal={arXiv preprint arXiv:2409.08264},
  year={2024}
}

@inproceedings{ben_mobile,
  title={Mobile-bench: An evaluation benchmark for llm-based mobile agents},
  author={Deng, Shihan and Xu, Weikai and Sun, Hongda and Liu, Wei and Tan, Tao and Liujianfeng, Liujianfeng and Li, Ang and Luan, Jian and Wang, Bin and Yan, Rui and others},
  booktitle={Proceedings of the 62nd Annual Meeting of the Association for Computational Linguistics (Volume 1: Long Papers)},
  pages={8813--8831},
  year={2024}
}

@inproceedings{ben_realweb,
  title={Realwebassist: A benchmark for long-horizon web assistance with real-world users},
  author={Ye, Suyu and Shi, Haojun and Shih, Darren and Yun, Hyokun and Roosta, Tanya G and Shu, Tianmin},
  booktitle={Proceedings of the AAAI Conference on Artificial Intelligence},
  volume={40},
  number={40},
  pages={34441--34449},
  year={2026}
}

@article{ben_vita,
  title={Vitabench: Benchmarking llm agents with versatile interactive tasks in real-world applications},
  author={He, Wei and Sun, Yueqing and Hao, Hongyan and Hao, Xueyuan and Xia, Zhikang and Gu, Qi and Han, Chengcheng and Zhao, Dengchang and Su, Hui and Zhang, Kefeng and others},
  journal={arXiv preprint arXiv:2509.26490},
  year={2025}
}

@article{ben_odyssey,
  title={Odysseybench: Evaluating llm agents on long-horizon complex office application workflows},
  author={Wang, Weixuan and Han, Dongge and Diaz, Daniel Madrigal and Xu, Jin and R{\"u}hle, Victor and Rajmohan, Saravan},
  journal={arXiv preprint arXiv:2508.09124},
  year={2025}
}

@article{ben_memagentbench,
  title={Evaluating memory in llm agents via incremental multi-turn interactions},
  author={Hu, Yuanzhe and Wang, Yu and McAuley, Julian},
  journal={arXiv preprint arXiv:2507.05257},
  year={2025}
}

@article{ben_color,
 title={ColorBench: Benchmarking Mobile Agents with Graph-Structured Framework for Complex Long-Horizon Tasks},
  author={Song, Yuanyi and Huang, Heyuan and Lin, Qiqiang and Zhao, Yin and Qu, Xiangmou and Wang, Jun and Lou, Xingyu and Liu, Weiwen and Zhang, Zhuosheng and Yu, Yong and others},
  journal={arXiv preprint arXiv:2510.14621},
  year={2025}
}

@inproceedings{ben_criticaltool,
  title={Critictool: Evaluating self-critique capabilities of large language models in tool-calling error scenarios},
  author={Huang, Shiting and Fang, Zhen and Chen, Zehui and Yuan, Siyu and Ye, Junjie and Zeng, Yu and Chen, Lin and Mao, Qi and Zhao, Feng},
  booktitle={Proceedings of the 2025 Conference on Empirical Methods in Natural Language Processing},
  pages={26683--26692},
  year={2025}
}

@article{ben_astra,
  title={ASTRA-bench: Evaluating Tool-Use Agent Reasoning and Action Planning with Personal User Context},
  author={Xiu, Zidi and Sun, David Q and Cheng, Kevin and Patel, Maitrik and Zhang, Yizhe and Lu, Jiarui and Attia, Omar and Vemulapalli, Raviteja and Tuzel, Oncel and Cao, Meng and others},
  journal={arXiv preprint arXiv:2603.01357},
  year={2026}
}

@inproceedings{ben_rot,
title={Rotbench: A multi-level benchmark for evaluating the robustness of large language models in tool learning},
  author={Ye, Junjie and Wu, Yilong and Gao, Songyang and Huang, Caishuang and Li, Sixian and Li, Guanyu and Fan, Xiaoran and Zhang, Qi and Gui, Tao and Huang, Xuan-Jing},
  booktitle={Proceedings of the 2024 conference on empirical methods in natural language processing},
  pages={313--333},
  year={2024}
}

@inproceedings{ben_toolsword,
  title={Toolsword: Unveiling safety issues of large language models in tool learning across three stages},
  author={Ye, Junjie and Li, Sixian and Li, Guanyu and Huang, Caishuang and Gao, Songyang and Wu, Yilong and Zhang, Qi and Gui, Tao and Huang, Xuan-Jing},
  booktitle={Proceedings of the 62nd Annual Meeting of the Association for Computational Linguistics (Volume 1: Long Papers)},
  pages={2181--2211},
  year={2024}
}

@inproceedings{app_visualwebarena,
  title={Visualwebarena: Evaluating multimodal agents on realistic visual web tasks},
  author={Koh, Jing Yu and Lo, Robert and Jang, Lawrence and Duvvur, Vikram and Lim, Ming and Huang, Po-Yu and Neubig, Graham and Zhou, Shuyan and Salakhutdinov, Russ and Fried, Daniel},
  booktitle={Proceedings of the 62nd Annual Meeting of the Association for Computational Linguistics (Volume 1: Long Papers)},
  pages={881--905},
  year={2024}
}

@article{app_som_gpt4v,
  title={Set-of-mark prompting unleashes extraordinary visual grounding in gpt-4v},
  author={Yang, Jianwei and Zhang, Hao and Li, Feng and Zou, Xueyan and Li, Chunyuan and Gao, Jianfeng},
  journal={arXiv preprint arXiv:2310.11441},
  year={2023}
}

@inproceedings{app_uground,
  title={Navigating the Digital World as Humans Do: Universal Visual Grounding for {GUI} Agents},
  author={Gou, Boyu and Wang, Ruohan and Zheng, Boyuan and Xie, Yanan and Chang, Cheng and Shu, Yiheng and Sun, Huan and Su, Yu},
  booktitle={International Conference on Learning Representations},
  year={2025},
  url={https://openreview.net/forum?id=kxnoqaisCT}
}

@article{app_mcpbench,
  title={MCP-Bench: Benchmarking Tool-Using LLM Agents with Complex Real-World Tasks via MCP Servers},
  author={Wang, Zhenting and Chang, Qi and Patel, Hemani and Biju, Shashank and Wu, Cheng-En and Liu, Quan and Ding, Aolin and Rezazadeh, Alireza and Shah, Ankit and Bao, Yujia and Siow, Eugene},
  journal={arXiv preprint arXiv:2508.20453},
  year={2025},
  url={https://arxiv.org/abs/2508.20453}
}

@article{app_stwebagentbench,
  title={ST-WebAgentBench: A Benchmark for Evaluating Safety and Trustworthiness in Web Agents},
  author={Levy, Ido and Wiesel, Ben and Marreed, Sami and Oved, Alon and Yaeli, Avi and Shlomov, Segev},
  journal={arXiv preprint arXiv:2410.06703},
  year={2024},
  url={https://arxiv.org/abs/2410.06703}
}

@inproceedings{app_webvoyager,
  title={Webvoyager: Building an end-to-end web agent with large multimodal models},
  author={He, Hongliang and Yao, Wenlin and Ma, Kaixin and Yu, Wenhao and Dai, Yong and Zhang, Hongming and Lan, Zhenzhong and Yu, Dong},
  booktitle={Proceedings of the 62nd Annual Meeting of the Association for Computational Linguistics (Volume 1: Long Papers)},
  pages={6864--6890},
  year={2024}
}

@article{app_seeact,
  title={Gpt-4v (ision) is a generalist web agent, if grounded},
  author={Zheng, Boyuan and Gou, Boyu and Kil, Jihyung and Sun, Huan and Su, Yu},
  journal={arXiv preprint arXiv:2401.01614},
  year={2024}
}

@inproceedings{app_screenagent,
  title={ScreenAgent: A Vision Language Model-driven Computer Control Agent},
  author={Niu, Runliang and Li, Jindong and Wang, Shiqi and Fu, Yali and Hu, Xiyu and Leng, Xueyuan and Kong, He and Chang, Yi and Wang, Qi},
  booktitle={Proceedings of the Thirty-Third International Joint Conference on Artificial Intelligence},
  pages={6433--6441},
  year={2024},
  url={https://www.ijcai.org/proceedings/2024/711}
}

@article{app_agente,
  title={Agent-E: From Autonomous Web Navigation to Foundational Design Principles in Agentic Systems},
  author={Abuelsaad, Tamer and Akkil, Deepak and Dey, Prasenjit and Jagmohan, Ashish and Vempaty, Aditya and Kokku, Ravi and Azam, Ruhana},
  journal={arXiv preprint arXiv:2407.13032},
  year={2024},
  url={https://arxiv.org/abs/2407.13032}
}

@inproceedings{app_agentoccam,
  title={AgentOccam: A Simple Yet Strong Baseline for LLM-Based Web Agents},
  author={Yang, Ke and Liu, Yao and Chaudhary, Sapana and Fakoor, Rasool and Chaudhari, Pratik and Karypis, George and Rangwala, Huzefa},
  booktitle={International Conference on Learning Representations},
  year={2025},
  url={https://openreview.net/forum?id=oWdzUpOlkX}
}

@inproceedings{app_weblinx,
  title={WebLINX: Real-World Website Navigation with Multi-Turn Dialogue},
  author={Lu, Xing Han and Kasner, Zdenek and Reddy, Siva},
  booktitle={Proceedings of the 41st International Conference on Machine Learning},
  pages={33007--33056},
  year={2024},
  url={https://proceedings.mlr.press/v235/lu24e.html}
}

@article{app_osworldhuman,
  title={OSWorld-Human: Benchmarking the Efficiency of Computer-Use Agents},
  author={Abhyankar, Reyna and Qi, Qi and Zhang, Yiying},
  journal={arXiv preprint arXiv:2506.16042},
  year={2025},
  url={https://arxiv.org/abs/2506.16042}
}

@misc{app_swebenchverified,
  title={Introducing {SWE}-bench Verified},
  author={Chowdhury, Neil and Aung, James and Shern, Chan Jun and Jaffe, Oliver and Sherburn, Dane and Starace, Giulio and Mays, Evan and Dias, Rachel and Aljubeh, Marwan and Glaese, Mia and Jimenez, Carlos E. and Yang, John and Ho, Leyton and Patwardhan, Tejal and Liu, Kevin and Madry, Aleksander},
  year={2024},
  url={https://openai.com/index/introducing-swe-bench-verified/}
}

@inproceedings{app_utboost,
  title={UTBoost: Rigorous Evaluation of Coding Agents on {SWE}-Bench},
  author={Yu, Boxi and Zhu, Yuxuan and He, Pinjia and Kang, Daniel},
  booktitle={Proceedings of the 63rd Annual Meeting of the Association for Computational Linguistics (Volume 1: Long Papers)},
  pages={3762--3774},
  year={2025},
  url={https://aclanthology.org/2025.acl-long.189/}
}

@article{app_agentless,
  title={Agentless: Demystifying {LLM}-based Software Engineering Agents},
  author={Xia, Chunqiu Steven and Deng, Yinlin and Dunn, Soren and Zhang, Lingming},
  journal={arXiv preprint arXiv:2407.01489},
  year={2024},
  url={https://arxiv.org/abs/2407.01489}
}

@article{app_repobench,
  title={RepoBench: Benchmarking Repository-Level Code Auto-Completion Systems},
  author={Liu, Tianyang and Xu, Canwen and McAuley, Julian},
  journal={arXiv preprint arXiv:2306.03091},
  year={2023},
  url={https://arxiv.org/abs/2306.03091}
}

@article{app_usebench,
  title={Unified Software Engineering Agent as AI Software Engineer},
  author={Applis, Leonhard and Zhang, Yuntong and Liang, Shanchao and Jiang, Nan and Tan, Lin and Roychoudhury, Abhik},
  journal={arXiv preprint arXiv:2506.14683},
  year={2025},
  url={https://arxiv.org/abs/2506.14683}
}

@article{app_swepolybench,
  title={Swe-polybench: A multi-language benchmark for repository level evaluation of coding agents},
  author={Rashid, Muhammad Shihab and Bock, Christian and Zhuang, Yuan and Buchholz, Alexander and Esler, Tim and Valentin, Simon and Franceschi, Luca and Wistuba, Martin and Sivaprasad, Prabhu Teja and Kim, Woo Jung and others},
  journal={arXiv preprint arXiv:2504.08703},
  year={2025}
}

@article{app_conveyor,
  title={Conveyor: Efficient Tool-aware LLM Serving with Tool Partial Execution},
  author={Xu, Yechen and Kong, Xinhao and Chen, Tingjun and Zhuo, Danyang},
  journal={arXiv preprint arXiv:2406.00059},
  year={2024},
  url={https://arxiv.org/abs/2406.00059}
}

@article{app_workarena,
  title={Workarena: How capable are web agents at solving common knowledge work tasks?},
  author={Drouin, Alexandre and Gasse, Maxime and Caccia, Massimo and Laradji, Issam H and Del Verme, Manuel and Marty, Tom and Boisvert, L{\'e}o and Thakkar, Megh and Cappart, Quentin and Vazquez, David and others},
  journal={arXiv preprint arXiv:2403.07718},
  year={2024}
}

@article{app_workarenapp,
  title={Workarena++: Towards compositional planning and reasoning-based common knowledge work tasks},
  author={Boisvert, L{\'e}o and Thakkar, Megh and Gasse, Maxime and Caccia, Massimo and De Chezelles, Thibault L and Cappart, Quentin and Chapados, Nicolas and Lacoste, Alexandre and Drouin, Alexandre},
  journal={Advances in Neural Information Processing Systems},
  volume={37},
  pages={5996--6051},
  year={2024}
}

@inproceedings{app_spider2,
  title={Spider 2.0: Evaluating Language Models on Real-World Enterprise Text-to-SQL Workflows},
  author={Lei, Fangyu and Chen, Jixuan and Ye, Yuxiao and Cao, Ruisheng and Shin, Dongchan and Su, Hongjin and Suo, Zhaoqing and Gao, Hongcheng and Hu, Wenjing and Yin, Pengcheng and Zhong, Victor and Xiong, Caiming and Sun, Ruoxi and Liu, Qian and Wang, Sida and Yu, Tao},
  booktitle={International Conference on Learning Representations},
  year={2025},
  url={https://openreview.net/forum?id=XmProj9cPs}
}

@article{app_wowbench,
  title={World of Workflows: A Benchmark for Bringing World Models to Enterprise Systems},
  author={Gupta, Lakshya and Li, Litao and Liu, Yizhe and Subramanian, Sriram Ganapathi and Suleman, Kaheer and Zhang, Zichen and Lu, Haoye and Pasupalak, Sumit},
  journal={arXiv preprint arXiv:2601.22130},
  year={2026},
  url={https://arxiv.org/abs/2601.22130}
}

@inproceedings{app_topclass,
  title={Top of the CLASS: Benchmarking LLM Agents on Real-World Enterprise Tasks},
  author={Wornow, Michael and Garodia, Vaishnav and Vassalos, Vasilis and Contractor, Utkarsh},
  booktitle={ICLR 2025 Workshop on Building Trust in LLMs and LLM Applications},
  year={2025},
  url={https://openreview.net/pdf?id=RQjUpeINII}
}

@article{app_mobileagentbench,
  title={MobileAgentBench: An Efficient and User-Friendly Benchmark for Mobile LLM Agents},
  author={Wang, Luyuan and Deng, Yongyu and Zha, Yiwei and Mao, Guodong and Wang, Qinmin and Min, Tianchen and Chen, Wei and Chen, Shoufa},
  journal={arXiv preprint arXiv:2406.08184},
  year={2024},
  url={https://arxiv.org/abs/2406.08184}
}

@inproceedings{app_androidarena,
  title={Understanding the weakness of large language model agents within a complex android environment},
  author={Xing, Mingzhe and Zhang, Rongkai and Xue, Hui and Chen, Qi and Yang, Fan and Xiao, Zhen},
  booktitle={Proceedings of the 30th ACM SIGKDD conference on knowledge discovery and data mining},
  pages={6061--6072},
  year={2024}
}

@inproceedings{app_androidlab,
  title={AndroidLab: Training and Systematic Benchmarking of Android Autonomous Agents},
  author={Xu, Yifan and Liu, Xiao and Sun, Xueqiao and Cheng, Siyi and Yu, Hao and Lai, Hanyu and Zhang, Shudan and Zhang, Dan and Tang, Jie and Dong, Yuxiao},
  booktitle={Proceedings of the 63rd Annual Meeting of the Association for Computational Linguistics (Volume 1: Long Papers)},
  pages={2144--2166},
  year={2025},
  url={https://aclanthology.org/2025.acl-long.107/}
}

@article{app_mobilebenchv2,
  title={Mobile-Bench-v2: A More Realistic and Comprehensive Benchmark for VLM-based Mobile Agents},
  author={Xu, Weikai and Jiang, Zhizheng and Liu, Yuxuan and Gao, Pengzhi and Liu, Wei and Luan, Jian and Li, Yuanchun and Liu, Yunxin and Wang, Bin and An, Bo},
  journal={arXiv preprint arXiv:2505.11891},
  year={2025},
  url={https://arxiv.org/abs/2505.11891}
}

@inproceedings{app_uinavbench,
  title={UINavBench: A Framework for Comprehensive Evaluation of Interactive Digital Agents},
  author={Agrawal, Harsh and Schoop, Eldon and Pan, Xinlei and Mahajan, Anuj and Seff, Ari and Feng, Di and Cheng, Ruijia and Teran, Andres Romero Mier Y and Gomez, Esteban and Sundararajan, Abhishek and others},
  booktitle={Proceedings of the IEEE/CVF International Conference on Computer Vision},
  pages={23353--23363},
  year={2025}
}

@inproceedings{app_stabletoolbench,
  title={Stabletoolbench: Towards stable large-scale benchmarking on tool learning of large language models},
  author={Guo, Zhicheng and Cheng, Sijie and Wang, Hao and Liang, Shihao and Qin, Yujia and Li, Peng and Liu, Zhiyuan and Sun, Maosong and Liu, Yang},
  booktitle={Findings of the Association for Computational Linguistics: ACL 2024},
  pages={11143--11156},
  year={2024}
}

@misc{app_mcp,
  title={Model Context Protocol (MCP) Specification},
  year={2025},
  url={https://modelcontextprotocol.io/specification/2025-06-18}
}

@misc{app_mcp_intro,
  title={Introducing the Model Context Protocol},
  year={2024},
  url={https://www.anthropic.com/news/model-context-protocol}
}

@inproceedings{app_injecagent,
  title={InjecAgent: Benchmarking Indirect Prompt Injections in Tool-Integrated Large Language Model Agents},
  author={Zhan, Qiusi and Liang, Zhixiang and Ying, Zifan and Kang, Daniel},
  booktitle={Findings of the Association for Computational Linguistics: ACL 2024},
  pages={10471--10506},
  year={2024},
  url={https://aclanthology.org/2024.findings-acl.624/}
}

@misc{app_owasp_llm_top10_2025,
  title={{OWASP} Top 10 for Large Language Model Applications (v2025)},
  year={2025},
  url={https://owasp.org/www-project-top-10-for-large-language-model-applications/assets/PDF/OWASP-Top-10-for-LLMs-v2025.pdf}
}

@techreport{app_nist_ai600_1,
  title={Artificial Intelligence Risk Management Framework: Generative Artificial Intelligence Profile},
  institution={National Institute of Standards and Technology},
  number={NIST AI 600-1},
  year={2024},
  url={https://nvlpubs.nist.gov/nistpubs/ai/NIST.AI.600-1.pdf}
}

@article{app_agentops,
  title={AgentOps: Enabling Observability of LLM Agents},
  author={Dong, Liming and Lu, Qinghua and Zhu, Liming},
  journal={arXiv preprint arXiv:2411.05285},
  year={2024},
  url={https://arxiv.org/abs/2411.05285}
}

@article{app_agentic_assess,
  title={Beyond Task Completion: An Assessment Framework for Evaluating Agentic AI Systems},
  author={Akshathala, Sreemaee and Adnan, Bassam and Ramesh, Mahisha and Vaidhyanathan, Karthik and Muhammed, Basil and Parthasarathy, Kannan},
  journal={arXiv preprint arXiv:2512.12791},
  year={2025},
  url={https://arxiv.org/abs/2512.12791}
}

@article{app_focusagent,
  title={FocusAgent: Simple Yet Effective Ways of Trimming the Large Context of Web Agents},
  author={Kerboua, Imene and Omidi Shayegan, Sahar and Thakkar, Megh and L{\`u}, Xing Han and Boisvert, L{\'e}o and Caccia, Massimo and Espinas, J{\'e}r{\'e}my and Aussem, Alexandre and Eglin, V{\'e}ronique and Lacoste, Alexandre},
  journal={arXiv preprint arXiv:2510.03204},
  year={2025},
  url={https://arxiv.org/abs/2510.03204}
}

@article{eff_sot,
  title={Skeleton-of-thought: Prompting LLMs for efficient parallel generation},
  author={Ning, Xuefei and Lin, Zinan and Zhou, Zixuan and Wang, Zifu and Yang, Huazhong and Wang, Yu},
  journal={arXiv preprint arXiv:2307.15337},
  year={2023}
}

@inproceedings{eff_LLMCompiler,
  title={An LLM Compiler for Parallel Function Calling},
  author={Kim, Sehoon and Moon, Suhong and Tabrizi, Ryan and Lee, Nicholas and Mahoney, Michael W. and Keutzer, Kurt and Gholami, Amir},
  booktitle={Proceedings of the 41st International Conference on Machine Learning},
  year={2024}
}

@article{eff_M1Parallel,
  title={Optimizing Sequential Multi-Step Tasks with Parallel LLM Agents},
  author={Zhang, Enhao and Zhu, Erkang and Bansal, Gagan and Fourney, Adam and Mozannar, Hussein and Gerrits, Jack},
  journal={arXiv preprint arXiv:2507.08944},
  year={2025}
}

@inproceedings{eff_para,
  title={Divide-then-aggregate: An efficient tool learning method via parallel tool invocation},
  author={Zhu, Dongsheng and Shi, Weixian and Shi, Zhengliang and Ren, Zhaochun and Wang, Shuaiqiang and Yan, Lingyong and Yin, Dawei},
  booktitle={Proceedings of the 63rd Annual Meeting of the Association for Computational Linguistics (Volume 1: Long Papers)},
  pages={28859--28875},
  year={2025}
}

@article{tune_anytool,
  title={Anytool: Self-reflective, hierarchical agents for large-scale api calls},
  author={Du, Yu and Wei, Fangyun and Zhang, Hongyang},
  journal={arXiv preprint arXiv:2402.04253},
  year={2024}
}

@inproceedings{tune_tooldreamer,
  title={Tooldreamer: Instilling llm reasoning into tool retrievers},
  author={Sengupta, Saptarshi and Zhou, Zhengyu and Araki, Jun and Wang, Xingbo and Wang, Bingqing and Wang, Suhang and Feng, Zhe},
  booktitle={Proceedings of the 19th Conference of the European Chapter of the Association for Computational Linguistics (Volume 1: Long Papers)},
  pages={5465--5482},
  year={2026}
}

@article{tune_mcp-zero,
  title={Mcp-zero: Active tool discovery for autonomous llm agents},
  author={Fei, Xiang and Zheng, Xiawu and Feng, Hao},
  journal={arXiv preprint arXiv:2506.01056},
  year={2025}
}

@inproceedings{tune_re-invoke,
  title={Re-invoke: Tool invocation rewriting for zero-shot tool retrieval},
  author={Chen, Yanfei and Yoon, Jinsung and Sachan, Devendra Singh and Wang, Qingze and Cohen-Addad, Vincent and Bateni, Mohammadhossein and Lee, Chen-Yu and Pfister, Tomas},
  booktitle={Findings of the Association for Computational Linguistics: EMNLP 2024},
  pages={4705--4726},
  year={2024}
}

@article{tune_pre-act,
  title={Pre-act: Multi-step planning and reasoning improves acting in llm agents},
  author={Rawat, Mrinal and Gupta, Ambuje and Goomer, Rushil and Di Bari, Alessandro and Gupta, Neha and Pieraccini, Roberto},
  journal={arXiv preprint arXiv:2505.09970},
  year={2025}
}

@inproceedings{tune_PLAY2PROMPT,
  title={Play2prompt: Zero-shot tool instruction optimization for llm agents via tool play},
  author={Fang, Wei and Zhang, Yang and Qian, Kaizhi and Glass, James and Zhu, Yada},
  booktitle={Findings of the Association for Computational Linguistics: ACL 2025},
  pages={26274--26290},
  year={2025}
}

@article{tune_Memento,
  title={Memento: Fine-tuning llm agents without fine-tuning llms},
  author={Zhou, Huichi and Chen, Yihang and Guo, Siyuan and Yan, Xue and Lee, Kin Hei and Wang, Zihan and Lee, Ka Yiu and Zhang, Guchun and Shao, Kun and Yang, Linyi and others},
  journal={arXiv preprint arXiv:2508.16153},
  year={2025}
}

@inproceedings{tune_gentool,
  title={GenTool: Enhancing Tool Generalization in Language Models through Zero-to-One and Weak-to-Strong Simulation},
  author={He, Jie and Neville, Jennifer and Wan, Mengting and Yang, Longqi and Liu, Hui and Xu, Xiaofeng and Song, Xia and Pan, Jeff Z and Zhou, Pei},
  booktitle={Findings of the Association for Computational Linguistics: ACL 2025},
  pages={1097--1122},
  year={2025}
}

@inproceedings{tune_luo,
  title={Self-training large language models for tool-use without demonstrations},
  author={Luo, Ne and Gema, Aryo Pradipta and He, Xuanli and Van Krieken, Emile and Lesci, Pietro and Minervini, Pasquale},
  booktitle={Findings of the Association for Computational Linguistics: NAACL 2025},
  pages={1253--1271},
  year={2025}
}

@article{tune_Button,
  title={Facilitating multi-turn function calling for llms via compositional instruction tuning},
  author={Chen, Mingyang and Sun, Haoze and Li, Tianpeng and Yang, Fan and Liang, Hao and Lu, Keer and Cui, Bin and Zhang, Wentao and Zhou, Zenan and Chen, Weipeng},
  journal={arXiv preprint arXiv:2410.12952},
  year={2024}
}

@article{tune_ASTRA,
  title={ASTRA: Automated Synthesis of agentic Trajectories and Reinforcement Arenas},
  author={Tian, Xiaoyu and Wang, Haotian and Chen, Shuaiting and Zhou, Hao and Yu, Kaichi and Zhang, Yudian and Ouyang, Jade and Yin, Junxi and Chen, Jiong and Guo, Baoyan and others},
  journal={arXiv preprint arXiv:2601.21558},
  year={2026}
}

@article{tune_apigen,
  title={Apigen: Automated pipeline for generating verifiable and diverse function-calling datasets},
  author={Liu, Zuxin and Hoang, Thai and Zhang, Jianguo and Zhu, Ming and Lan, Tian and Kokane, Shirley and Tan, Juntao and Yao, Weiran and Liu, Zhiwei and Feng, Yihao and others},
  journal={Advances in Neural Information Processing Systems},
  volume={37},
  pages={54463--54482},
  year={2024}
}

@article{tune_toolace-dev,
  title={Toolace-dev: Self-improving tool learning via decomposition and evolution},
  author={Huang, Xu and Liu, Weiwen and Zeng, Xingshan and Huang, Yuefeng and Hao, Xinlong and Wang, Yuxian and Zeng, Yirong and Wu, Chuhan and Wang, Yasheng and Tang, Ruiming and others},
  journal={arXiv preprint arXiv:2505.07512},
  year={2025}
}

@inproceedings{tune_toolace-r,
  title={ToolACE-R: Model-aware Iterative Training and Adaptive Refinement for Tool Learning},
  author={Zeng, Xingshan and Liu, Weiwen and Huang, Xu and Wang, Zezhong and Wang, Lingzhi and Li, Liangyou and Wang, Yasheng and Shang, Lifeng and Jiang, Xin and Tang, Ruiming and others},
  booktitle={Proceedings of the AAAI Conference on Artificial Intelligence},
  volume={40},
  number={41},
  pages={34593--34601},
  year={2026}
}

@article{tune_toolmind,
  title={ToolMind Technical Report: A Large-Scale, Reasoning-Enhanced Tool-Use Dataset},
  author={Yang, Chen and Le, Ran and Xing, Yun and An, Zhenwei and Chen, Zongchao and Zhao, Wayne Xin and Song, Yang and Zhang, Tao},
  journal={arXiv preprint arXiv:2511.15718},
  year={2025}
}

@article{tune_looptool,
  title={LoopTool: Closing the Data-Training Loop for Robust LLM Tool Calls},
  author={Zhang, Kangning and Jiao, Wenxiang and Du, Kounianhua and Lu, Yuan and Liu, Weiwen and Zhang, Weinan and Yu, Yong},
  journal={arXiv preprint arXiv:2511.09148},
  year={2025}
}

@article{tune_orchdag,
  title={OrchDAG: Complex tool orchestration in multi-turn interactions with plan DAGs},
  author={Lu, Yifu and Liu, Shengjie and Dong, Li},
  journal={arXiv preprint arXiv:2510.24663},
  year={2025}
}

@article{tune_gorilla,
  title={Gorilla: Large language model connected with massive apis},
  author={Patil, Shishir G and Zhang, Tianjun and Wang, Xin and Gonzalez, Joseph E},
  journal={Advances in Neural Information Processing Systems},
  volume={37},
  pages={126544--126565},
  year={2024}
}

@article{tune_GPT4Tools,
  title={Gpt4tools: Teaching large language model to use tools via self-instruction},
  author={Yang, Rui and Song, Lin and Li, Yanwei and Zhao, Sijie and Ge, Yixiao and Li, Xiu and Shan, Ying},
  journal={Advances in Neural Information Processing Systems},
  volume={36},
  pages={71995--72007},
  year={2023}
}

@article{tune_hammer,
  title={Hammer: Robust function-calling for on-device language models via function masking},
  author={Lin, Qiqiang and Wen, Muning and Peng, Qiuying and Nie, Guanyu and Liao, Junwei and Wang, Jun and Mo, Xiaoyun and Zhou, Jiamu and Cheng, Cheng and Zhao, Yin and others},
  journal={arXiv preprint arXiv:2410.04587},
  year={2024}
}

@inproceedings{tune_chain_of_abstraction,
  title={Efficient tool use with chain-of-abstraction reasoning},
  author={Gao, Silin and Dwivedi-Yu, Jane and Yu, Ping and Tan, Xiaoqing Ellen and Pasunuru, Ramakanth and Golovneva, Olga and Sinha, Koustuv and Celikyilmaz, Asli and Bosselut, Antoine and Wang, Tianlu},
  booktitle={Proceedings of the 31st International Conference on Computational Linguistics},
  pages={2727--2743},
  year={2025}
}

@article{tune_toolgen,
  title={Toolgen: Unified tool retrieval and calling via generation},
  author={Wang, Renxi and Han, Xudong and Ji, Lei and Wang, Shu and Baldwin, Timothy and Li, Haonan},
  journal={arXiv preprint arXiv:2410.03439},
  year={2024}
}

@inproceedings{tune_granite-function,
  title={Granite-function calling model: Introducing function calling abilities via multi-task learning of granular tasks},
  author={Abdelaziz, Ibrahim and Basu, Kinjal and Agarwal, Mayank and Kumaravel, Sadhana and Stallone, Matthew and Panda, Rameswar and Rizk, Yara and Bhargav, GP Shrivatsa and Crouse, Maxwell and Gunasekara, Chulaka and others},
  booktitle={Proceedings of the 2024 Conference on Empirical Methods in Natural Language Processing: Industry Track},
  pages={1131--1139},
  year={2024}
}

@article{tune_portool,
  title={PORTool: Tool-Use LLM Training with Rewarded Tree},
  author={Wu, Feijie and Zhu, Weiwu and Zhang, Yuxiang and Chatterjee, Soumya and Zhu, Jiarong and Mo, Fan and Luo, Rodin and Gao, Jing},
  journal={arXiv preprint arXiv:2510.26020},
  year={2025}
}

@article{tune_tool-star,
  title={Tool-star: Empowering llm-brained multi-tool reasoner via reinforcement learning},
  author={Dong, Guanting and Chen, Yifei and Li, Xiaoxi and Jin, Jiajie and Qian, Hongjin and Zhu, Yutao and Mao, Hangyu and Zhou, Guorui and Dou, Zhicheng and Wen, Ji-Rong},
  journal={arXiv preprint arXiv:2505.16410},
  year={2025}
}

@article{tune_toolrl,
  title={Toolrl: Reward is all tool learning needs},
  author={Qian, Cheng and Acikgoz, Emre Can and He, Qi and Wang, Hongru and Chen, Xiusi and Hakkani-T{\"u}r, Dilek and Tur, Gokhan and Ji, Heng},
  journal={arXiv preprint arXiv:2504.13958},
  year={2025}
}

@article{tune_toolrm,
  title={Toolrm: Outcome reward models for tool-calling large language models},
  author={Agarwal, Mayank and Abdelaziz, Ibrahim and Basu, Kinjal and Unuvar, Merve and Lastras, Luis A and Rizk, Yara and Kapanipathi, Pavan},
  journal={arXiv preprint arXiv:2509.11963},
  year={2025}
}

@article{tune_wei,
  title={Reinforcing multi-turn reasoning in llm agents via turn-level reward design},
  author={Wei, Quan and Zeng, Siliang and Li, Chenliang and Brown, William and Frunza, Oana and Deng, Wei and Schneider, Anderson and Nevmyvaka, Yuriy and Zhao, Yang Katie and Garcia, Alfredo and others},
  journal={arXiv preprint arXiv:2505.11821},
  year={2025}
}

@article{tune_gtpo,
  title={Empowering Multi-Turn Tool-Integrated Reasoning with Group Turn Policy Optimization},
  author={Ding, Yifeng and Le, Hung and Han, Songyang and Ruan, Kangrui and Jin, Zhenghui and Kumar, Varun and Wang, Zijian and Deoras, Anoop},
  journal={arXiv preprint arXiv:2511.14846},
  year={2025}
}

@article{tune_arpo,
  title={Agentic reinforced policy optimization},
  author={Dong, Guanting and Mao, Hangyu and Ma, Kai and Bao, Licheng and Chen, Yifei and Wang, Zhongyuan and Chen, Zhongxia and Du, Jiazhen and Wang, Huiyang and Zhang, Fuzheng and others},
  journal={arXiv preprint arXiv:2507.19849},
  year={2025}
}

@article{tune_ragen,
  title={Ragen: Understanding self-evolution in llm agents via multi-turn reinforcement learning},
  author={Wang, Zihan and Wang, Kangrui and Wang, Qineng and Zhang, Pingyue and Li, Linjie and Yang, Zhengyuan and Jin, Xing and Yu, Kefan and Nguyen, Minh Nhat and Liu, Licheng and others},
  journal={arXiv preprint arXiv:2504.20073},
  year={2025}
}

@article{tune_torl,
  title={Torl: Scaling tool-integrated rl},
  author={Li, Xuefeng and Zou, Haoyang and Liu, Pengfei},
  journal={arXiv preprint arXiv:2503.23383},
  year={2025}
}

@article{eff_5ai,
  title={Ai meets brain: Memory systems from cognitive neuroscience to autonomous agents},
  author={Liang, Jiafeng and Li, Hao and Li, Chang and Zhou, Jiaqi and Jiang, Shixin and Wang, Zekun and Ji, Changkai and Zhu, Zhihao and Liu, Runxuan and Ren, Tao and others},
  journal={arXiv preprint arXiv:2512.23343},
  year={2025}
}

@article{tune_agent-r1,
  title={Agent-r1: Training powerful llm agents with end-to-end reinforcement learning},
  author={Cheng, Mingyue and Ouyang, Jie and Yu, Shuo and Yan, Ruiran and Luo, Yucong and Liu, Zirui and Wang, Daoyu and Liu, Qi and Chen, Enhong},
  journal={arXiv preprint arXiv:2511.14460},
  year={2025}
}

@inproceedings{safe_not_what,
  title={Not what you've signed up for: Compromising real-world llm-integrated applications with indirect prompt injection},
  author={Greshake, Kai and Abdelnabi, Sahar and Mishra, Shailesh and Endres, Christoph and Holz, Thorsten and Fritz, Mario},
  booktitle={Proceedings of the 16th ACM workshop on artificial intelligence and security},
  pages={79--90},
  year={2023}
}

@article{safe_prompt_flow,
  title={Prompt flow integrity to prevent privilege escalation in llm agents},
  author={Kim, Juhee and Choi, Woohyuk and Lee, Byoungyoung},
  journal={arXiv preprint arXiv:2503.15547},
  year={2025}
}

@article{safe_design_pattern,
  title={Design patterns for securing llm agents against prompt injections},
  author={Beurer-Kellner, Luca and Buesser, Beat and Cre{\c{t}}u, Ana-Maria and Debenedetti, Edoardo and Dobos, Daniel and Fabian, Daniel and Fischer, Marc and Froelicher, David and Grosse, Kathrin and Naeff, Daniel and others},
  journal={arXiv preprint arXiv:2506.08837},
  year={2025}
}

@article{safe_survey_on_agentic_security,
  title={A Survey on Agentic Security: Applications, Threats and Defenses},
  author={Shahriar, Asif and Rahman, Md Nafiu and Ahmed, Sadif and Sadeque, Farig and Parvez, Md Rizwan},
  journal={arXiv preprint arXiv:2510.06445},
  year={2025}
}

@article{safe_Les_Dissonances,
  title={Les dissonances: Cross-tool harvesting and polluting in multi-tool empowered llm agents},
  author={Li, Zichuan and Cui, Jian and Liao, Xiaojing and Xing, Luyi},
  journal={arXiv e-prints},
  pages={arXiv--2504},
  year={2025}
}

@inproceedings{safe_Multi-Turn,
  title={Rethinking Stateful Tool Use in Multi-Turn Dialogues: Benchmarks and Challenges},
  author={Wang, Hongru and Huang, Wenyu and Wang, Yufei and Xi, Yuanhao and Lu, Jianqiao and Zhang, Huan and Hu, Nan and Liu, Zeming and Pan, Jeff Z and Wong, Kam-Fai},
  booktitle={Findings of the Association for Computational Linguistics: ACL 2025},
  pages={5433--5453},
  year={2025}
}

@article{safe_SagaLLM,
  title={SagaLLM: context management, validation, and transaction guarantees for multi-agent LLM planning},
  author={Chang, Edward Y and Geng, Longling},
  journal={arXiv preprint arXiv:2503.11951},
  year={2025}
}

@inproceedings{safe_TRICE,
  title={Making language models better tool learners with execution feedback},
  author={Qiao, Shuofei and Gui, Honghao and Lv, Chengfei and Jia, Qianghuai and Chen, Huajun and Zhang, Ningyu},
  booktitle={Proceedings of the 2024 Conference of the North American Chapter of the Association for Computational Linguistics: Human Language Technologies (Volume 1: Long Papers)},
  pages={3550--3568},
  year={2024}
}

@article{safe_ARRM,
  title={Autonomous Action Runtime Management (AARM): A System Specification for Securing AI-Driven Actions at Runtime},
  author={Errico, Herman},
  journal={arXiv preprint arXiv:2602.09433},
  year={2026}
}

@article{safe_agentspec,
  title={Agentspec: Customizable runtime enforcement for safe and reliable llm agents},
  author={Wang, Haoyu and Poskitt, Christopher M and Sun, Jun},
  journal={arXiv preprint arXiv:2503.18666},
  year={2025}
}

@article{safe_sampling,
  title={Sample, Predict, then Proceed: Self-Verification Sampling for Tool Use of LLMs},
  author={Guo, Shangmin and Domingues, Omar Darwiche and Avalos, Rapha{\"e}l and Courville, Aaron and Strub, Florian},
  year={2025}
}

@article{safe_Atomix,
  title={Atomix: Timely, Transactional Tool Use for Reliable Agentic Workflows},
  author={Mohammadi, Bardia and Potamitis, Nearchos and Klein, Lars and Arora, Akhil and Bindschaedler, Laurent},
  journal={arXiv preprint arXiv:2602.14849},
  year={2026}
}

@inproceedings{safe_Generator-Assistant,
  title={Generator-assistant stepwise rollback framework for large language model agent},
  author={Li, Xingzuo and Chen, Kehai and Long, Yunfei and Bai, Xuefeng and Xu, Yong and Zhang, Min},
  booktitle={Proceedings of the 2025 Conference on Empirical Methods in Natural Language Processing},
  pages={17694--17711},
  year={2025}
}

@article{safe_CRITIC,
  title={Critic: Large language models can self-correct with tool-interactive critiquing},
  author={Gou, Zhibin and Shao, Zhihong and Gong, Yeyun and Shen, Yelong and Yang, Yujiu and Duan, Nan and Chen, Weizhu},
  journal={arXiv preprint arXiv:2305.11738},
  year={2023}
}

@article{safe_verifiagent,
  title={Verifiagent: a unified verification agent in language model reasoning},
  author={Han, Jiuzhou and Buntine, Wray and Shareghi, Ehsan},
  journal={ArXiv preprint, abs/2504.00406},
  year={2025}
}

@inproceedings{safe_DVR,
  title={Divide-Verify-Refine: Can LLMs Self-Align with Complex Instructions?},
  author={Zhang, Xianren and Tang, Xianfeng and Liu, Hui and Wu, Zongyu and He, Qi and Lee, Dongwon and Wang, Suhang},
  booktitle={Findings of the Association for Computational Linguistics: ACL 2025},
  pages={13783--13800},
  year={2025}
}

@article{safe_minja,
  title={Memory Injection Attacks on LLM Agents via Query-Only Interaction},
  author={Dong, Shen and Xu, Shaochen and He, Pengfei and Li, Yige and Tang, Jiliang and Liu, Tianming and Liu, Hui and Xiang, Zhen},
  journal={arXiv preprint arXiv:2503.03704},
  year={2025}
}

@article{safe_deeper_harm,
  title={Deep Research Brings Deeper Harm},
  author={Chen, Shuo and Li, Zonggen and Han, Zhen and He, Bailan and Liu, Tong and Chen, Haokun and Groh, Georg and Torr, Philip and Tresp, Volker and Gu, Jindong},
  journal={arXiv preprint arXiv:2510.11851},
  year={2025}
}

@article{safe_AgentDoG,
  title={AgentDoG: A Diagnostic Guardrail Framework for AI Agent Safety and Security},
  author={Liu, Dongrui and Ren, Qihan and Qian, Chen and Shao, Shuai and Xie, Yuejin and Li, Yu and Yang, Zhonghao and Luo, Haoyu and Wang, Peng and Liu, Qingyu and others},
  journal={arXiv preprint arXiv:2601.18491},
  year={2026}
}

@article{safe_think,
  title={Think twice before you act: Enhancing agent behavioral safety with thought correction},
  author={Jiang, Changyue and Pan, Xudong and Yang, Min},
  journal={arXiv preprint arXiv:2505.11063},
  year={2025}
}

@article{safe_agent_safety,
  title={Agent Safety Alignment via Reinforcement Learning},
  author={Sha, Zeyang and Tian, Hanling and Xu, Zhuoer and Cui, Shiwen and Meng, Changhua and Wang, Weiqiang},
  journal={arXiv preprint arXiv:2507.08270},
  year={2025}
}

@article{safe_Butterfly,
  title={Butterfly effects in toolchains: A comprehensive analysis of failed parameter filling in llm tool-agent systems},
  author={Xiong, Qian and Huang, Yuekai and Jiang, Ziyou and Chang, Zhiyuan and Zheng, Yujia and Li, Tianhao and Li, Mingyang},
  journal={arXiv preprint arXiv:2507.15296},
  year={2025}
}

@inproceedings{safe_tool-mvr,
  title={Advancing tool-augmented large language models via meta-verification and reflection learning},
  author={Ma, Zhiyuan and Liu, Jiayu and Luo, Xianzhen and Huang, Zhenya and Zhu, Qingfu and Che, Wanxiang},
  booktitle={Proceedings of the 31st ACM SIGKDD Conference on Knowledge Discovery and Data Mining V. 2},
  pages={2078--2089},
  year={2025}
}

@inproceedings{safe_ToolReflection,
  title={ToolReflection: Improving Large Language Models for Real-World API Calls with Self-Generated Data},
  author={Polyakov, Gregory and Alimova, Ilseyar and Abulkhanov, Dmitry and Sedykh, Ivan and Bout, Andrey and Nikolenko, Sergey and Piontkovskaya, Irina},
  booktitle={Proceedings of the 1st Workshop for Research on Agent Language Models (REALM 2025)},
  pages={184--199},
  year={2025}
}

@article{safe_LATS,
  title={Language agent tree search unifies reasoning acting and planning in language models},
  author={Zhou, Andy and Yan, Kai and Shlapentokh-Rothman, Michal and Wang, Haohan and Wang, Yu-Xiong},
  journal={arXiv preprint arXiv:2310.04406},
  year={2023}
}

@article{eff_ppl,
  title={Parallelized planning-acting for efficient LLM-based multi-agent systems},
  author={Li, Yaoru and Liu, Shunyu and Zheng, Tongya and Song, Mingli},
  journal={arXiv preprint arXiv:2503.03505},
  year={2025}
}

@article{eff_maci,
  title={Maci: Multi-agent collaborative intelligence for adaptive reasoning and temporal planning},
  author={Chang, Edward Y},
  journal={arXiv preprint arXiv:2501.16689},
  year={2025}
}

@article{eff_incal,
  title={Incalmo: An Autonomous LLM-assisted System for Red Teaming Multi-Host Networks},
  author={Singer, Brian and Lucas, Keane and Adiga, Lakshmi and Jain, Meghna and Bauer, Lujo and Sekar, Vyas},
  journal={arXiv preprint arXiv:2501.16466},
  year={2025}
}

@article{eff_temac,
  title={Temac: Multi-agent collaboration for automated web gui testing},
  author={Liu, Chenxu and Gu, Zhiyu and Wu, Guoquan and Zhang, Ying and Wei, Jun and Xie, Tao},
  journal={arXiv preprint arXiv:2506.00520},
  year={2025}
}

@inproceedings{eff_spedec,
  title={Fast inference from transformers via speculative decoding},
  author={Leviathan, Yaniv and Kalman, Matan and Matias, Yossi},
  booktitle={International Conference on Machine Learning},
  pages={19274--19286},
  year={2023},
  organization={PMLR}
}

@article{eff_rewoo,
  title={Rewoo: Decoupling reasoning from observations for efficient augmented language models},
  author={Xu, Binfeng and Peng, Zhiyuan and Lei, Bowen and Mukherjee, Subhabrata and Liu, Yuchen and Xu, Dongkuan},
  journal={arXiv preprint arXiv:2305.18323},
  year={2023}
}

@article{eff_tot,
  title={Tree of thoughts: Deliberate problem solving with large language models},
  author={Yao, Shunyu and Yu, Dian and Zhao, Jeffrey and Shafran, Izhak and Griffiths, Tom and Cao, Yuan and Narasimhan, Karthik},
  journal={Advances in neural information processing systems},
  volume={36},
  pages={11809--11822},
  year={2023}
}

@inproceedings{eff_rest,
  title={Rest: Retrieval-based speculative decoding},
  author={He, Zhenyu and Zhong, Zexuan and Cai, Tianle and Lee, Jason and He, Di},
  booktitle={Proceedings of the 2024 Conference of the North American Chapter of the Association for Computational Linguistics: Human Language Technologies (Volume 1: Long Papers)},
  pages={1582--1595},
  year={2024}
}

@article{eff_dspy,
  title={Dspy: Compiling declarative language model calls into self-improving pipelines},
  author={Khattab, Omar and Singhvi, Arnav and Maheshwari, Paridhi and Zhang, Zhiyuan and Santhanam, Keshav and Vardhamanan, Sri and Haq, Saiful and Sharma, Ashutosh and Joshi, Thomas T and Moazam, Hanna and others},
  journal={arXiv preprint arXiv:2310.03714},
  year={2023}
}

@article{eff_eco,
  title={Ecoassistant: Using llm assistant more affordably and accurately},
  author={Zhang, Jieyu and Krishna, Ranjay and Awadallah, Ahmed H and Wang, Chi},
  journal={arXiv preprint arXiv:2310.03046},
  year={2023}
}

@article{eff_lost,
  title={Lost in the middle: How language models use long contexts},
  author={Liu, Nelson F and Lin, Kevin and Hewitt, John and Paranjape, Ashwin and Bevilacqua, Michele and Petroni, Fabio and Liang, Percy},
  journal={Transactions of the association for computational linguistics},
  volume={12},
  pages={157--173},
  year={2024}
}

@article{eff_toolkengpt,
  title={Toolkengpt: Augmenting frozen language models with massive tools via tool embeddings},
  author={Hao, Shibo and Liu, Tianyang and Wang, Zhen and Hu, Zhiting},
  journal={Advances in neural information processing systems},
  volume={36},
  pages={45870--45894},
  year={2023}
}

@article{eff_frugalgpt,
  title={Frugalgpt: How to use large language models while reducing cost and improving performance},
  author={Chen, Lingjiao and Zaharia, Matei and Zou, James},
  journal={arXiv preprint arXiv:2305.05176},
  year={2023}
}

@article{eff_swiftsage,
  title={Swiftsage: A generative agent with fast and slow thinking for complex interactive tasks},
  author={Lin, Bill Yuchen and Fu, Yicheng and Yang, Karina and Brahman, Faeze and Huang, Shiyu and Bhagavatula, Chandra and Ammanabrolu, Prithviraj and Choi, Yejin and Ren, Xiang},
  journal={Advances in Neural Information Processing Systems},
  volume={36},
  pages={23813--23825},
  year={2023}
}

@article{eff_orche,
  title={Orchestrating Intelligence: Confidence-Aware Routing for Efficient Multi-Agent Collaboration across Multi-Scale Models},
  author={Wang, Jingbo and Zhao, Sendong and Liu, Jiatong and Wang, Haochun and Li, Wanting and Qin, Bing and Liu, Ting},
  journal={arXiv preprint arXiv:2601.04861},
  year={2026}
}

@article{eff_memgpt,
  title={MemGPT: towards LLMs as operating systems.},
  author={Packer, Charles and Fang, Vivian and Patil, Shishir\_G and Lin, Kevin and Wooders, Sarah and Gonzalez, Joseph\_E},
  year={2023},
  publisher={ArXiv}
}

@inproceedings{eff_gptcache,
  title={Gptcache: An open-source semantic cache for llm applications enabling faster answers and cost savings},
  author={Bang, Fu},
  booktitle={Proceedings of the 3rd Workshop for Natural Language Processing Open Source Software (NLP-OSS 2023)},
  pages={212--218},
  year={2023}
}

@inproceedings{com_failtalms,
  title={Benchmarking failures in tool-augmented language models},
  author={Trevi{\~n}o, Eduardo and Contant, Hugo and Ngai, James and Neubig, Graham and Wang, Zora Zhiruo},
  booktitle={Proceedings of the 2025 Conference of the Nations of the Americas Chapter of the Association for Computational Linguistics: Human Language Technologies (Volume 1: Long Papers)},
  pages={2916--2934},
  year={2025}
}

@article{com_toolhaystack,
  title={Toolhaystack: Stress-testing tool-augmented language models in realistic long-term interactions},
  author={Kwak, Beong-woo and Kim, Minju and Lim, Dongha and Chae, Hyungjoo and Kang, Dongjin and Kim, Sunghwan and Yang, Dongil and Yeo, Jinyoung},
  journal={arXiv preprint arXiv:2505.23662},
  year={2025}
}

@inproceedings{com_asktoact,
  title={Asktoact: Enhancing llms tool use via self-correcting clarification},
  author={Zhang, Xuan and Shen, Yongliang and Zheng, Zhe and Wu, Linjuan and Zhang, Wenqi and Yan, Yuchen and Peng, Qiuying and Wang, Jun and Lu, Weiming},
  booktitle={Proceedings of the 2025 Conference on Empirical Methods in Natural Language Processing},
  pages={13495--13522},
  year={2025}
}

@inproceedings{com_creator,
  title={Creator: Tool creation for disentangling abstract and concrete reasoning of large language models},
  author={Qian, Cheng and Han, Chi and Fung, Yi and Qin, Yujia and Liu, Zhiyuan and Ji, Heng},
  booktitle={Findings of the Association for Computational Linguistics: EMNLP 2023},
  pages={6922--6939},
  year={2023}
}

@article{com_restgpt,
  title={Restgpt: Connecting large language models with real-world applications via restful apis. CoRR, abs/2306.06624, 2023. doi: 10.48550},
  author={Song, Yifan and Xiong, Weimin and Zhu, Dawei and Li, Cheng and Wang, Ke and Tian, Ye and Li, Sujian},
  journal={arXiv preprint arXiv.2306.06624},
  year={2023}
}

@inproceedings{com_toolmaker,
  title={Llm agents making agent tools},
  author={W{\"o}lflein, Georg and Ferber, Dyke and Truhn, Daniel and Arandjelovic, Ognjen and Kather, Jakob Nikolas},
  booktitle={Proceedings of the 63rd Annual Meeting of the Association for Computational Linguistics (Volume 1: Long Papers)},
  pages={26092--26130},
  year={2025}
}

@article{com_latm,
  title={Large language models as tool makers},
  author={Cai, Tianle and Wang, Xuezhi and Ma, Tengyu and Chen, Xinyun and Zhou, Denny},
  journal={arXiv preprint arXiv:2305.17126},
  year={2023}
}

@article{com_voyager,
  title={Voyager: An open-ended embodied agent with large language models},
  author={Wang, Guanzhi and Xie, Yuqi and Jiang, Yunfan and Mandlekar, Ajay and Xiao, Chaowei and Zhu, Yuke and Fan, Linxi and Anandkumar, Anima},
  journal={arXiv preprint arXiv:2305.16291},
  year={2023}
}

@inproceedings{com_expel,
  title={Expel: Llm agents are experiential learners},
  author={Zhao, Andrew and Huang, Daniel and Xu, Quentin and Lin, Matthieu and Liu, Yong-Jin and Huang, Gao},
  booktitle={Proceedings of the AAAI Conference on Artificial Intelligence},
  volume={38},
  number={17},
  pages={19632--19642},
  year={2024}
}

@inproceedings{com_appagent,
  title={Appagent: Multimodal agents as smartphone users},
  author={Zhang, Chi and Yang, Zhao and Liu, Jiaxuan and Li, Yanda and Han, Yucheng and Chen, Xin and Huang, Zebiao and Fu, Bin and Yu, Gang},
  booktitle={Proceedings of the 2025 CHI Conference on Human Factors in Computing Systems},
  pages={1--20},
  year={2025}
}

@article{com_review,
  title={Lifelong learning of large language model based agents: A roadmap},
  author={Zheng, Junhao and Shi, Chengming and Cai, Xidi and Li, Qiuke and Zhang, Duzhen and Li, Chenxing and Yu, Dong and Ma, Qianli},
  journal={IEEE Transactions on Pattern Analysis and Machine Intelligence},
  year={2026},
  publisher={IEEE}
}
